\documentclass[traditabstract]{aa} 

\usepackage{times,epsfig,amssymb,amsmath,natbib}

\newcommand{\xmm}{{\it XMM-Newton}}

\usepackage{placeins}
\usepackage{subfig}

 \usepackage{url}

\usepackage{float}
\usepackage[toc,page]{appendix}
\usepackage{tabularx}
\usepackage{braket}
        
\usepackage{natbib,twoopt}
\usepackage{verbatim}

\usepackage[]{hyperref}
\hypersetup{pdftex,colorlinks=true,allcolors=blue}
\usepackage{hypcap}
\usepackage{txfonts}
\usepackage{color}
\usepackage[]{amsmath}

\begin{document}

\title{The eROSITA Final Equatorial-Depth Survey (eFEDS):}
\subtitle{Characterization of Morphological Properties of Galaxy Groups and Clusters}
\titlerunning{eFEDS properties}
\mail{vittorio@mpe.mpg.de}

\author{Vittorio Ghirardini\inst{1}\thanks{e-mail: \href{mailto:vittorio@mpe.mpg.de}{\tt vittorio@mpe.mpg.de}},
Y.~Emre~Bahar\inst{1}, 
Esra~Bulbul\inst{1},
Ang~Liu\inst{1},
Nicolas~Clerc\inst{2}, 
Florian~Pacaud\inst{3},
Johan~Comparat\inst{1}, 
Teng~Liu\inst{1},
Miriam~E.~Ramos-Ceja\inst{1},
Duy~Hoang\inst{5},
Jacob~Ider-Chitham\inst{1},
Matthias~Klein\inst{1,4,6},
Andrea~Merloni\inst{1}, 
Kirpal~Nandra\inst{1}, 
Naomi~Ota\inst{3,7},
Peter~Predehl\inst{1},
Thomas~Reiprich\inst{3}
Jeremy~Sanders\inst{1},
Tim~Schrabback\inst{3}
}

\authorrunning{V. Ghirardini et al.}

\institute{
\inst{1}Max-Planck-Institut für extraterrestrische Physik, Giessenbachstraße 1, D-85748 Garching, Germany \\
\inst{2}{IRAP, Université de Toulouse, CNRS, UPS, CNES, Toulouse, France}\\
\inst{3}{Argelander-Institut f{\"{u}}r Astronomie (AIfA), Universit{\"{a}}t Bonn, Auf dem H{\"{u}}gel 71, 53121 Bonn, Germany}\\
\inst{4}{Universitaets-Sternwarte Muenchen, Fakultaet fuer Physik, LMU Munich, Scheinerstr. 1, 81679 Munich, Germany}\\
\inst{5}{Hamburger Sternwarte, University of Hamburg, Gojenbergsweg 112, 21029 Hamburg, Germany }\\
\inst{6}{Faculty of Physics, Ludwig-Maximilians-Universit{\"a}t, Scheinerstr. 1, 81679, Munich, Germany}\\
\inst{7}{ Department of Physics, Nara Women's University, Kitauoyanishi-machi, Nara, 630-8506, Japan}
}

\abstract
{Morphological parameters are the estimators for the dynamical state of clusters of galaxies. Surveys performed at different wavelengths through their selection effects may be biased towards different populations of clusters, for example X-ray surveys are biased to detecting cool-core clusters as opposed to SZ surveys being more biased towards non-cool-core systems. Understanding the underlying population of clusters of galaxies in surveys is of the utmost importance for using these samples in both astrophysical and cosmological studies. }
{We present an in-depth analysis of the X-ray morphological parameters of the galaxy clusters and groups detected in the eROSITA Final Equatorial-Depth Survey (eFEDS). The eFEDS Survey completed during the Performance Verification phase of the Spectrum-Roentgen-Gamma(SRG)/eROSITA telescope is designed to provide the first eROSITA X-ray selected sample of galaxy clusters and groups.   }
{We study the eROSITA X-ray imaging data for a sample of 325 clusters and groups that are significantly detected in the eFEDS field. We characterize their dynamical properties by measuring a number of dynamical estimators: concentration, central density, cuspiness, ellipticity, power-ratios, photon asymmetry, and Gini coefficient. The galaxy clusters and groups detected in eFEDS, covering a luminosity range of more than three orders of magnitude and large redshift range out to 1.2 provide an ideal sample for studying the redshift and luminosity evolution of the morphological parameters and characterization of the underlying dynamical state of the sample. Based on these measurements we construct a new dynamical indicator, relaxation score, for all the clusters in the sample.  }
{We find no evidence for bimodality in the distribution of morphological parameters of our clusters, rather we observe a smooth transition from the cool-core to non-cool-core and from relaxed to disturbed states with preference for skewed distributions or log-normal distributions. A significant evolution in redshift and luminosity is also observed in the morphological parameters examined in this study after carefully taking into account the selection effects.  }
{ We determine that our eFEDS-selected cluster sample, differently than ROSAT-based cluster samples, is not biased toward cool-core clusters, but contains a similar fraction of cool-cores as SZ surveys.
}

\keywords{Galaxies: clusters: intracluster medium -- Galaxies: clusters: general -- X-rays: galaxies: clusters } 

\maketitle 

\section{Introduction}

Clusters of galaxies are the most massive virialized systems in the Universe, located at the densest nodes of the cosmic web. Clusters have the majority of their mass in the form of a dark matter halo, whose mass ranges from $\sim10^{13}~M_{\odot}$ (low-mass groups) to $\sim10^{15}~M_{\odot}$ (massive clusters). The baryonic mass is dominated by the so-called intracluster medium (ICM), which is a diffuse gas heated up to millions of degrees Kelvin ($\sim$ keV scale) by the deep gravitational potential well, and therefore emits X-rays mainly through bremsstrahlung. 

According to the hierarchical structure formation scenario, clusters form from and evolve through multiple accretion and merging processes which take place frequently during their lifetime \citep[see][for a review]{kravtsov12}. Such processes leave their imprints in the dynamical state of a cluster, and often manifest themselves as a disturbed ICM morphology with observing features such as shocks \citep[e.g.,][]{Markevitch2002, Markevitch2005, Russell2010}, cold fronts \citep{Vikhlinin2001, Markevitch2007}, and substructures/clumps \citep{Eckert2015, Parekh2015, ghirardini18}. These features have been observed and studied mostly in recent years, thanks to the growth in the volume and depth of X-ray cluster surveys and the good spatial resolution of focusing X-ray telescopes. Apart from these prominent features, an elongated shape of a relaxed cluster can simply serve as an immediate indication of past mergers. Therefore, ICM morphology and its evolution with cosmic time and/or cluster mass can be used as a tracer for the formation history of clusters, and consequently put constraints on the evolution of the large-scale structure \citep[e.g.,][]{Evrard1993, Mohr1995, Suwa2003, Ho2006, weissmann2013}. On smaller scales, feedback activities of the central Active Galactic Nuclei (AGN) can also affect the morphology of the ICM by producing X-ray cavities and regulating the core properties \citep{Fabian+94, Mcnamara2007, Fabian2012}. In fact, several morphological parameters, such as the concentration and central gas density, have been widely used as cool-core/non-cool-core (CC/NCC) indicators \citep[e.g.,][]{hudson+10, santos+08}. Since the measurements of other thermodynamic observables rely on much more expensive X-ray spectroscopic data, cluster morphology is a practical and effective probe to trace the evolution of cool cores up to high redshifts \citep[see, e.g.,][]{santos+08,santos+10, Santos2012}, and therefore provides important clues to reveal the cycle of baryons in the center of galaxy clusters.

Many attempts have been made on the morphological study of galaxy clusters, introducing various morphological parameters \citep[see, e.g,][]{santos+08, rasia+13b, lovisari17, Yuan+20}. However, since the X-ray emission of a cluster can 
 extend up to several Mpc, different morphological parameters evaluate cluster dynamical state at different scales and are reflecting different physical properties. For example, concentration, central density, and cuspiness, are most sensitive to core properties; Ellipticity characterises the distribution of the ICM at large scales, and can be used to estimate how long it will take for the main halo to virialize; Power ratio and Gini coefficient reflect the fluctuation in the surface brightness distribution, related with the level of stochastic gas motions; Symmetry/asymmetry and the separation between the X-ray center and the Brightest Central Galaxy (BCG) mainly indicate the offset of the cluster core with respect to the center of the gravitational potential well, usually associated with features such as core-sloshing, induced by off-center mergers.

Studies on morphological parameters that investigate how they evolve with cluster mass and redshift, or whether these parameters follow a relaxed/disturbed or cool-core/non-cool-core bimodal distribution, are obviously dependent on selection effects of the cluster sample. 
Early studies based on X-ray selected cluster samples mainly focused their attention to possible presence of two distinct cluster populations, cool-cores and non-cool-cores.
While first results did indeed find that clusters can be divided in two populations \citep[e.g.][]{Sanderson+09,cavagnolo+09,hudson+10}, following works did not observe this bimodal distinction \citep[e.g.][]{santos+10,pratt+10,ghirardini_a+17,Yuan+20}.
In the last decade, the detection of the Sunyaev-Zel'dovich \citep[SZ,][]{Sunyaev+72} effect of galaxy clusters opened a new window to observe and study the morphological properties of clusters. Thanks to its largely redshift independent signal, recent SZ surveys conducted by \emph{Planck} \citep{PESZ,PSZ1,PSZ2}, the South Pole Telescope \citep[SPT,][]{Bleem+20}, and the Atacama Cosmology Telescope \citep[ACT,][]{Hilton+21} have particularly increased the volume of ICM-based cluster samples at high redshift. Since X-ray and SZ signals have different dependencies on cluster thermodynamic properties, X-ray and SZ samples may not represent the same underlying cluster population. 
In particular, X-ray surveys are found to be prone to detect more relaxed clusters, while SZ surveys detect more disturbed clusters \citep[see, e.g.,][]{rossetti16,rossetti17,santos+17,lovisari17}. On the other hand, other works report no significant difference between X-ray and SZ samples in their morphology distribution \citep[][]{nurgaliev+13,Mantz+15,nurgaliev+17,mcdonald17}. The quantitative comparison between these results is not possible due to the issue that both the samples and morphological parameters used in these works differ from each other. Therefore, a coherent picture describing the morphological evolution of clusters is still missing. The best approach to establishing such a picture is through a more unified investigation on a more complete cluster sample that spans a large mass and redshift range.

Recently, the extended ROentgen Survey with an Imaging Telescope Array \citep[eROSITA,][]{predehl+21} onboard the Spectrum-Roentgen-Gamma (SRG) mission started its X-ray all-sky survey with an unprecedented sensitivity \citep{sunyaev+21}. 
eROSITA has been designed to provide unique survey science capabilities that will enable key cosmological studies with clusters of galaxies. In fact, with its large effective area (1365~cm$^2$ at 1~keV), good spatial resolution (half energy width of 26~arcsec averaged over the field of view at 1.49~keV) and good spectral resolution ($\sim 80$~eV full width half maximum at 1~keV), and large field-of-view (1~deg diameter) it is able to scan quickly and efficiently large areas of the sky \citep{predehl+21}. 

The eROSITA Final Equatorial Depth Survey (eFEDS) is uniquely designed to test and demonstrate these survey capabilities. Covering an area of $\sim$140~deg$^2$, it has been observed during the Performance Verification (PV) phase at a depth of $\sim$2.2~ks (non vignetted), which is slightly deeper than the exposure of the final All-Sky Survey in equatorial fields.
eFEDS will enable calibration of key mass scaling relations, combining X-ray properties with weak lensing masses of detected groups and clusters, obtained from the detailed analysis of Hyper-Suprime-Camera (HSC) data on the Subaru telescope (Chiu et al. submitted).

In this work, we study the morphological properties of the clusters and groups detected by eROSITA in the eFEDS field. We measure multiple morphological parameters using the eROSITA X-ray imaging data, we study the correlations in these parameters, and investigate their possible evolution with cosmic time and cluster luminosity after accounting for the selection effects of the sample. The paper is organized as follows. In Sect.~\ref{sec:data_analysis}, we briefly introduce our eFEDS cluster sample and eROSITA data analysis; In Sect.~\ref{sec:morph_description}, we describe the morphological parameters that are studied in this work; In Sect.~\ref{sec:result}, we present our results regarding 
the dependence of morphological parameterson redshift and luminosity, we introduce a new metric to measure morphological properties, and provide the comparison with previous literature results. Our conclusions are summarized and discussed in Sect.~\ref{sec:conclusion}. Throughout this paper, we assume a concordance $\Lambda$CDM cosmology with $H_{0} = 70$~km~s$^{-1}$~Mpc$^{-1}$, $\Omega_{\rm m} = 0.3$, and $\Omega_\Lambda = 0.7$. Error bars correspond to the 1$\sigma$ confidence level, unless noted otherwise.

\section{Data analysis}
\label{sec:data_analysis}
The eROSITA Standard Analysis Software System (eSASS, Brunner et al. submitted) is used to process the eFEDS data. The {\tt eSASS} ({\tt eSASSusers\_201009}) software consists of a pipeline whose end products are calibrated event lists for each eROSITA telescope module (TM), applying pattern recognition and energy calibration, with determination of good time intervals (GTI), dead times, flagging of corrupted events, frames, and bad pixels.
Celestial coordinates (i.e. equatorial R.A. and Dec.) are assigned to each reconstructed event using star-tracker and gyro data. This allows to project the photons onto the sky, and thus enables the production of images and exposure maps.
In this work, we select all valid pixel patterns, namely, single, double, triple, and quadruple events, and we remove events in the corners of the square CCDs, which are events detected with off-axis angles $\gtrsim$30~arcmin, where the vignetting and point spread function (PSF) calibration is currently less accurate.

The source detection procedure is performed on the merged data from all seven eROSITA telescope modules. The detection is based on a sliding-cell method. In a first step, the algorithm scans the X-ray image with a local sliding window, which identifies enhancements above a certain threshold. The detected candidate objects are then excised from the images. The resulting source-free images are used to create background maps via adaptive filtering. The sliding window detection is then repeated, but this time using the created background map to search for signal excess with respect to it, producing another candidate source list. For each source candidate, a maximum likelihood point spread function fitting algorithm determines the best-fit source parameters, detection and extent likelihoods (Brunner et al. submitted). Applying this algorithm on the eFEDS data, using images in the $0.2-2.3$~keV energy band, an extent likelihood threshold of 6, a detection likelihood threshold of 5, and source extension threshold of $60$~arcsec, we detect $542$~extended sources in total. For further details on the construction of the cluster catalog and some of the properties of the detected clusters from the eFEDS survey we refer the reader to Liu A. et al. (submitted).

\begin{figure*}
\includegraphics[width=\textwidth]{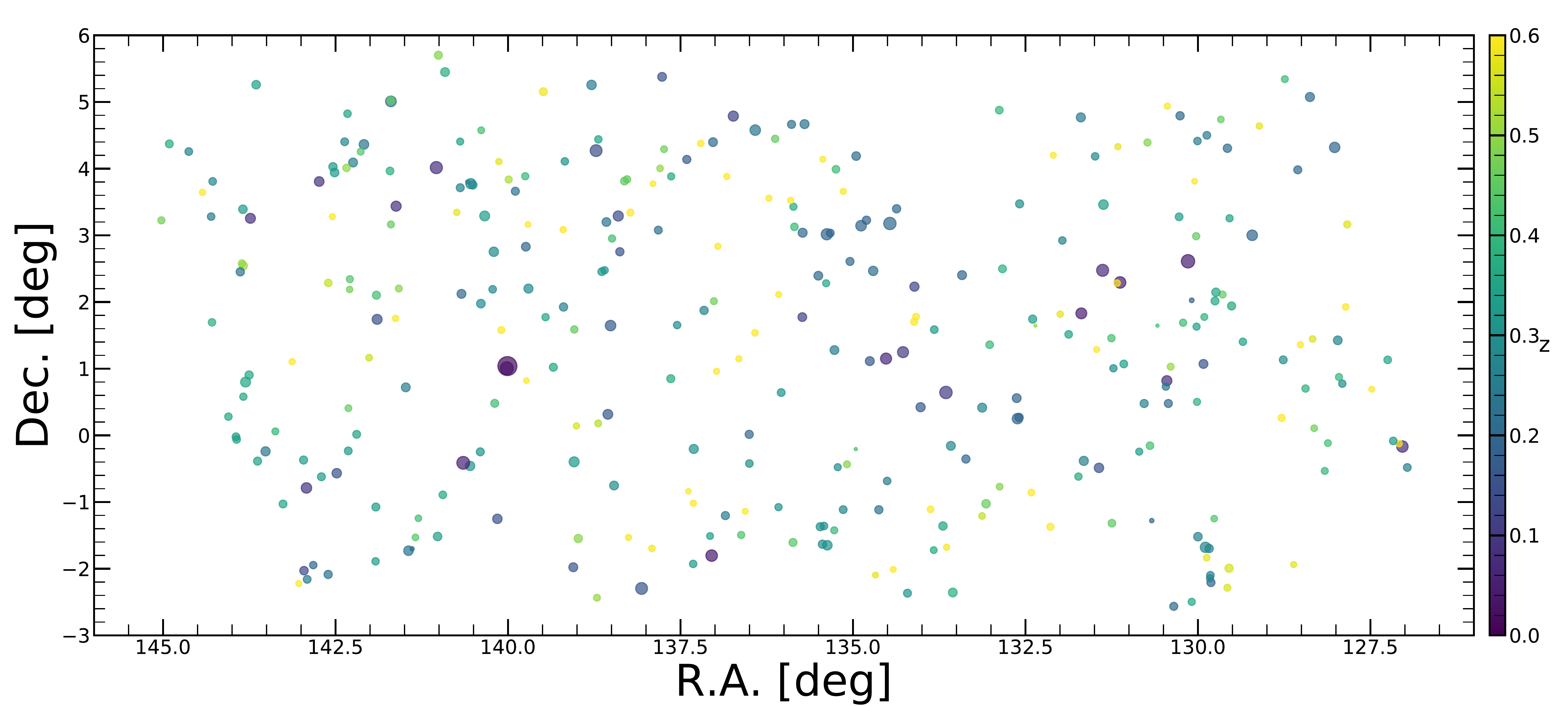}
\caption{Sky image showing the location of the eFEDS clusters studied in this work. Data points are color coded according with their redshift (for clarity reasons we have restricted the redshift range in the color-bar), and the size of the points is proportional to their $R_{500}$ in arcmin.}
\label{fig:radec}
\end{figure*}

\begin{figure}
\includegraphics[width=0.5\textwidth]{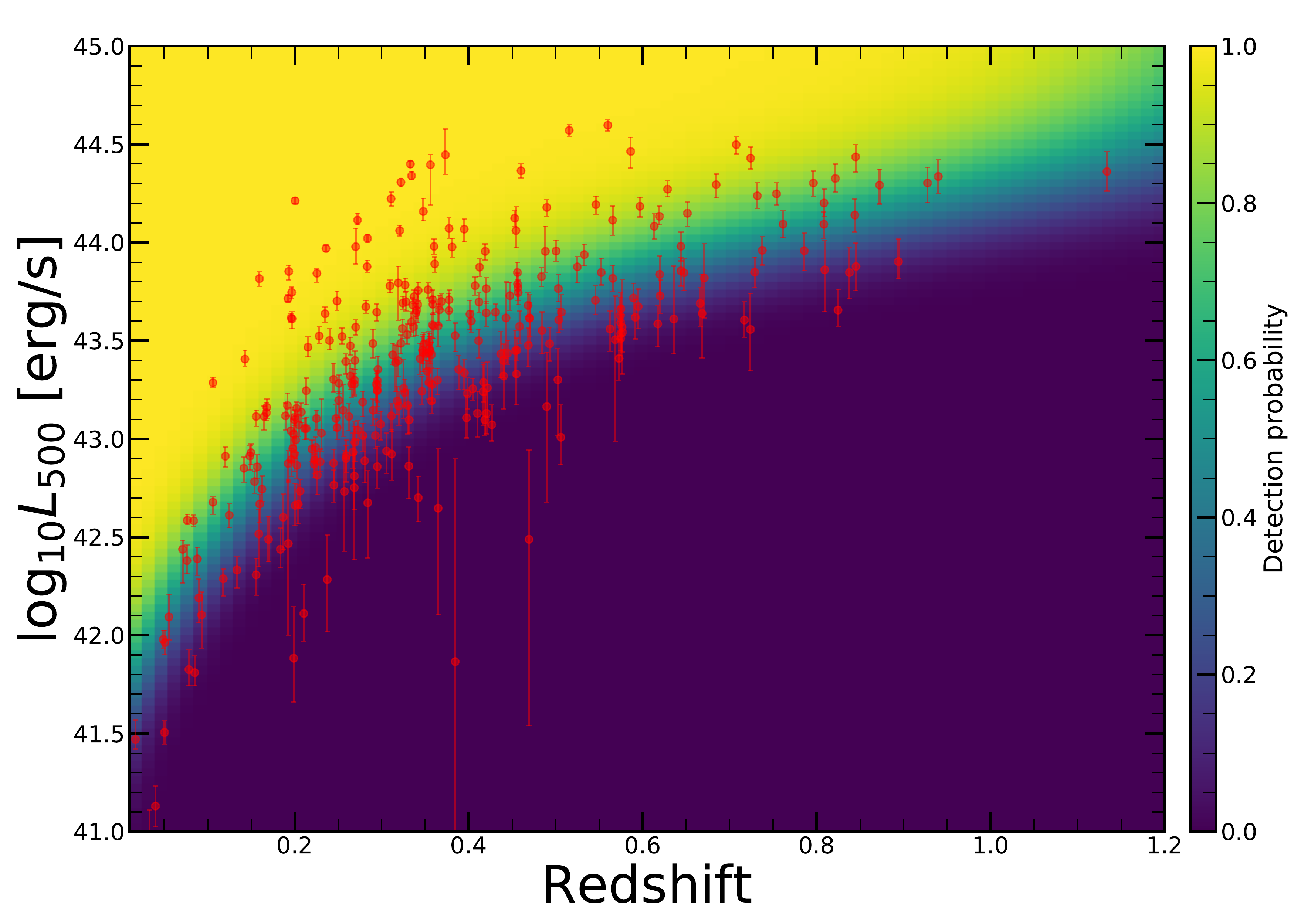}
\caption{Luminosity redshift distribution of the eFEDS clusters subsample studied in this work. As a background color reference we show the eFEDS selection function with the same cuts according to extent and detection likelihoods.}
\label{fig:Lz}
\end{figure}

In the full extent selected catalog, our dedicated realistic simulations of the field, presented in Liu T. et al (submitted), predict a contamination level of $\sim$20\%. To obtain a cleaner sample, we further apply selection criteria of extent likelihood and detection likelihood values larger than 12. This selection reduces the fraction of spurious clusters to $\sim$14\% in the eFEDS sample, while decreasing the sample size to a total of 325 clusters.
In Fig.~\ref{fig:radec}, we show the resulting sky image showing the location of these clusters where data points are color coded with redshift, and circle radius is proportional to $R_{500}$\footnote{$R_{500}$ is defined as the region within which the mean cluster density is 500 times the critical density of the Universe} in arcmin (Chiu et al. submitted). Fig.~\ref{fig:Lz} shows the luminosity (Bahar et al. in preparation) versus redshift distribution of this sample. The luminosity covers a range from $9 \times 10^{40}$ erg/s to $4 \times 10^{44}$ erg/s and the redshift of the sample ranges from 0.017 out to 1.1 (Klein et al. submitted) using the recently developed code MCMF \citep[Multi-Component Matched Filter,][]{Klein+2018,Klein+2019}.

\subsection{eROSITA imaging analysis}
\label{sec:imaging}
For all the clusters in our sample we apply the procedure presented in \citet{ghirardini+21} and Liu A. et al. (submitted) in order to recover surface brightness and density profiles. 
While we direct the reader to \citet{ghirardini+21} for details, we here provide a short summary of the major steps of the analysis.
We start from the clean event files, we extract images and exposure maps in the 0.5--2.0~keV energy band using {\tt eSASS} tasks \texttt{evtool} and \texttt{expmap} respectively with size as large as 4 times $R_{500}$ around each cluster position. 
We directly fit the 2D distribution of the X-ray photons in the produced images of these clusters, allowing for multiple cluster fits at the same time (when more than one cluster is present in the image), for fitting of cluster centers, and for modeling of point sources detected.
As in our previous works we model only bright point sources (with more than 0.1 count per second), because modelling all of them would be computationally too expensive. These point sources are modelled as delta functions convolved with the PSF, thus taking into account effects caused by PSF wings.
Faint point sources are masked by removing a circular region around them whose radius is large enough to assure that point source signal outside is consistent with the background level.
The clusters are modelled in 2D using the \citet{vikhlini+06} model:
\begin{equation}
n_e^2(r) = n_0^2 \cdot \left( \frac{r}{r_c} \right)^{-\alpha} \cdot \left( 1 + \left( \frac{r}{r_c} \right)^2 \right)^{-3\beta+\alpha/2} \cdot \left(  1 + \left( \frac{r}{r_s} \right)^3 \right)^{-\epsilon/3} ,
\end{equation}
where the priors on our parameters are $\epsilon < 5$ \citep[as suggested by][]{vikhlini+06}, $\beta > 1/3$, and $\alpha > 0$, and we freeze $r_s = r_c$. If more clusters are present in the image, then this model above is also used for these other clusters. The center of the clusters is not fixed, but allowed to vary using a gaussian prior centered on the detection location and $\sigma$ of 20 arcsec. The resulting model cluster images are convolved with the PSF of eROSITA. The instrumental background, see \citet{Freyberg+20}, (particle-induced background and camera noise) model folded with the un-vignetted exposure map, and the sky background (including the cosmic X-ray background and the soft background component from the galactic halo and local bubble) folded with the vignetted exposure map are added to the total model. 

We then fit the image obtained from the eROSITA observations with the  model image in 2D using the Monte Carlo Markov Chain (MCMC) code \textit{emcee} \citep{emcee} to find the best-fit model parameters. We assume a Poisson log-likelihood function $\sum N_i - \mu_i \log N_i$, where $N_i$ are the model predicted counts, and $\mu_i$ are the observed counts in each pixel of the image. 

The fitting of the images can be interpreted as a true density once we consider the emissivity of the gas, which is determined, as in our previous works, by making use of the spectral information. Similarly as in \citet{ghirardini+21} and Liu A. et al. (submitted) we fit cluster spectra within $R_{500}$ to obtain the conversion factor from count rate to emissivity, thus determining the physical properties of the clusters. In this work, we make use of only the products of imaging data, that are only density profiles, surface brightness profiles, and luminosity within $R_{500}$. The final luminosities used in this work are provided in Bahar et al. in preparation.

\section{Description of the Morphological Parameters}
\label{sec:morph_description}
In this section, we describe the different morphological parameters used in work, which are the concentration parameter, central density, cuspiness, ellipticity, power ratios, photon asymmetry, and Gini coefficient. We also provide details on how they are measured, and how instrumental factors affects these parameters. All these parameters were used in previous works to characterize the dynamical states of the underlying cluster samples in X-ray and SZ surveys \citep[e.g.][]{santos+10,rossetti17,nurgaliev+17,lovisari17}. 

For each of the computed parameters we calculate the posterior distribution, from which we can compute the median value and asymmetric errors using 16$^{\rm th}$, 50$^{\rm th}$, and 84$^{\rm th}$ percentiles of the distribution. We further point out that the entire distribution that we have calculated is used in both the fitting processes and the error propagation.
In cases where the parameter is computed directly from a previous MCMC chain, like concentration, central density, ellipticity, and cuspiness, the distribution is computed directly by randomizing the choice of the parameter set in the second half of the chain. While in cases where parameters are computed directly from the images, like Gini coefficient, power ratios, and photon asymmetry, their distribution is computed by using Monte Carlo process, by randomizing the pixel values of the observed cluster images.

\subsection{Central density}
The central value of the gas density, $n_0$, is another indicator of the relaxation state of clusters, since relaxed systems are expected to have higher central densities \citep{hudson+10}. We use the value of the electron density computed at 0.02 $R_{500}$.
This value is quite close to the center, and for the vast majority of eFEDS clusters this is well within the field-of-view (FoV) averaged PSF value of eROSITA. However, as we also specified for the concentration parameter, the electron density profile is deconvolved by eROSITA's PSF, see Sect.~\ref{sec:imaging}. Therefore, the $n_0$ we present in this work is PSF-independent, and can be used to compare with previous results.
We do not consider the self-similar evolution of the density profile \citep{mcdonald17, Ghirardini+2020} as a redshift correction on this morphological parameter because we will model it separately in order to identify the actual evolution of the central density with redshift.

\subsection{Concentration}
The concentration parameter, $c_{\rm SB}$, has been introduced by \cite{santos+08} as an indicator of the presence of a peaked X-ray surface brightness, which has been shown to correlate with the relaxation state of clusters. It is defined as the ratio between the integrated surface brightness in two different circular apertures. In this work we use two definitions: the original one introduced by \cite{santos+08}, 
\begin{equation}
\quad \quad  c_{\rm SB, \ 40-400 \rm kpc} = \frac{S_B(< 40 \rm kpc)}{S_B(< 400 \rm kpc)},
\end{equation}
and one scaled with $R_{500}$  \citep{maughan+12}
\begin{equation}
\quad \quad  c_{\rm SB, \ R_{500}} = \frac{S_B(< 0.1 R_{500})}{S_B(R_{500})}.
\end{equation}

When calculating the concentration parameter for our sample, the effects caused by eROSITA's PSF are fully taken into account. In fact, the concentration is not simply the count ratio in the two apertures, but comes from the two dimensional image fitting we have described in Sect.~\ref{sec:imaging} which produces PSF-deconvolved surface brightness profiles. 
This also means that the concentration we measure here, being corrected for instrumental effects, can be directly compared with previous results with different instruments.

\subsection{Ellipticity}
The ellipticity parameter, $\epsilon$, is defined as the ratio between the minor and major axes. 
Contrary to several past studies, we do not compute the two axes using the second-order
moments (variance) of the flux distribution in the cluster image because statistically the variance is the ``width'' of the distribution only when the distribution is gaussian. Rather, it is known that the photon distribution in cluster images is not gaussian but is similar to a $\beta$-model.
Therefore we measure the ellipticity by extending the analysis described in Sect.~\ref{sec:imaging}, allowing the density profile to be elliptical and rotated on the plane-of-the sky. This is accomplished by introducing a rotation angle and the ellipticity in the model construction of the image.
In short, the 2 dimensional density profile along the \emph{x}-axis is squeezed by changing the core radius as $r_{c,x} = \epsilon \times r_{c,y}$, and then it is rotated by an angle $\theta$. The best fitting parameter $\epsilon$ is the ellipticity we adopt in this work. We note that on these two parameters we put uniform prior between 0 and 1 on the ellipticity, and uniform prior between 0 and $\pi$ on the rotation angle, not between 0 and 2$\pi$ to avoid problems caused by the $\pi$ rotational symmetry of the ellipse.

\subsection{Cuspiness}
The cuspiness parameter, $\alpha$, introduced by \citep{Vikhlinin+07}, measures the slope of the density profile at a specific radius. We use the value of the slope at 0.04 $R_{500}$ as:
\begin{equation}
\quad \quad \alpha = - \frac{d \log \rho_g(r)}{d \log r} \Bigg|_{r = 0.04 R_{500}},
\end{equation}
\noindent where $\rho_g(r)$ is the gas mass density profile. This particular radius has been already used in the literature \citep[e.g.,][]{lovisari17}, due to the fact that it is close enough to the core to have cooling play an important effect, and is far enough to avoid the flattening of the profile caused by the AGN outflows.
Also in this case, by following the procedure highlighted in Sect.~\ref{sec:imaging} we recover PSF deconvolved value for this slope, thus its value can be directly compared with results in the literature.

\subsection{Power ratios}
The power ratios $P_{m}$ consist of a 2-dimensional decomposition of the surface brightness distribution within a specific aperture, of $R_{500}$, and they account for radial fluctuations since the higher the order of the power ratio the higher the sensitivity to smaller fluctuations. They are introduced by \citet{buote+95}, and are defined as $P_{m0}=P_{m}/P_0$, where
\begin{equation}
\quad \quad  P_0 = [a_0 \log (R_{500})]^2 ,
\end{equation}
and 
\begin{equation}
\quad \quad  P_m = \frac{1}{2 m^2 R_{500}^{2m}} (a_m^2 + b_m^2) .
\end{equation}
The values of $a_m$ and $b_m$ are calculated as
\begin{align}
\quad \quad  a_m(R) = & \int_{r<R} S_B(x) r^m \cos(m \phi) d^2 x , \\
\quad \quad  b_m(R) = & \int_{r<R} S_B(x) r^m \sin(m \phi) d^2 x , 
\end{align}
where the X-ray surface brightness $S_B(x)$ is calculated at the position expressed in polar coordinates $x=(r,\phi)$. In this work, we include only the power ratios up to order 4, that is $P_{10}$, $P_{20}$, $P_{30}$, and $P_{40}$.
We remind the reader that power ratios can only be directly computed from the images, therefore it is not possible to be corrected for eROSITA's PSF. Since the results will be affected by the PSF, comparison with the literature will be challenging due to instrumental differences. For this reason we do not compare the obtained values with previous studies.

\subsection{Photon asymmetry }
Introduced by \cite{nurgaliev+13}, the photon asymmetry, $A_{\rm phot}$, quantifies the degree of rotational symmetry in the emission of an object. 
Interestingly, \cite{nurgaliev+13} investigated how does the photon asymmetry evolve when a cluster is moved to a different redshift. They find that this parameter is insensitive to the redshift of the cluster, thus it is quite useful in determining the cluster morphology for cluster samples spanning a large redshift range.
Indeed, it has been introduced to study SPT-selected clusters, which span a large redshift range, from 0.2 to 1.2, hence need a cluster morphological parameter independent of redshift.
To compute it, we first use Watson's test \citep{watson+61}, which compares two cumulative distributions (similarly to the Kolmogorov-Smirnov test), evaluating the distance between the two:
\begin{equation}
\quad \quad  U_{N}^{2}[F_N, G] = N \cdot \underset{\phi_0}{\rm min} \int (F_N - G)^2 dG ,
\end{equation}
\noindent where $N$ is the total counts in an annulus, $F_N$ is the observed cumulative distribution of the angle of all the photons located in a given annulus, $G$ is the expected cumulative distribution assuming an axis-symmetric cluster, and $\phi_0$ is the starting angle for the cumulative distribution. In short, $U_{N}^{2}$ is the minimum value of the integrate squared difference between the observed and expected cumulative distribution over all possible starting points for the starting angles. If $C$ is the number of cluster counts in the annulus, then the distance between the two distribution is
\begin{equation}
\quad \quad  \hat{d}_{N,C} = \frac{N}{C^2} \left( U_N^2 - \frac{1}{12} \right).
\end{equation}
Finally, we can write the photon asymmetry as
\begin{equation}
\quad \quad  A_{\rm phot} = 100 \sum_{k=1}^4 C_k \hat{d}_{N_k,C_k} / \sum_{k=1}^4 C_k ,
\end{equation}
\noindent where the sum is performed over a set of annuli, which have the edges of 0.05, 0.12, 0.2, 0.3, 1 in units of $R_{500}$.
This parameter, being calculated directly from images, is sensitive to eROSITA's PSF. It, therefore, should not be directly compared with previous works in the literature, since the PSF might affect the redshift evolution.

\subsection{Gini coefficient}Distribution of morphological parameters for the eFEDS clusters
The Gini coefficient is a standard measure of income inequality in economy. It was introduced in astronomy by \cite{Abraham+03}. However here we use the definition in \cite{Lotz+04} to measure the X-ray flux inhomogeneities in galaxy clusters:
\begin{equation}
\quad \quad G = \frac{1}{| \bar{K} | n (n-1)} \sum_i (2i -n -1) | K_i | ,
\end{equation}
\noindent where $K_{i}$ is the pixel value of the image in the $i$-th pixel,$n$ is the total number of pixels, and $\bar{K}$ is the mean of the absolute values of all the pixels in the image. We point out that the Gini coefficient has a small dependency on surface brightness, and does not require a precise center. Similarly, being a parameter calculated directly from images, it is dependent on eROSITA's PSF, and therefore cannot be directly compared with previous works.

\begin{figure*}
\includegraphics[width=\textwidth]{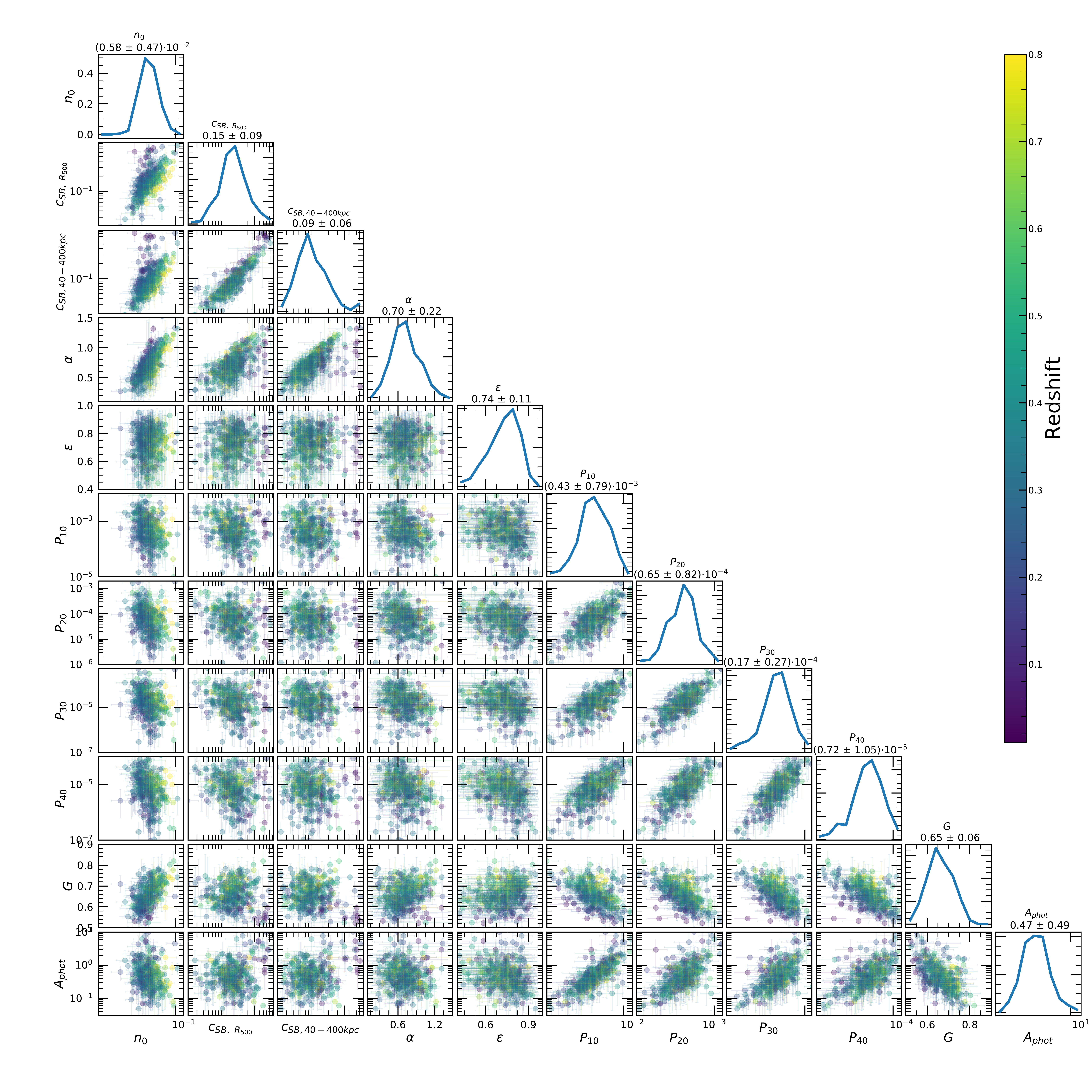}
\caption{Distribution of morphological parameters for the eFEDS clusters.}
\label{fig:morph}
\end{figure*}

\begin{figure}
\includegraphics[width=0.5\textwidth]{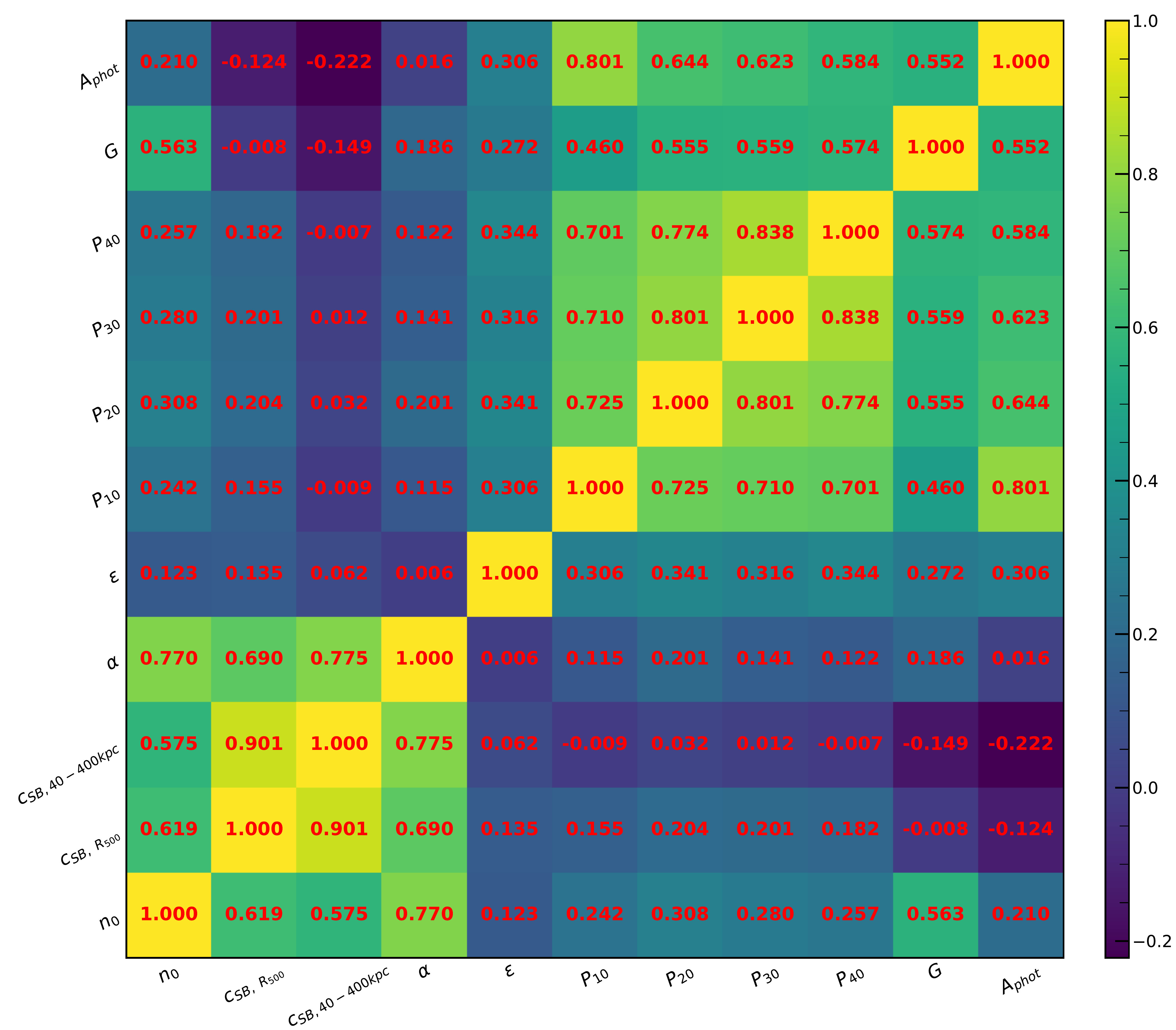}
\caption{Covariance between the morphological parameters in the eFEDS cluster sample.}
\label{fig:morph_cov}
\end{figure}

\vspace{0.5 cm}

In Fig.~\ref{fig:morph} we show the parameter-parameter planes and in Fig.~\ref{fig:morph_cov} we show the correlation matrix between the parameters. As observed in previous works \citep[e.g.][]{lovisari17} there is strong correlation in-between core sensitive parameters, like central density and concentration, and in-between core insensitive parameters, like power ratios and photon asymmetry.

\section{Results}
\label{sec:result}

The eFEDS galaxy cluster and group sample covers the largest range in both luminosity and redshift among the current X-ray and SZ selected samples and provide a unique opportunity to investigate the evolution of the several morphological parameters with redshift and luminosity. In this section, we provide the results of the evolution analysis and compare our result with the similar studies in the literature based on different selections. However, one must be careful when comparing these results as the recently launched eROSITA's detector properties, for example PSF and effective area in particular, are quite different than the other instruments launched more than 20 years ago, like \emph{Chandra} and \emph{XMM-Newton}. A direct comparison with previous works on the morphological properties of clusters needs to be carefully interpreted, and a fair comparison is only possible for those parameters for which the instrumental effects have been corrected, namely concentration, central density, and cuspiness.

\subsection{Analysis of the Redshift and Luminosity Quartiles}
\label{sec:evo}

\begin{figure*}
    \centering
    \includegraphics[width=0.33\textwidth]{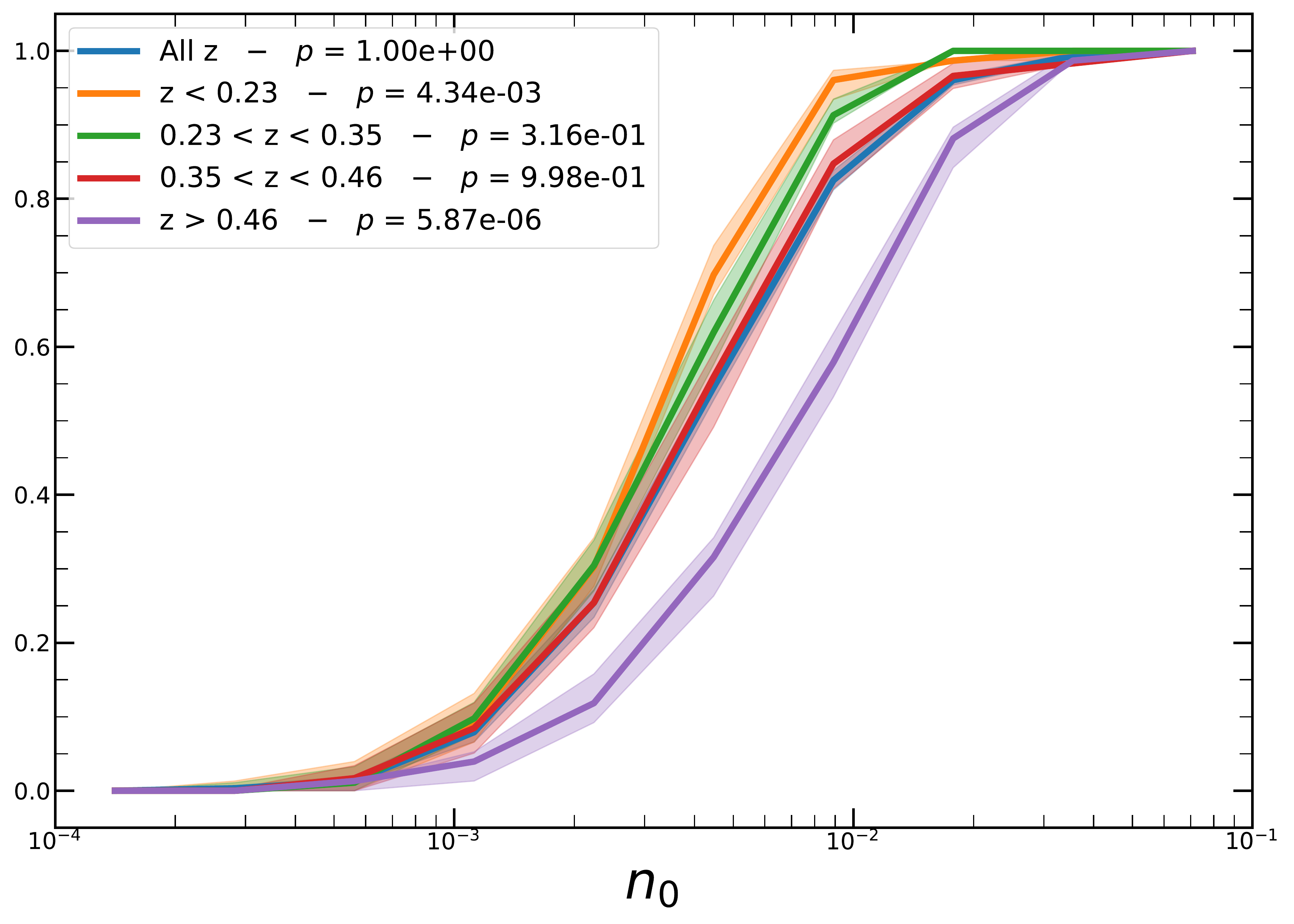}~
    \includegraphics[width=0.33\textwidth]{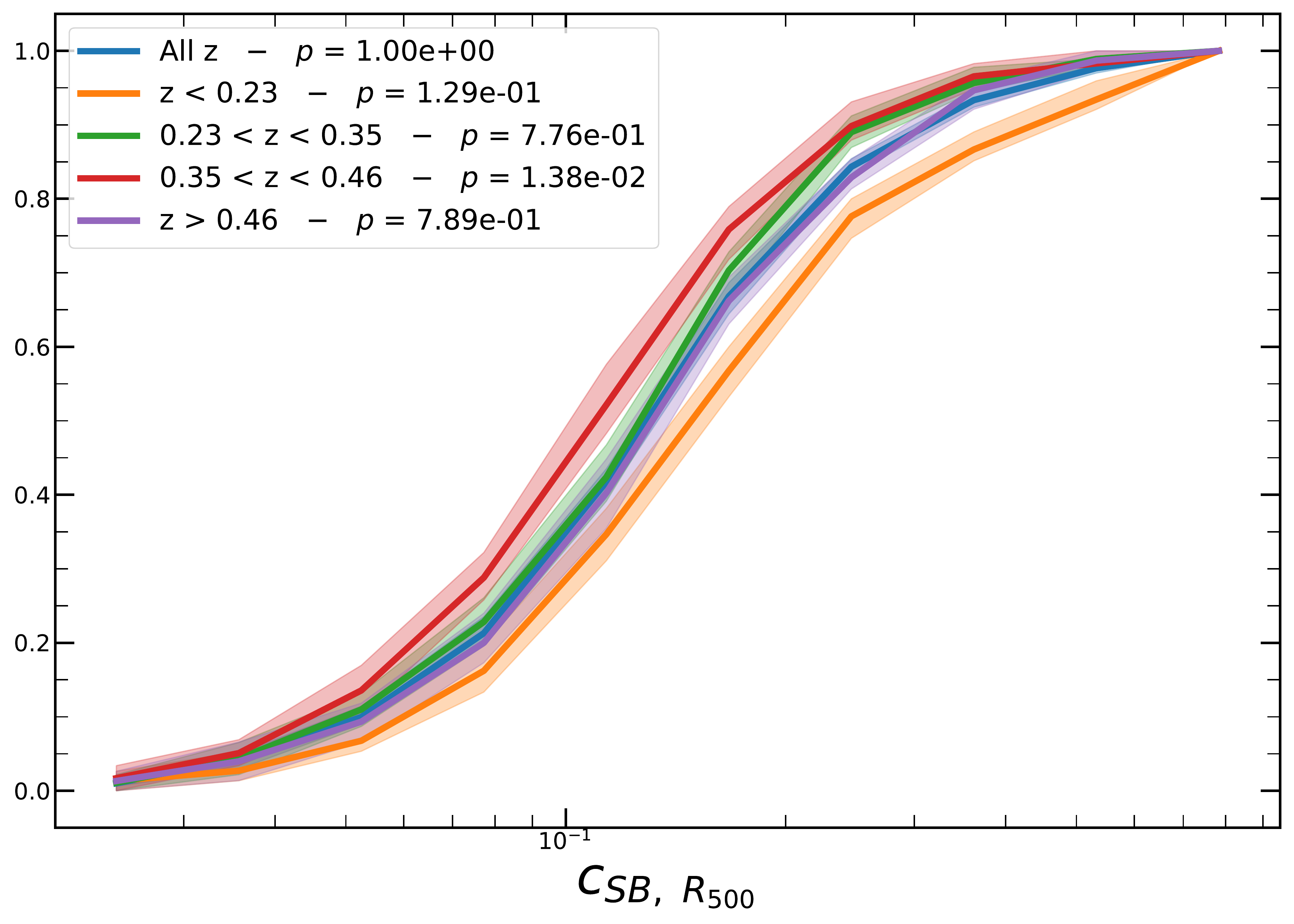}~
    \includegraphics[width=0.33\textwidth]{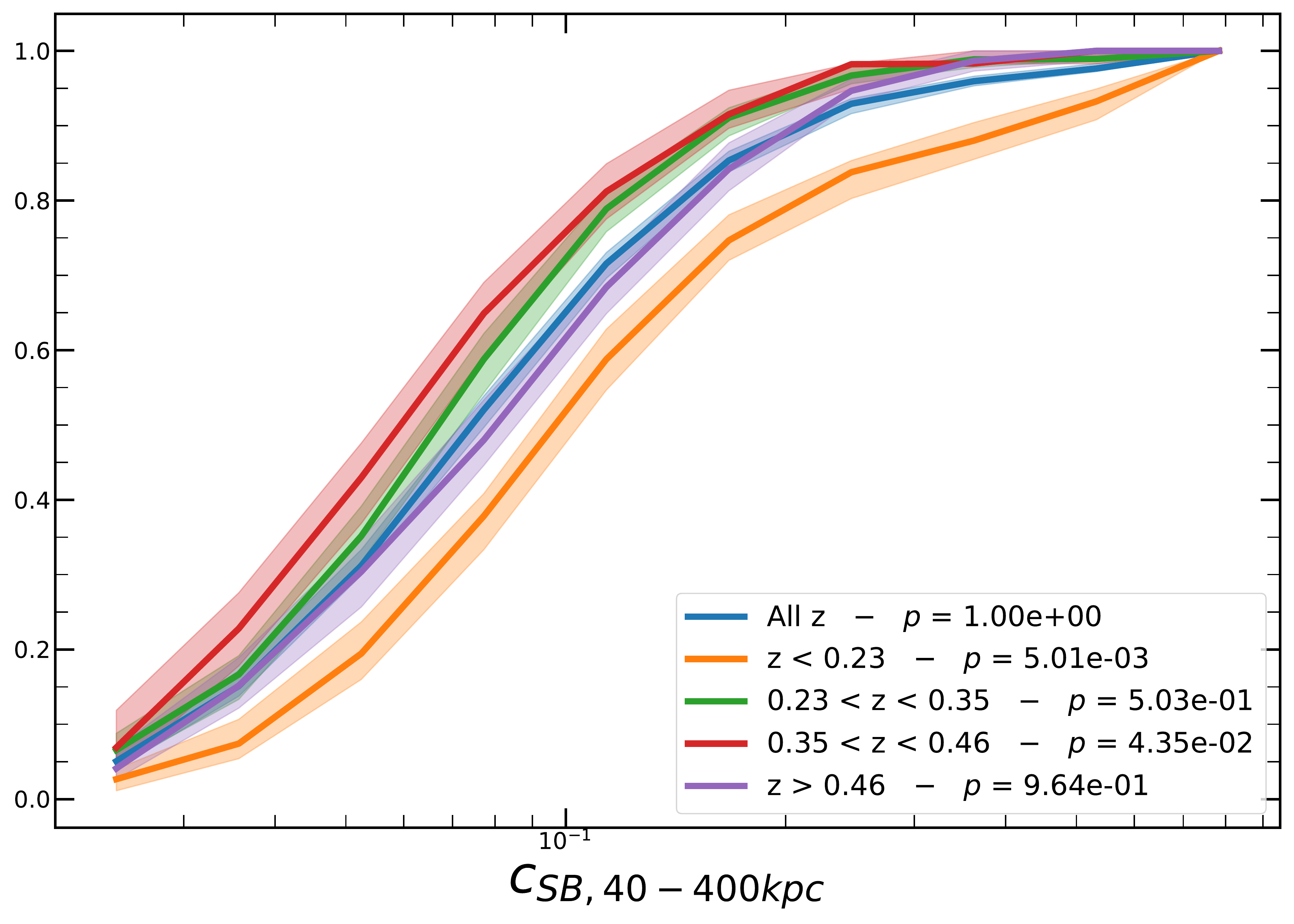}
    
    \includegraphics[width=0.33\textwidth]{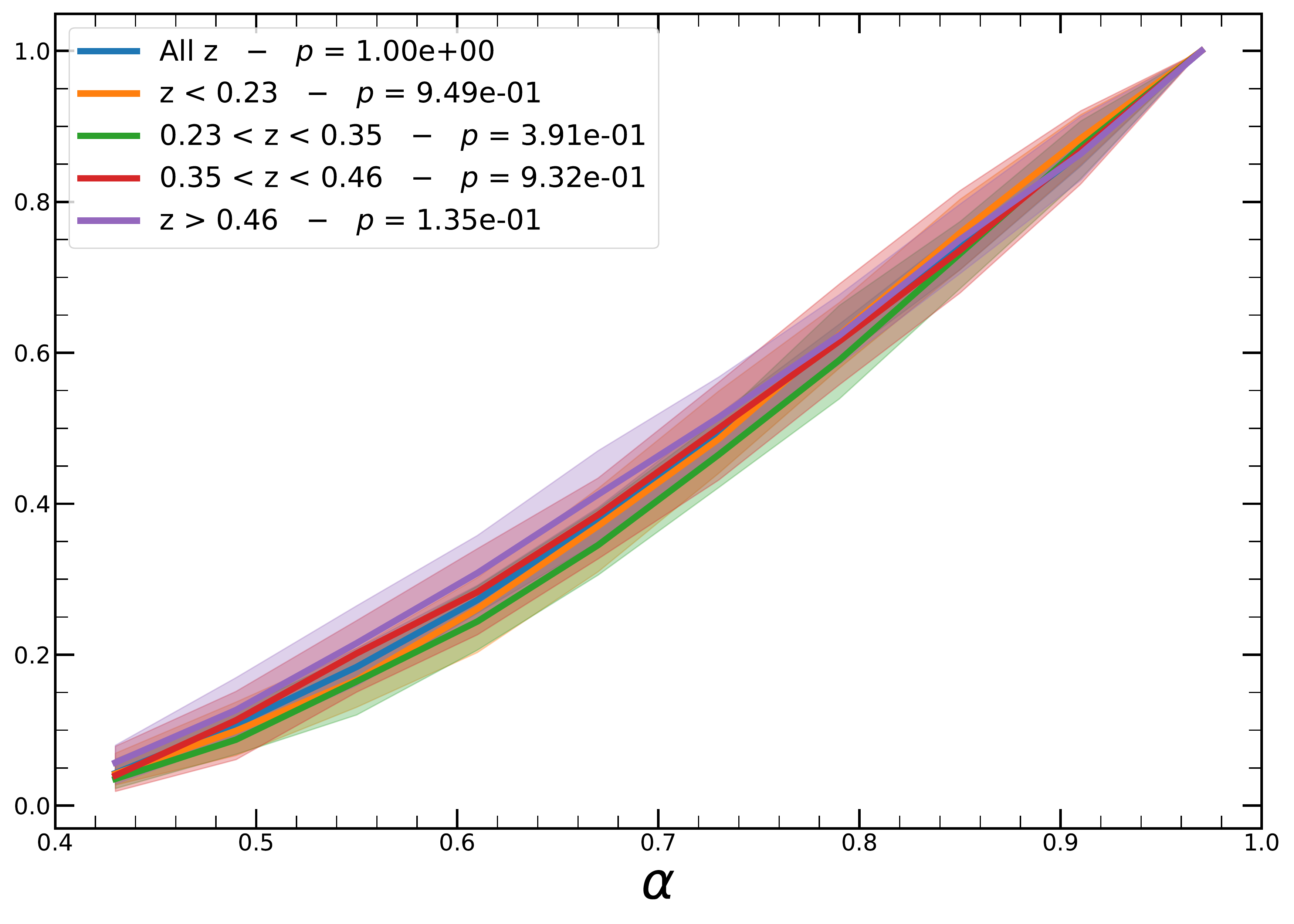}~
    \includegraphics[width=0.33\textwidth]{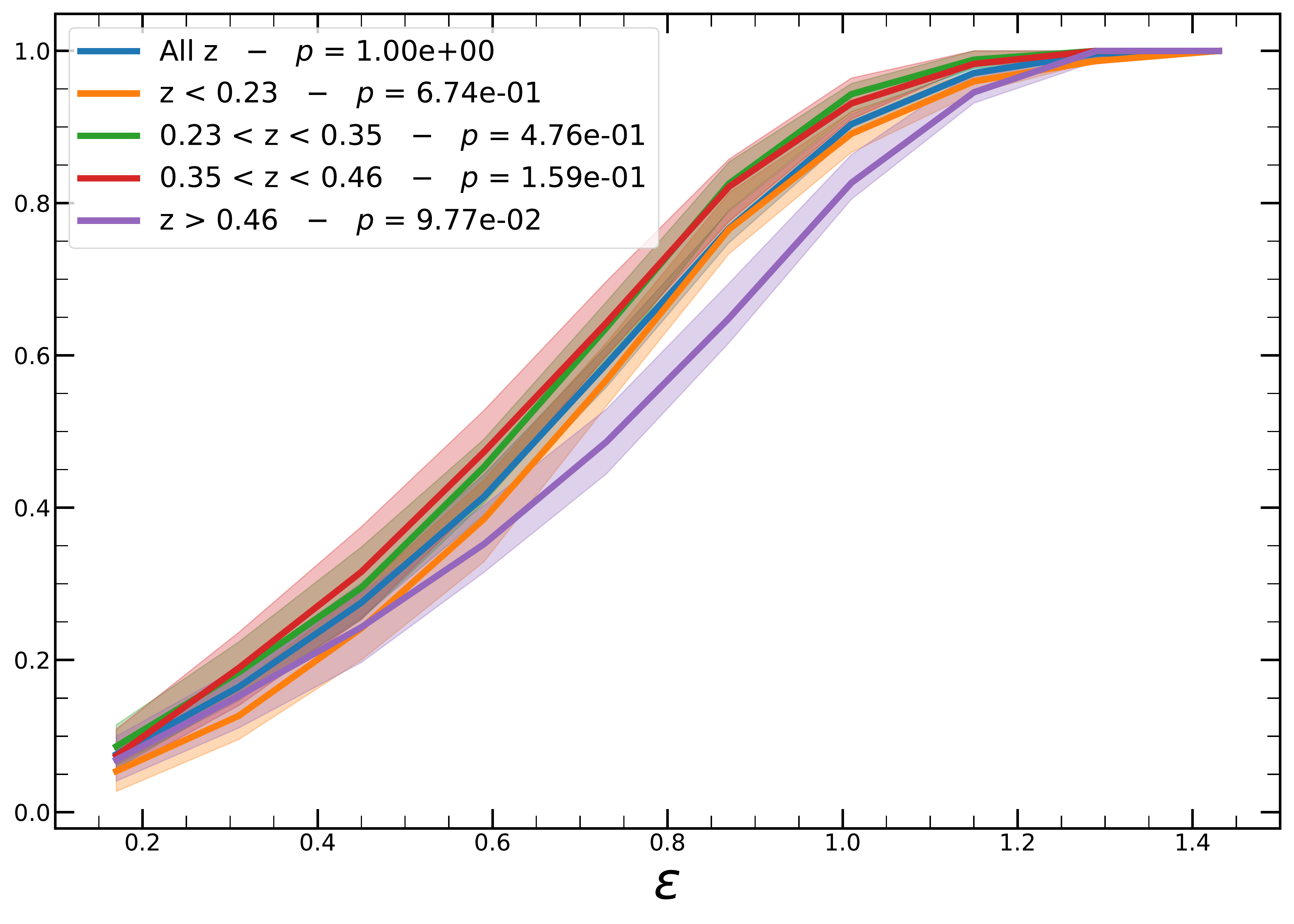}~
    \includegraphics[width=0.33\textwidth]{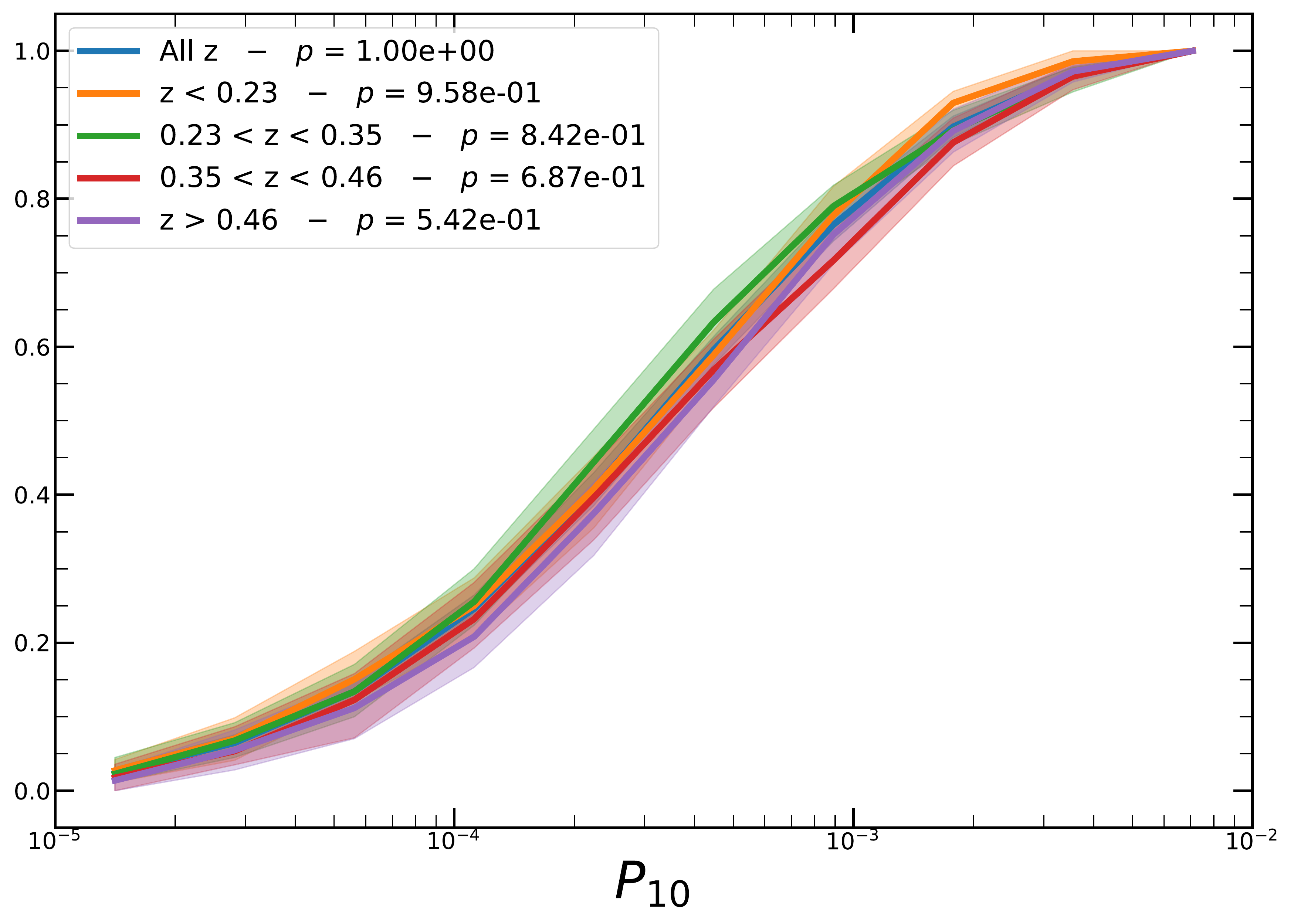}

    \includegraphics[width=0.33\textwidth]{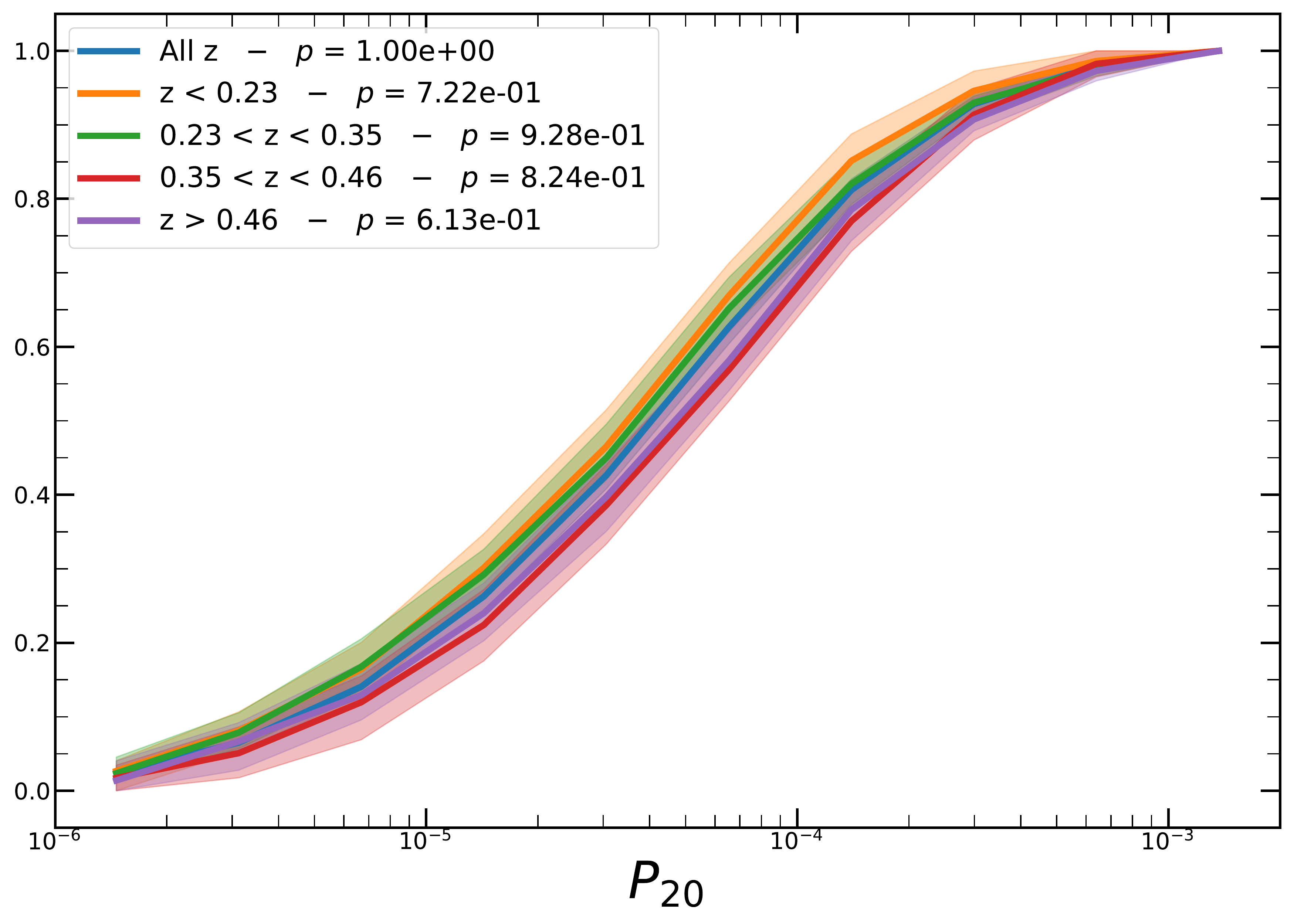}~
    \includegraphics[width=0.33\textwidth]{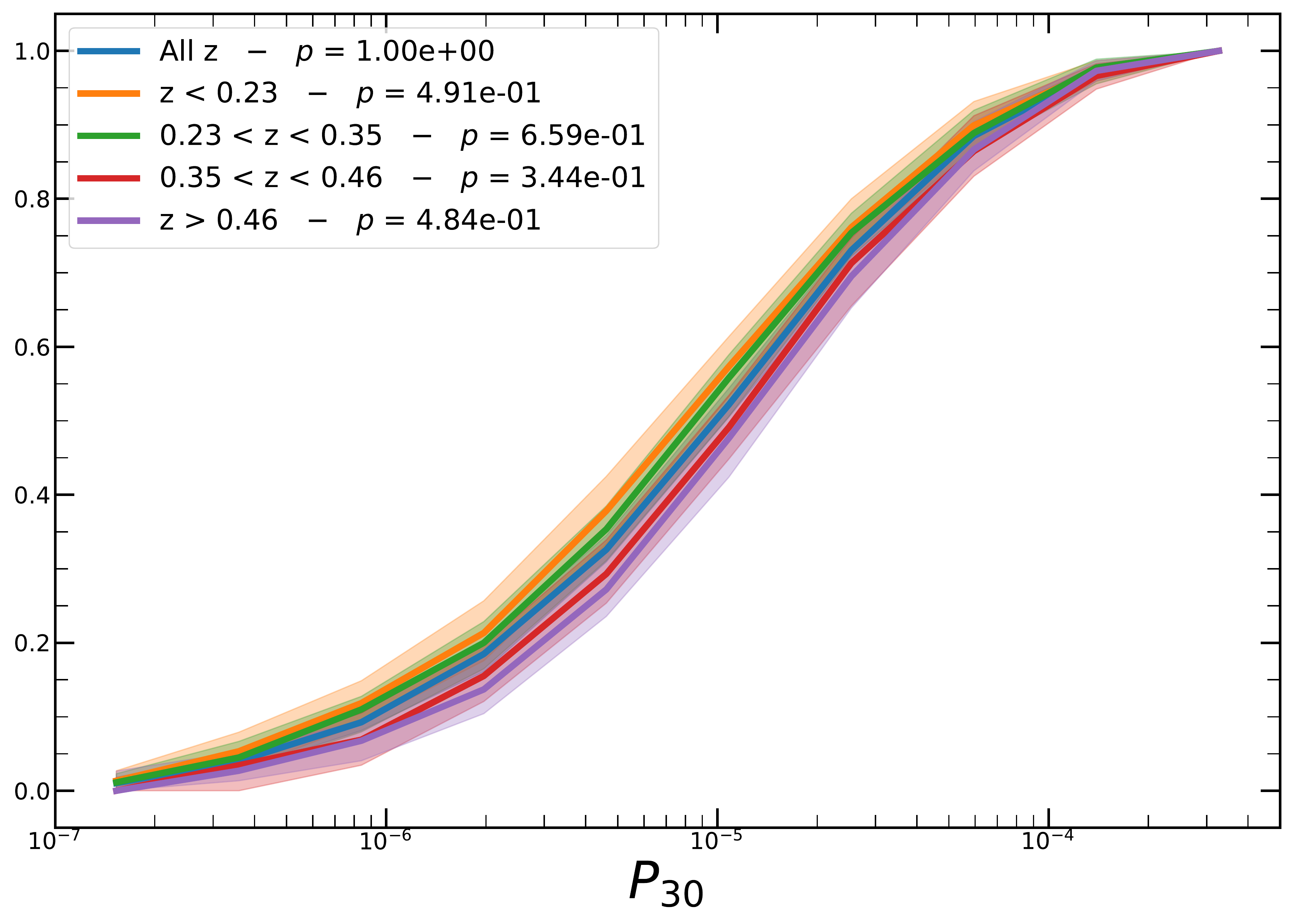}~
    \includegraphics[width=0.33\textwidth]{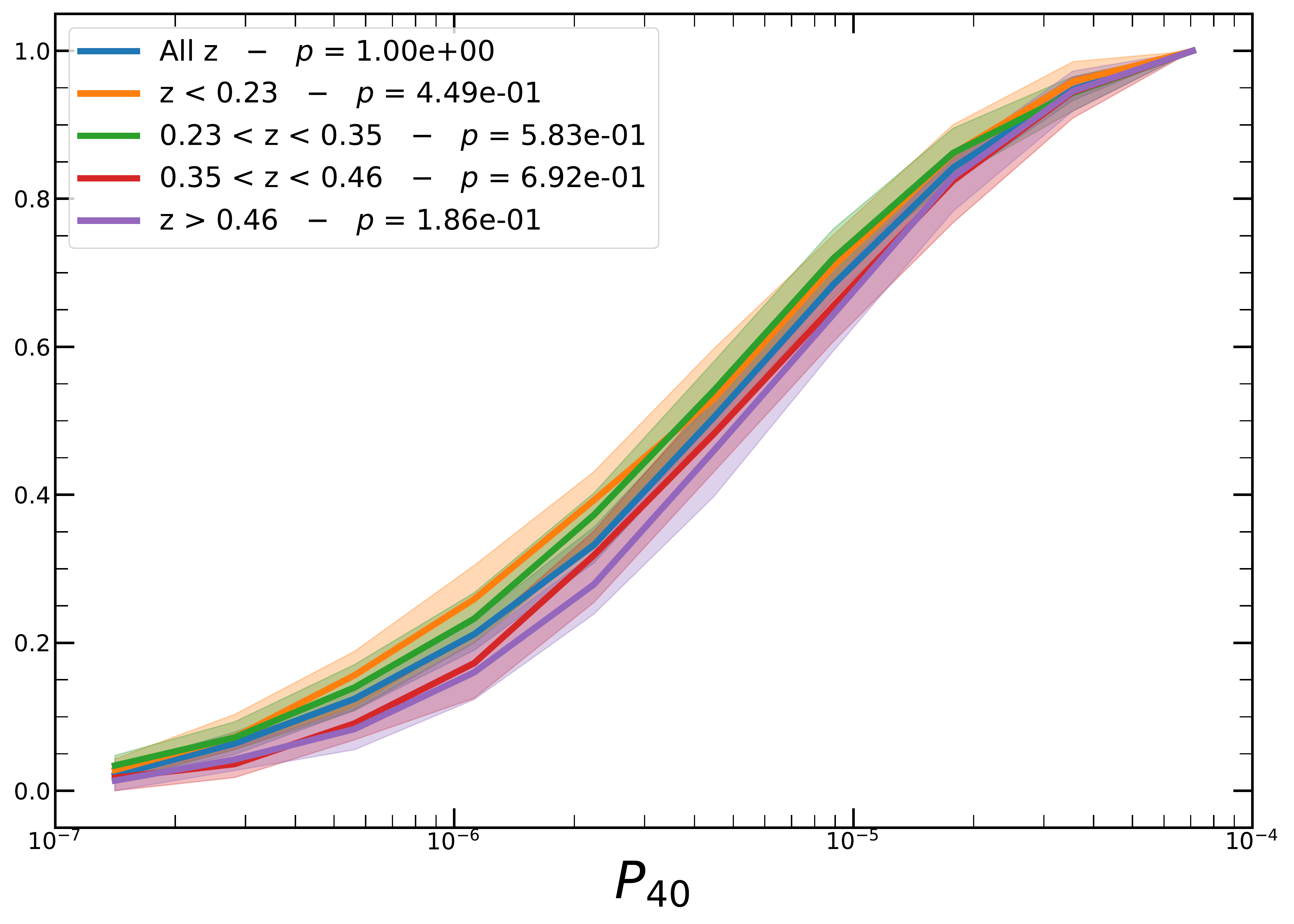}~

    \includegraphics[width=0.33\textwidth]{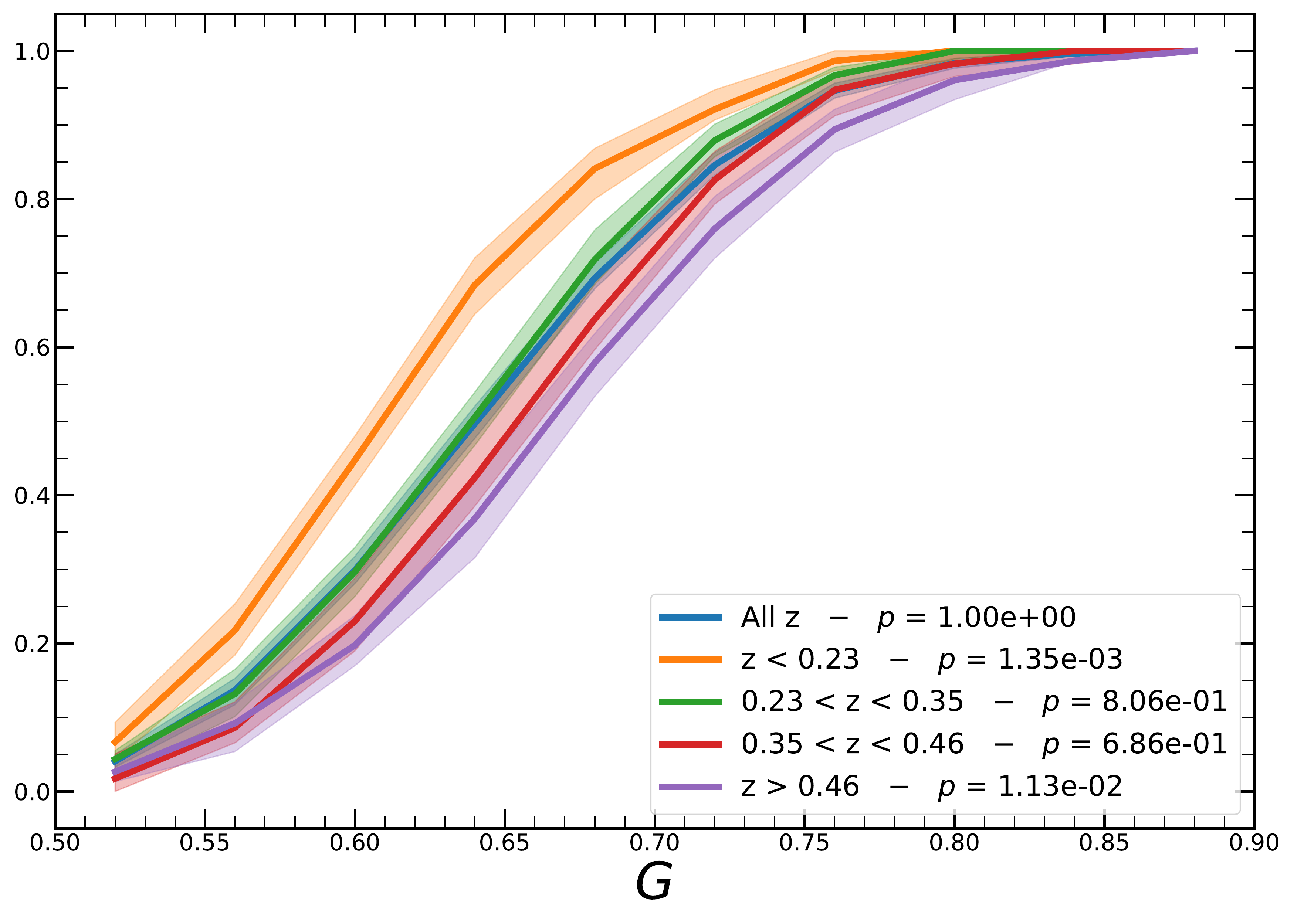}~
    \includegraphics[width=0.33\textwidth]{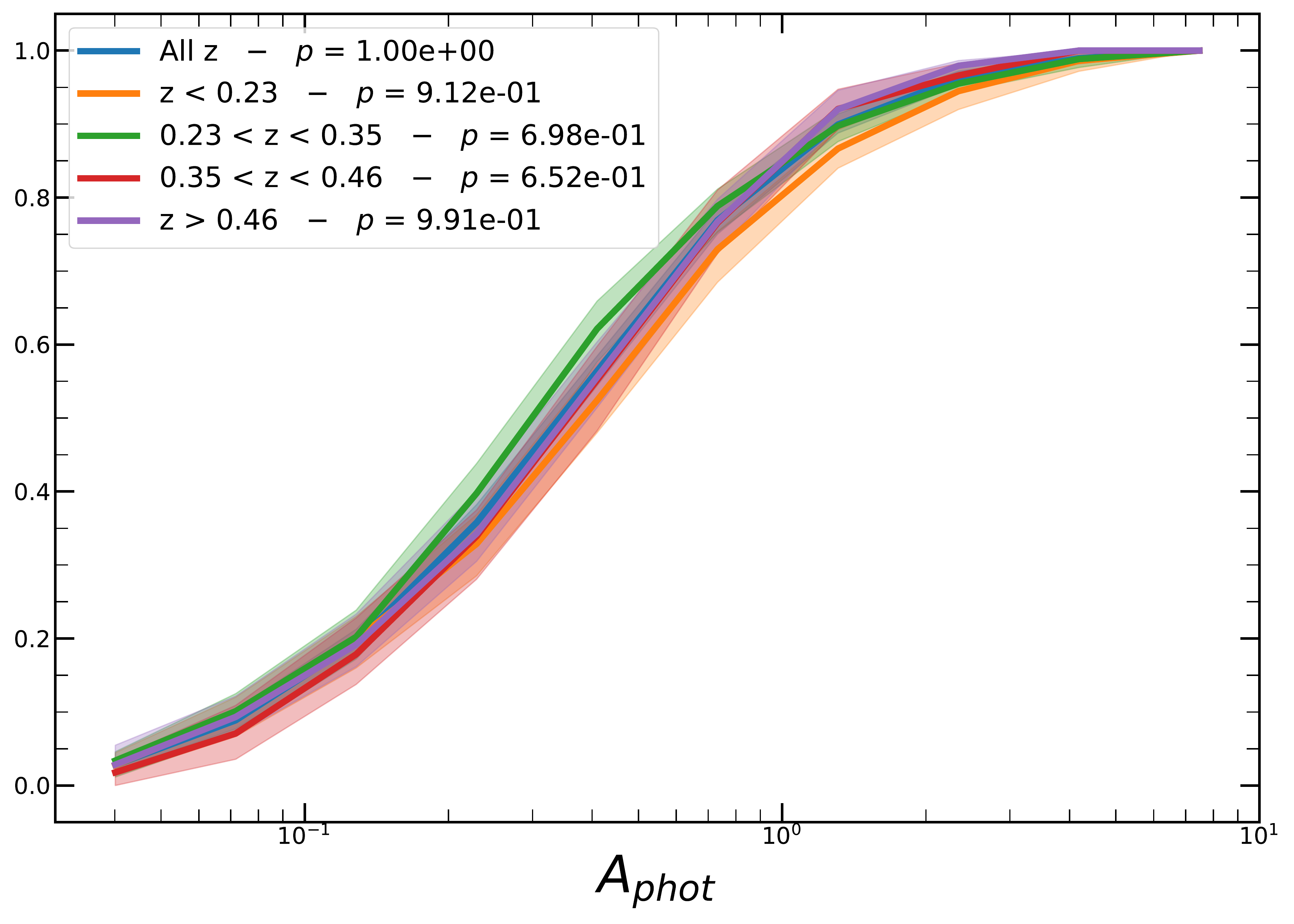}

    \caption{Cumulative distribution of the morphological parameters divided in 4 redshift bins. We indicate the $p$-value of the probability to be drawn from the same underlying distribution.
    }
    \label{fig:redshift_evo_cumulative}
\end{figure*}

In order to investigate the redshift dependence of the morphological parameters, 
we first divide our sample into four homogeneous sub-samples, selecting the cluster population according to their quartile in the redshift distribution, $z < 0.23$, $0.23 < z < 0.35$, $0.35 < z < 0.46$, and $z > 0.46$, therefore with same number of clusters in each bin. 

\begin{figure*}
    \centering
    \includegraphics[width=0.33\textwidth]{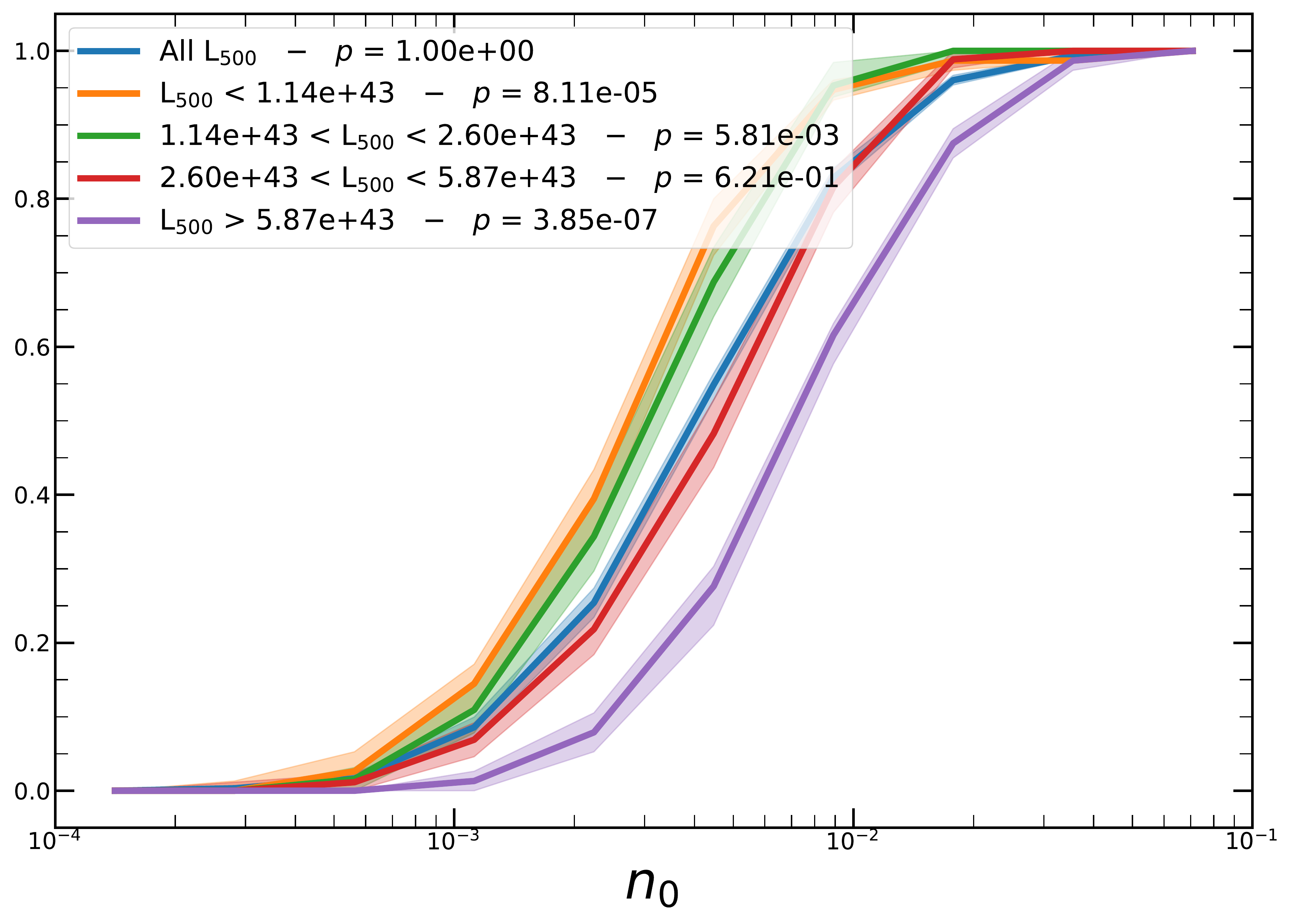}~
    \includegraphics[width=0.33\textwidth]{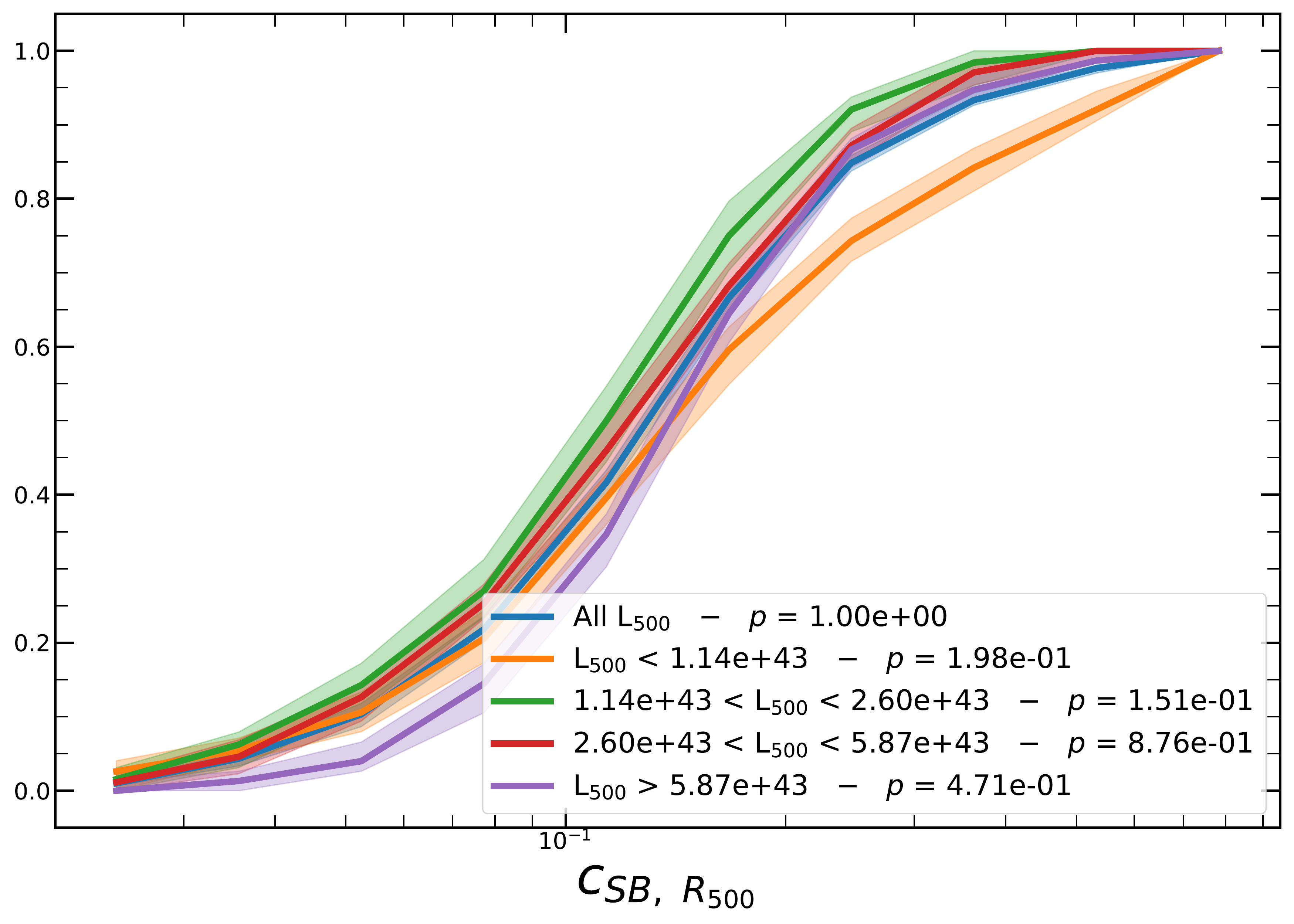}~
    \includegraphics[width=0.33\textwidth]{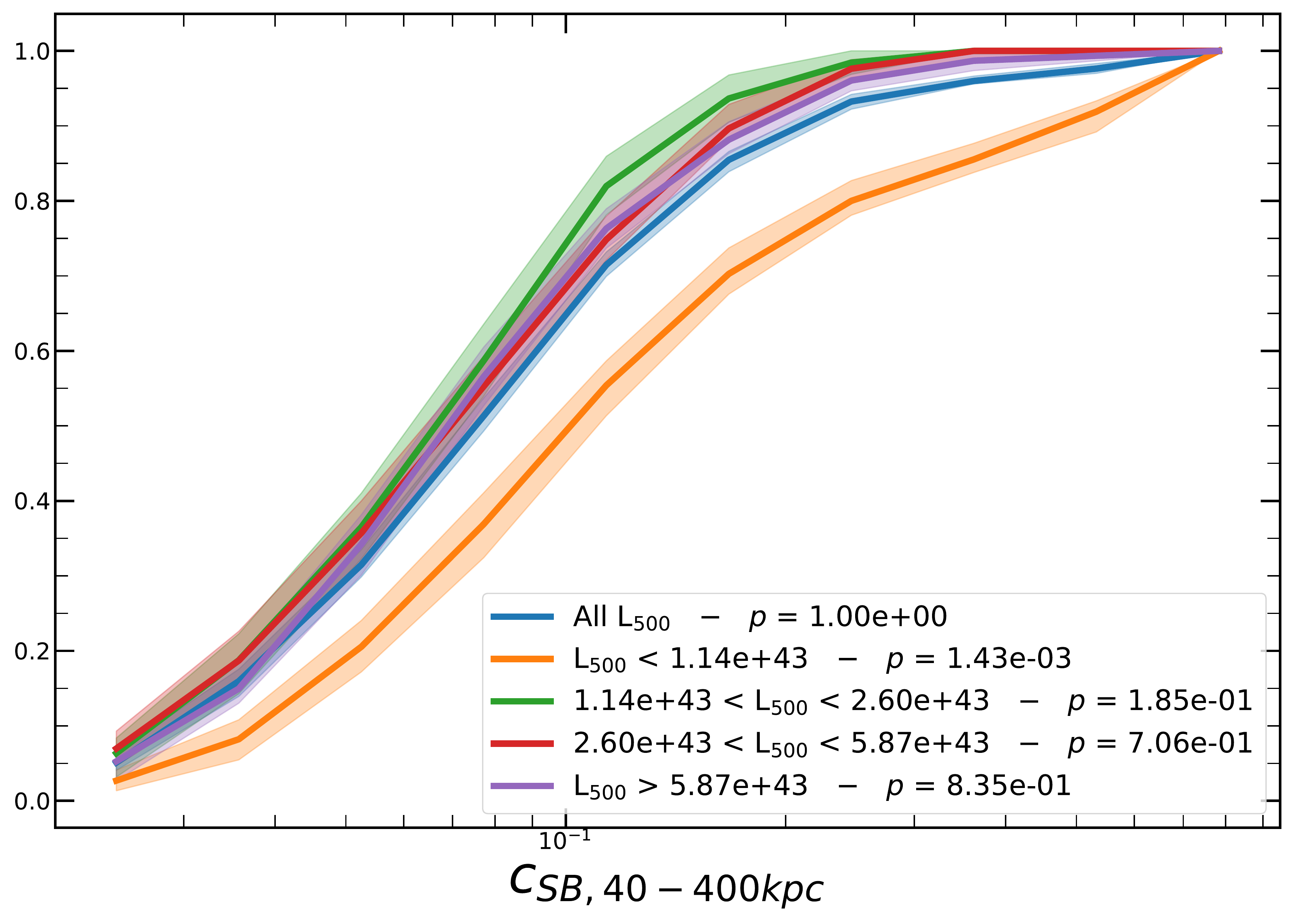}
    
     \includegraphics[width=0.33\textwidth]{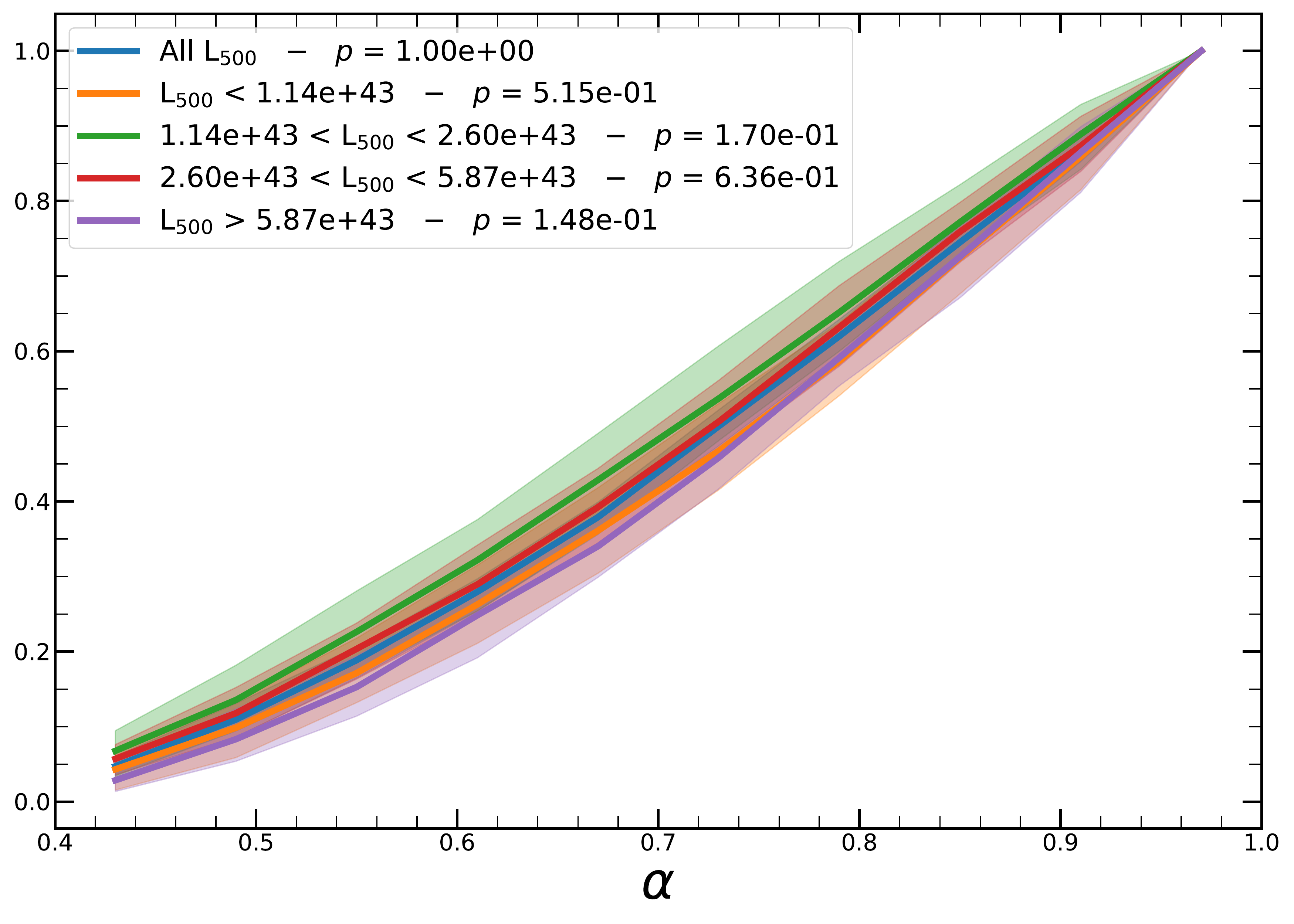}~
    \includegraphics[width=0.33\textwidth]{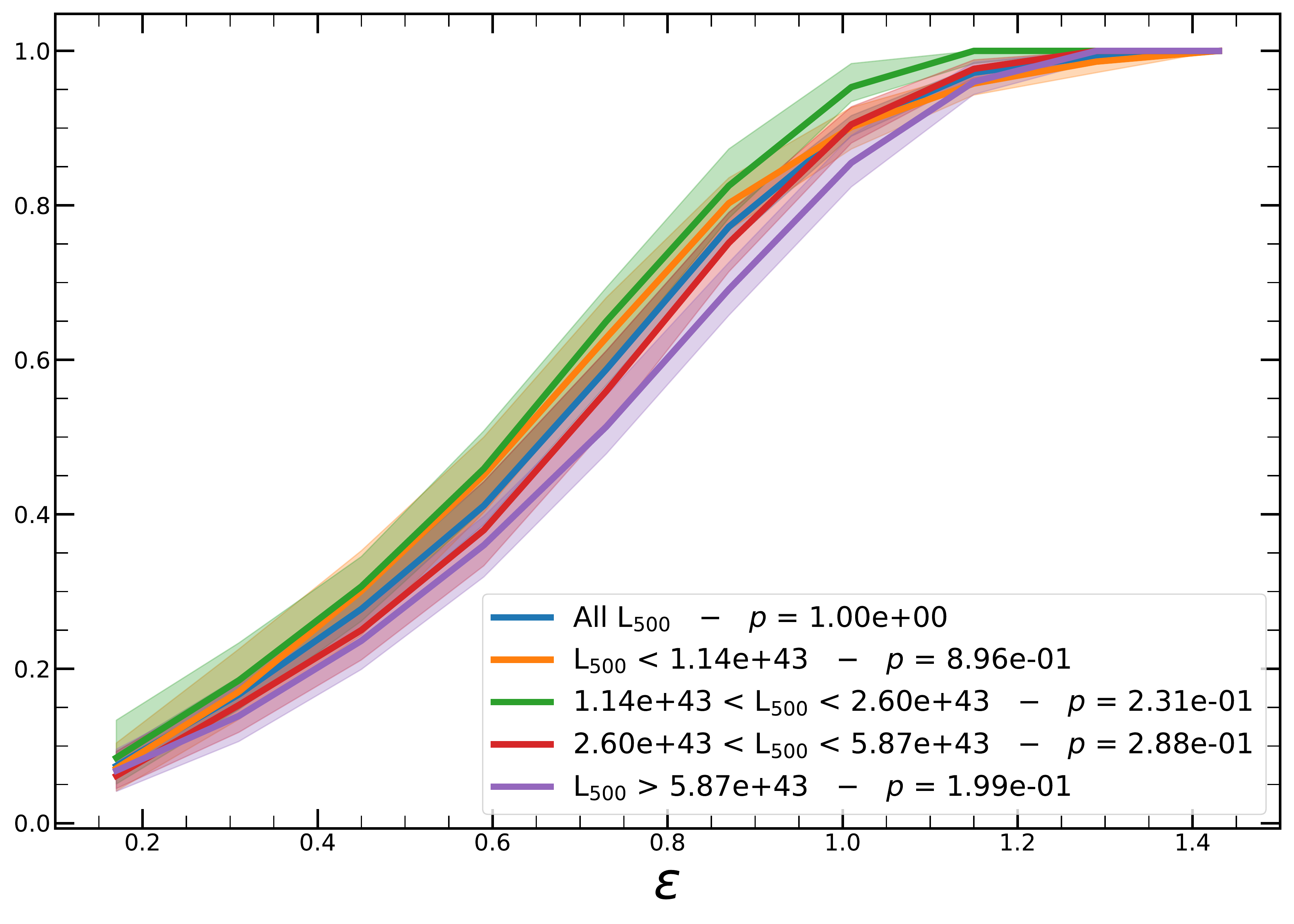}~
    \includegraphics[width=0.33\textwidth]{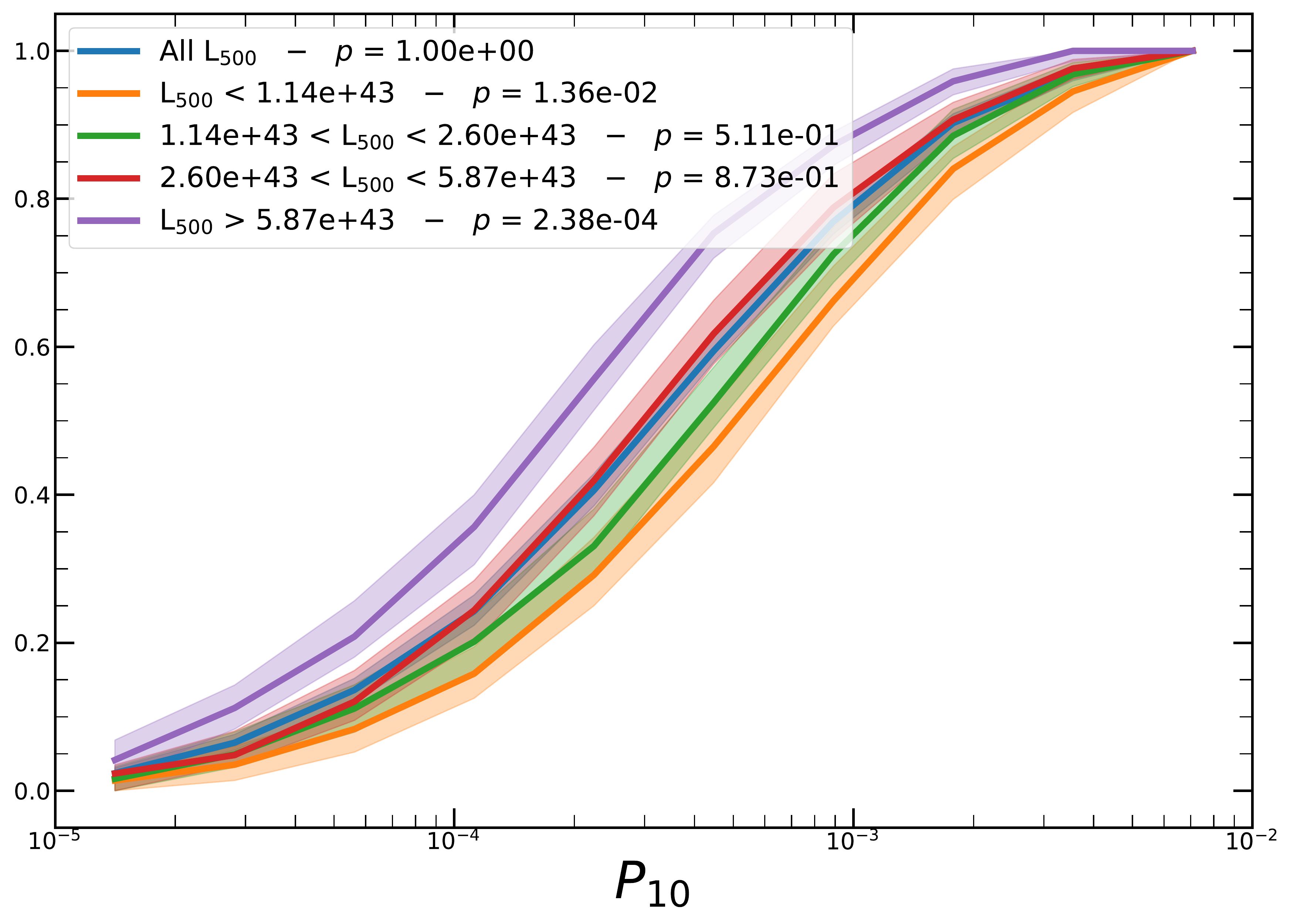}

    \includegraphics[width=0.33\textwidth]{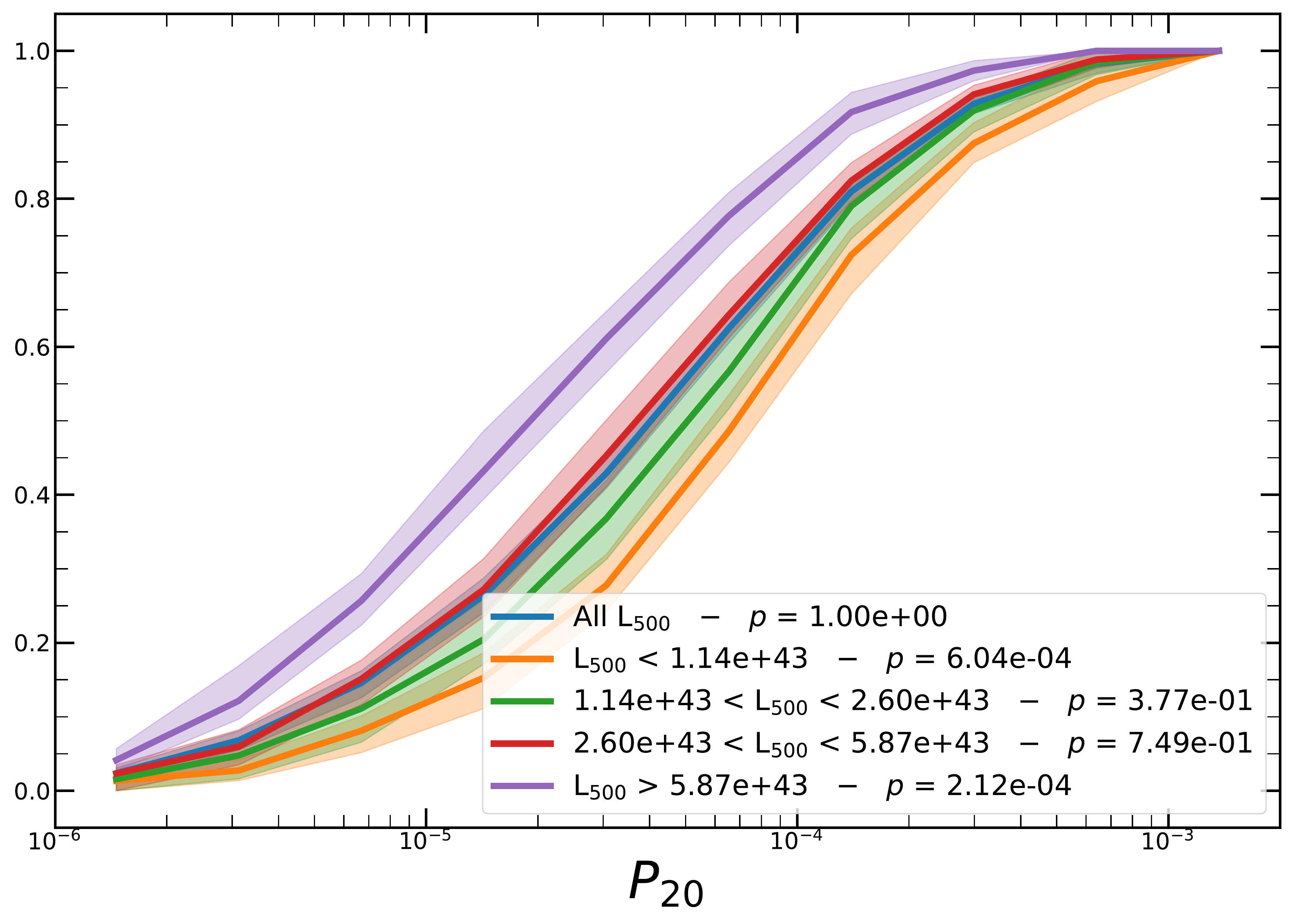}~
    \includegraphics[width=0.33\textwidth]{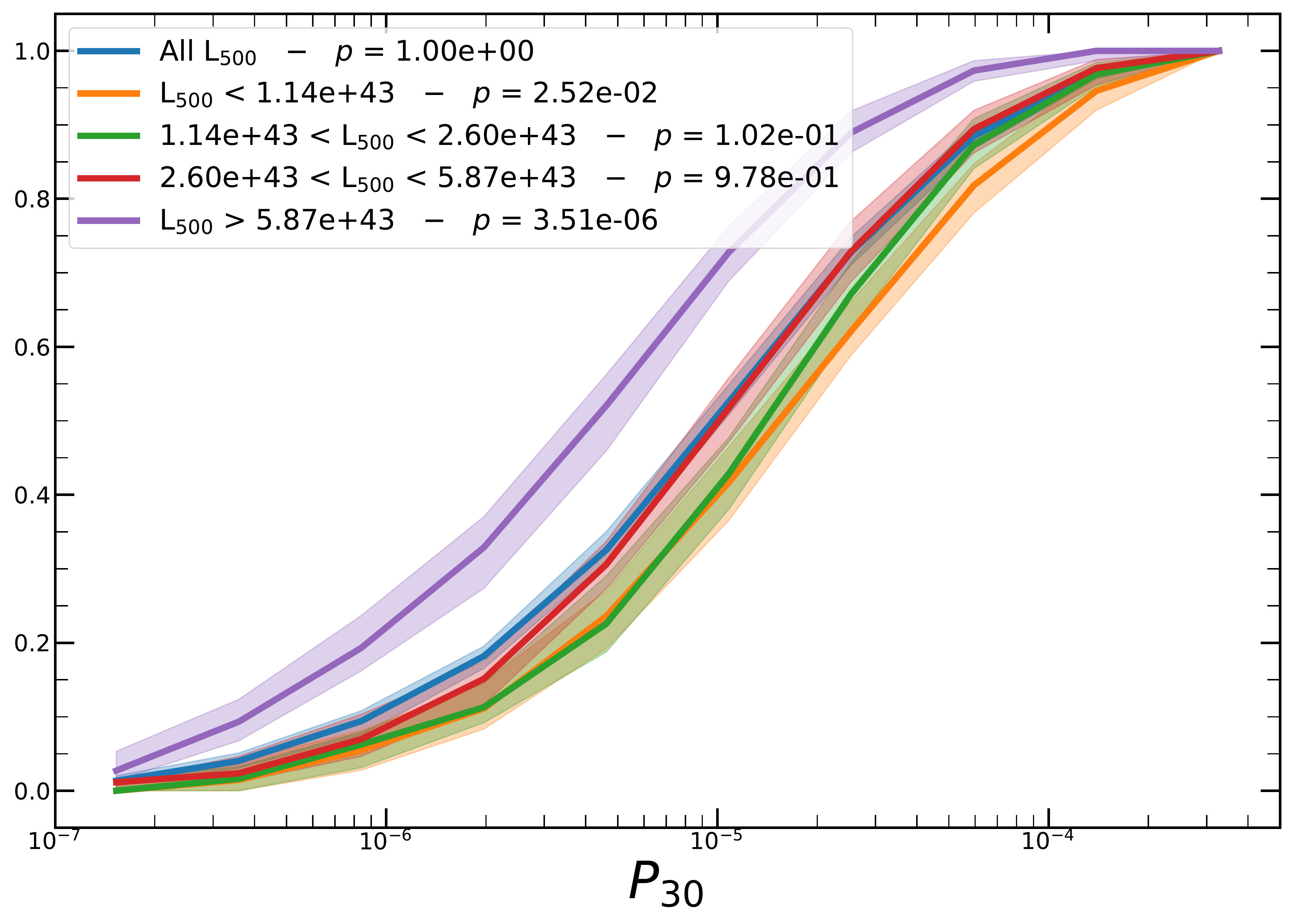}~
    \includegraphics[width=0.33\textwidth]{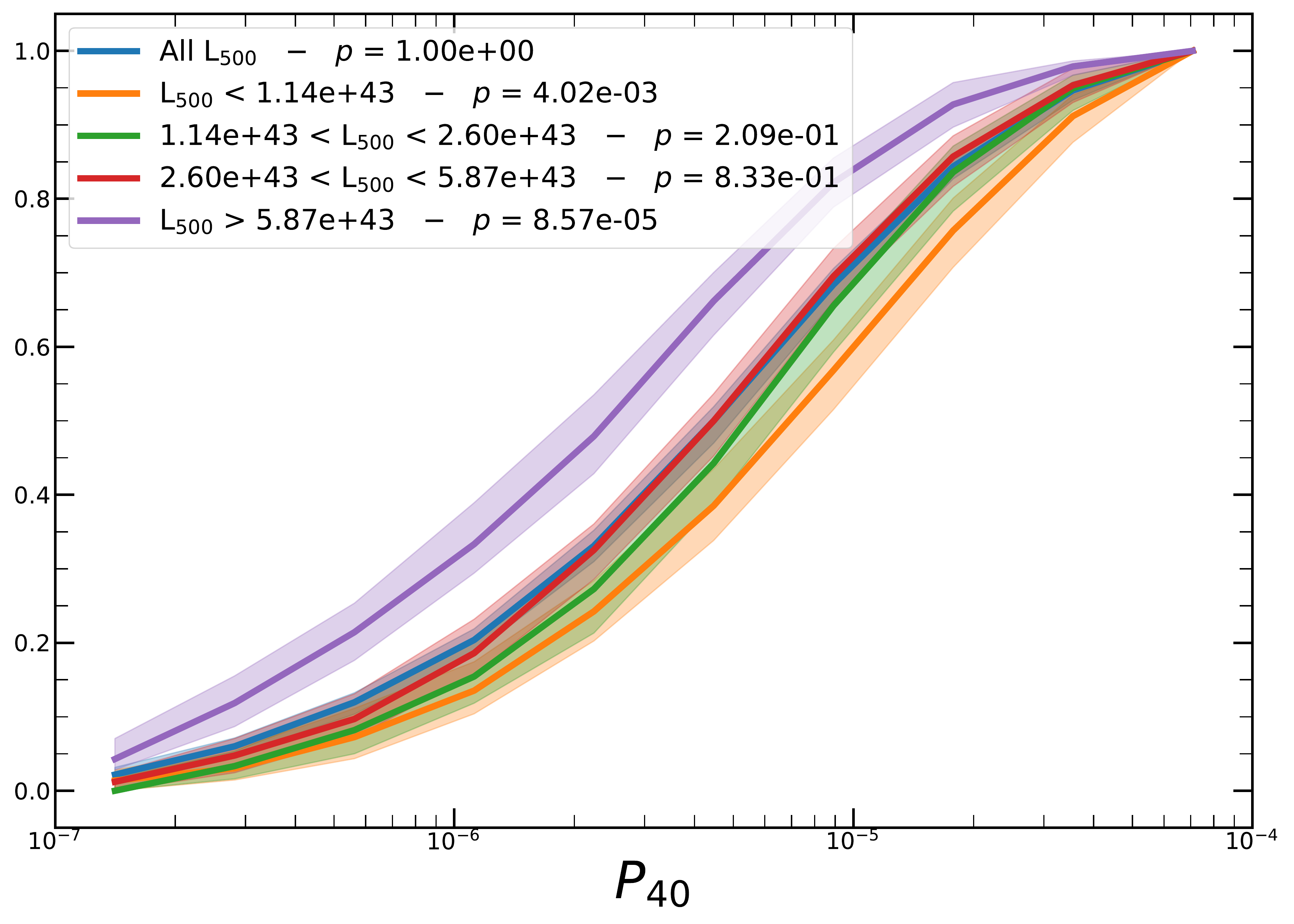}
    
    \includegraphics[width=0.33\textwidth]{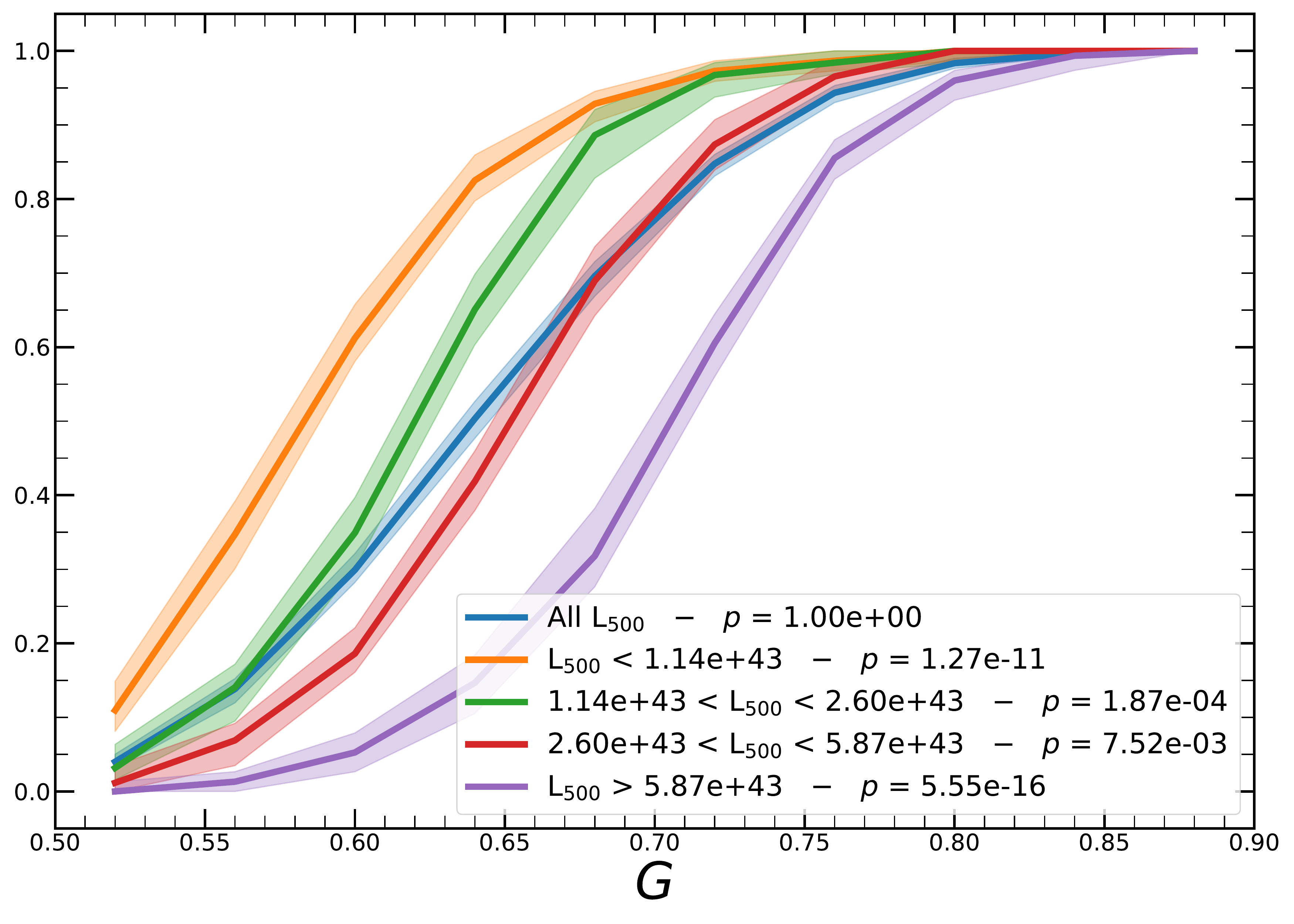}~
    \includegraphics[width=0.33\textwidth]{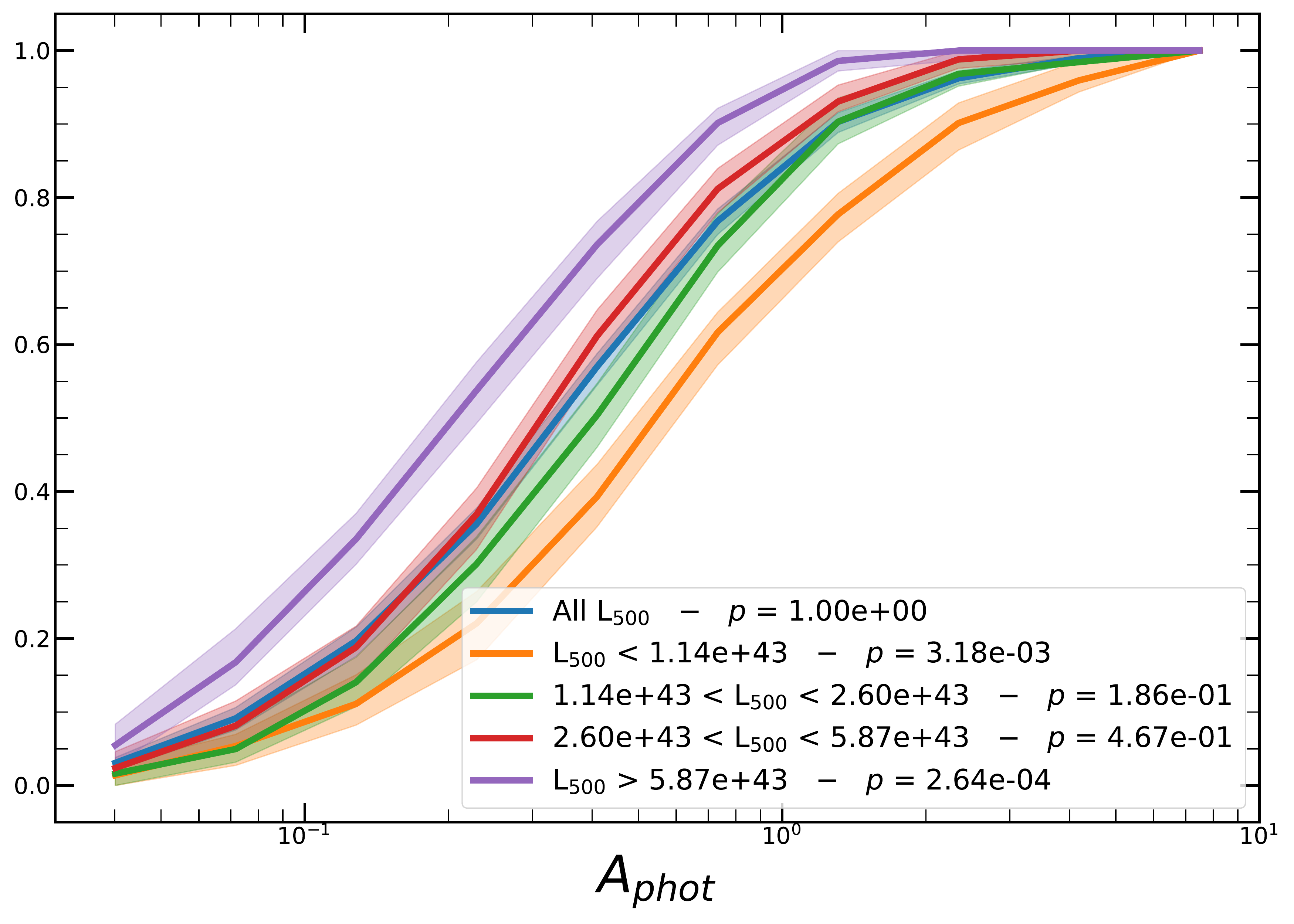}
   
    \caption{Same as Figure~\ref{fig:redshift_evo_cumulative} but the cluster sample is divided in 4 luminosity bins.
    }
    \label{fig:luminosity_evo_cumulative}    
\end{figure*}

It must be noted that the detection (or non-detection) of a significant redshift evolution of samples covering a large redshift and luminosity span can be caused by the significant correlation between luminosity and redshift. 
Therefore, the redshift and luminosity evolution should be simultaneously modelled. In fact, as shown in Fig.~\ref{fig:Lz}, we observe a clear trend where at higher redshifts we observe a larger number of luminous clusters due to eROSITA's selection function. 
Therefore, we further divide the sample based on their luminosity as in the following: 
$L_{500} < 1.1 \times 10^{43}$ erg/s, 
$1.1 \times 10^{43}$ erg/s < $L_{500} < 2.6 \times 10^{43}$ erg/s,
$2.6 \times 10^{43}$ erg/s < $L_{500} < 5.9 \times 10^{43}$ erg/s,
and $L_{500} > 5.9 \times 10^{43}$ erg/s, where also in this case we have used the quartile of the distribution to have approximately the same number of clusters in each bin.
The cumulative distributions of the morphological parameters for the sample of eFEDS clusters, split based on the above criteria, are shown in Fig.~\ref{fig:redshift_evo_cumulative} and Fig.~\ref{fig:luminosity_evo_cumulative}.

We perform the Kolmogorov-Smirnov statistical test (KS test) to check whether the morphological parameters in each redshift (and luminosity) bin are drawn from the same parent distribution as the unbinned cluster sample and whether there are clear indications for evolution with redshift.
We compute the $p$-value from the KS test, which gives the probability for clusters in each redshift quartile to be indistinguishable from the unbinned cluster sample. Small value indicates that the morphological parameters in a specific redshift bin are clearly different than the unbinned distribution, while large values indicate they are indistinguishable. We find that the central density and the Gini coefficient have very small $p$-values, indicating that these two parameters might significantly evolve with redshift, while for the other parameters, with large $p$-values, such evidence is not present. On the other hand, all the morphological parameters, except for the cuspiness, ellipticity, and concentration, show a significant luminosity dependence. However, we remind that such dependence might be mimicked by or canceled out by the degeneracy between luminosity and redshift, therefore conclusive analysis on the redshift evolution or luminosity dependence can be derived only when we model at the same time both dependencies, considering eROSITA selection function and luminosity function. In the next section, we provide the detailed analysis of this modeling.

\begin{table}[]
\centering
\resizebox{0.5\textwidth}{!}{  
    \begin{tabular}{ c c c c c}
    Parameter & $\gamma$ & $P(\gamma \neq 0)$ & $\beta$ & $P(\beta \neq 0)$ \\
\hline
\hline
$                 n_0$ & $0.163_{-0.038}^{+0.039}$ & $4.2$ & $1.610_{-0.467}^{+0.487}$  & $3.4$  \\
$   c_{\rm SB, \ R_{500}}$ & $-0.031_{-0.035}^{+0.035}$ & $0.9$ & $-0.358_{-0.403}^{+0.392}$  & $0.9$  \\
$c_{\rm SB, \ 40-400 kpc}$ & $-0.231_{-0.038}^{+0.038}$ & $6.1$ & $0.607_{-0.449}^{+0.496}$  & $1.4$  \\
$            \epsilon$ & $-0.017_{-0.008}^{+0.008}$ & $2.1$ & $0.270_{-0.096}^{+0.081}$  & $2.8$  \\
$              \alpha$ & $-0.028_{-0.015}^{+0.015}$ & $1.9$ & $0.473_{-0.181}^{+0.172}$  & $2.6$  \\
$              P_{10}$ & $-0.683_{-0.067}^{+0.064}$ & $10.7$ & $7.441_{-0.753}^{+0.773}$  & $9.9$  \\
$              P_{20}$ & $-0.814_{-0.063}^{+0.062}$ & $13.0$ & $8.774_{-0.715}^{+0.769}$  & $12.3$  \\
$              P_{30}$ & $-0.871_{-0.067}^{+0.067}$ & $13.0$ & $9.262_{-0.802}^{+0.782}$  & $11.6$  \\
$              P_{40}$ & $-0.901_{-0.065}^{+0.065}$ & $13.8$ & $9.759_{-0.788}^{+0.779}$  & $12.4$  \\
$                   G$ & $0.063_{-0.003}^{+0.003}$ & $23.4$ & $-0.306_{-0.045}^{+0.044}$  & $6.9$  \\
$            A_{phot}$ & $-0.527_{-0.049}^{+0.045}$ & $11.8$ & $4.296_{-0.541}^{+0.525}$  & $7.9$  \\
\hline
\hline
    \end{tabular}
}
    \caption{Best-fit values for the luminosity and redshift dependence, $\gamma$ and $\beta$ respectively. We also indicate the significance of these values being inconsistent with 0 in terms of number of $\sigma$.}
    \label{tab:alpha_beta_evo}
\end{table}

\begin{figure*}
    \centering
    \includegraphics[width=0.5\textwidth]{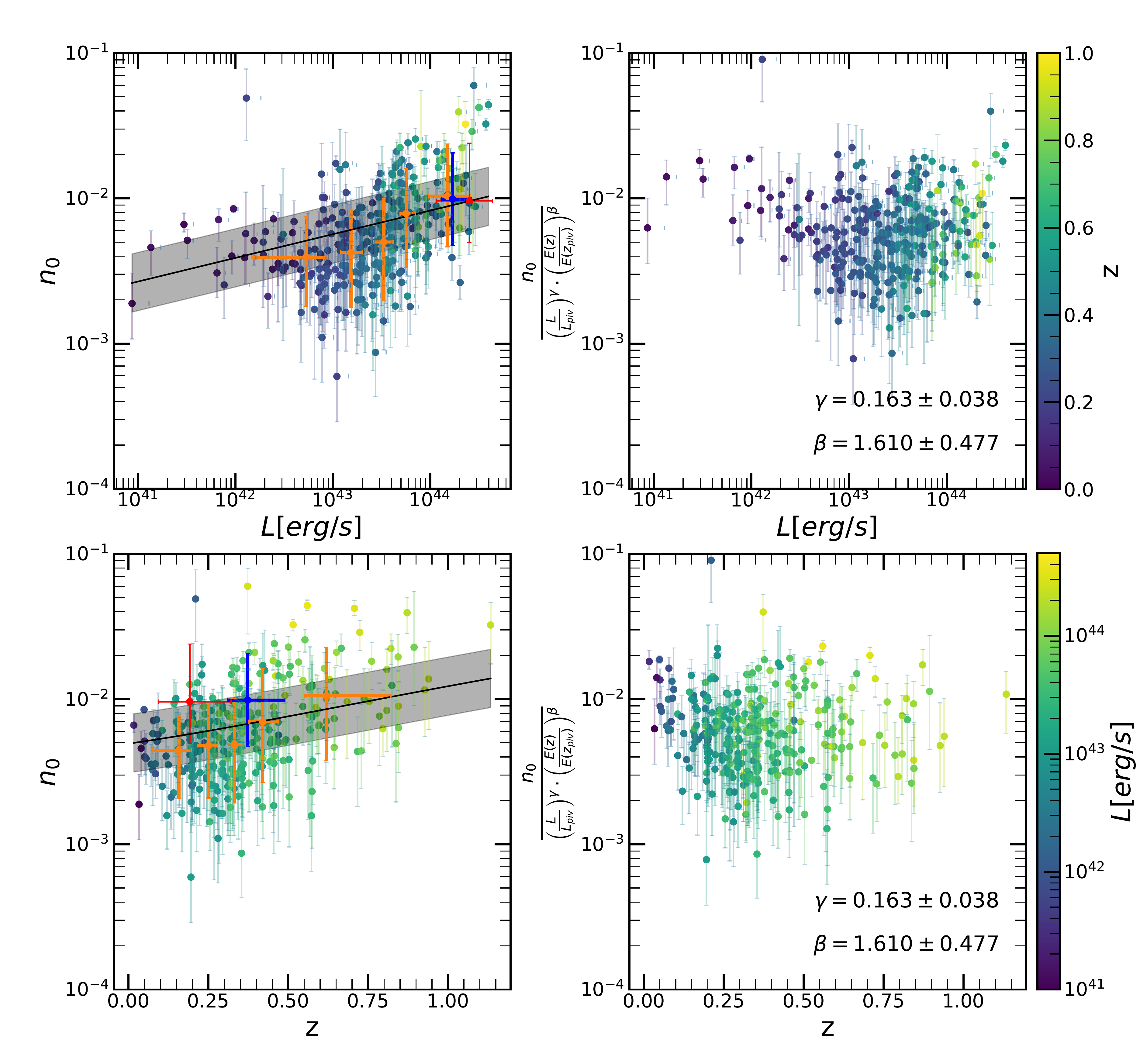}~
    \includegraphics[width=0.5\textwidth]{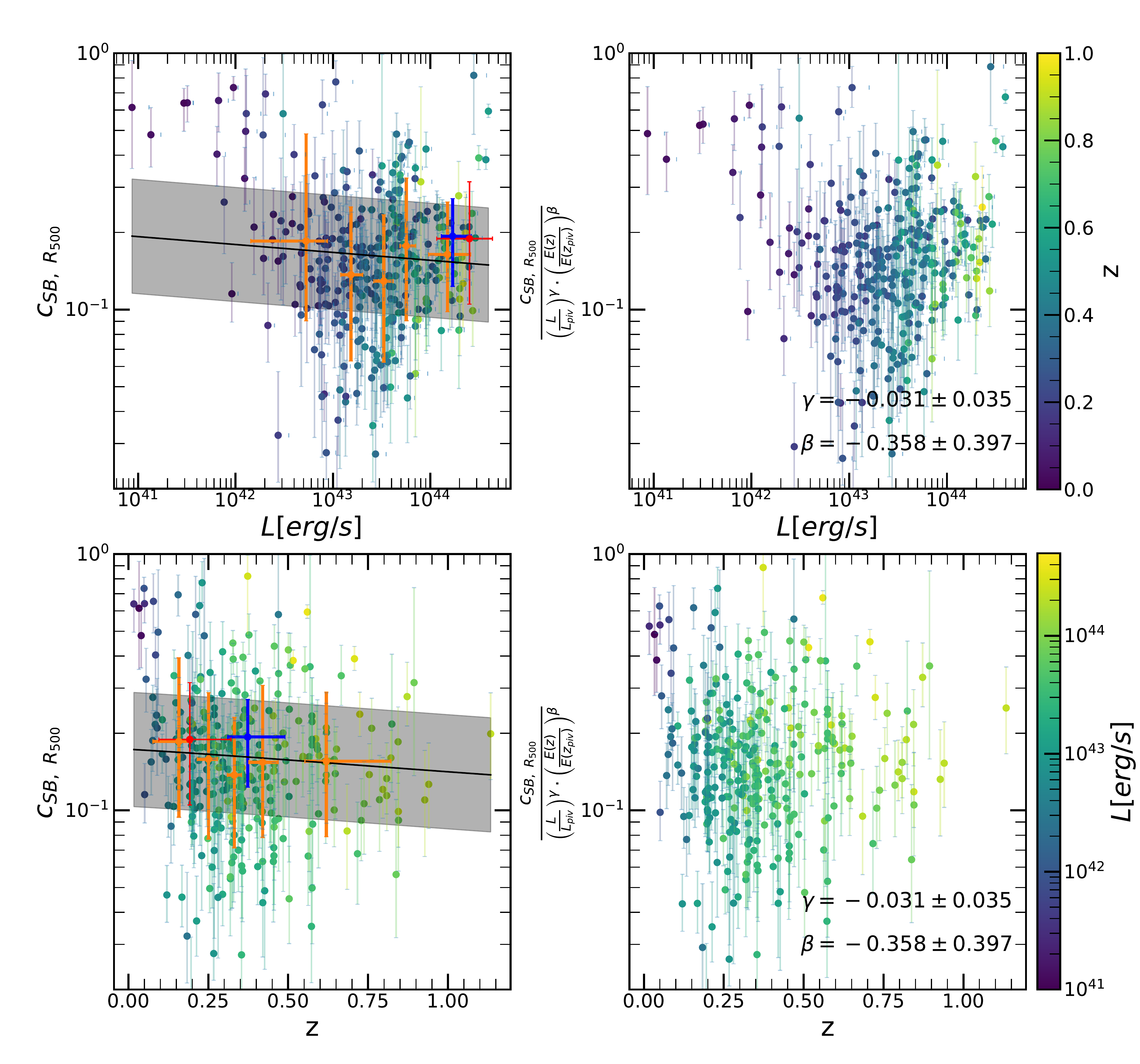}
    
     \includegraphics[width=0.5\textwidth]{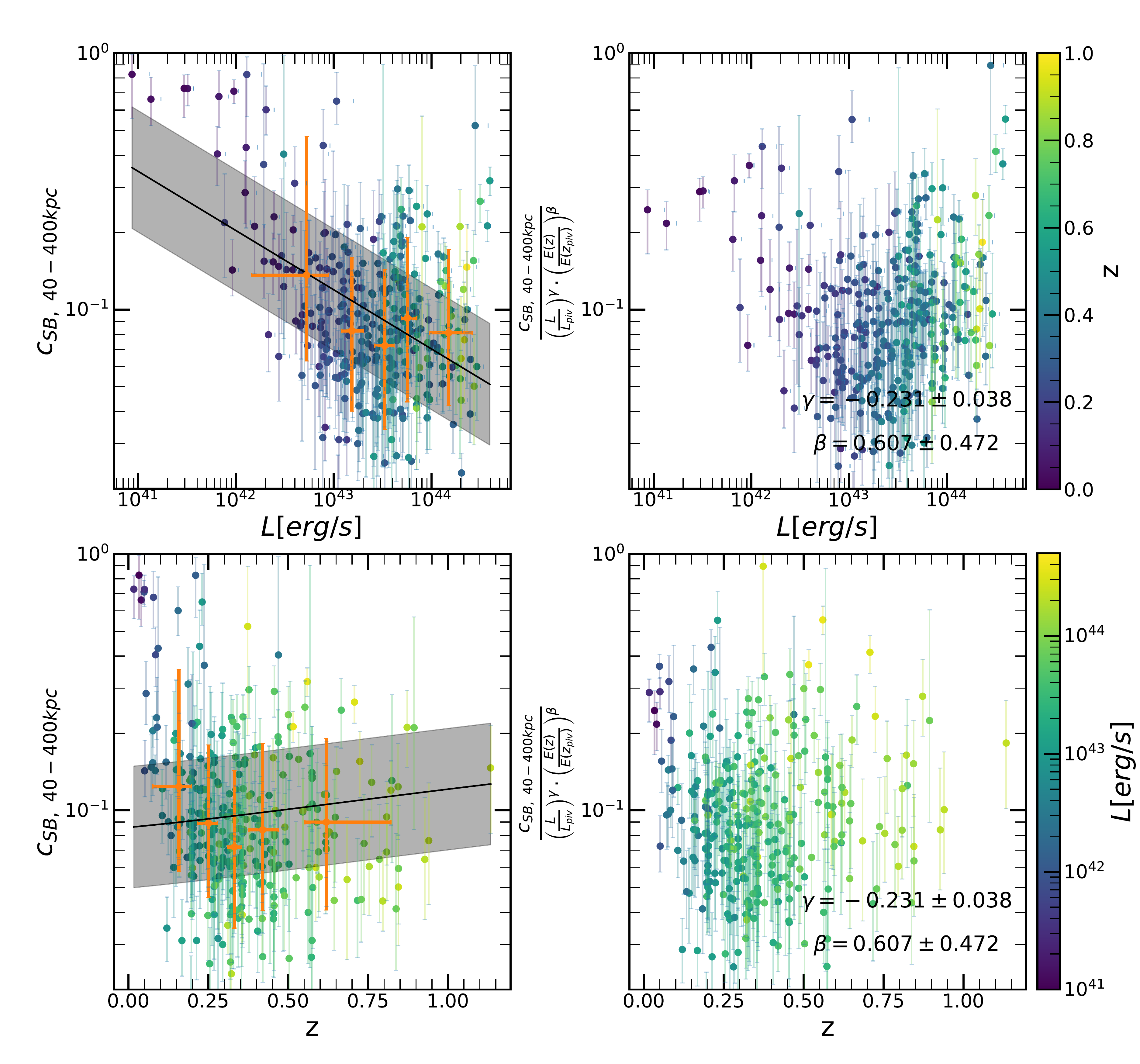}~
     \includegraphics[width=0.5\textwidth]{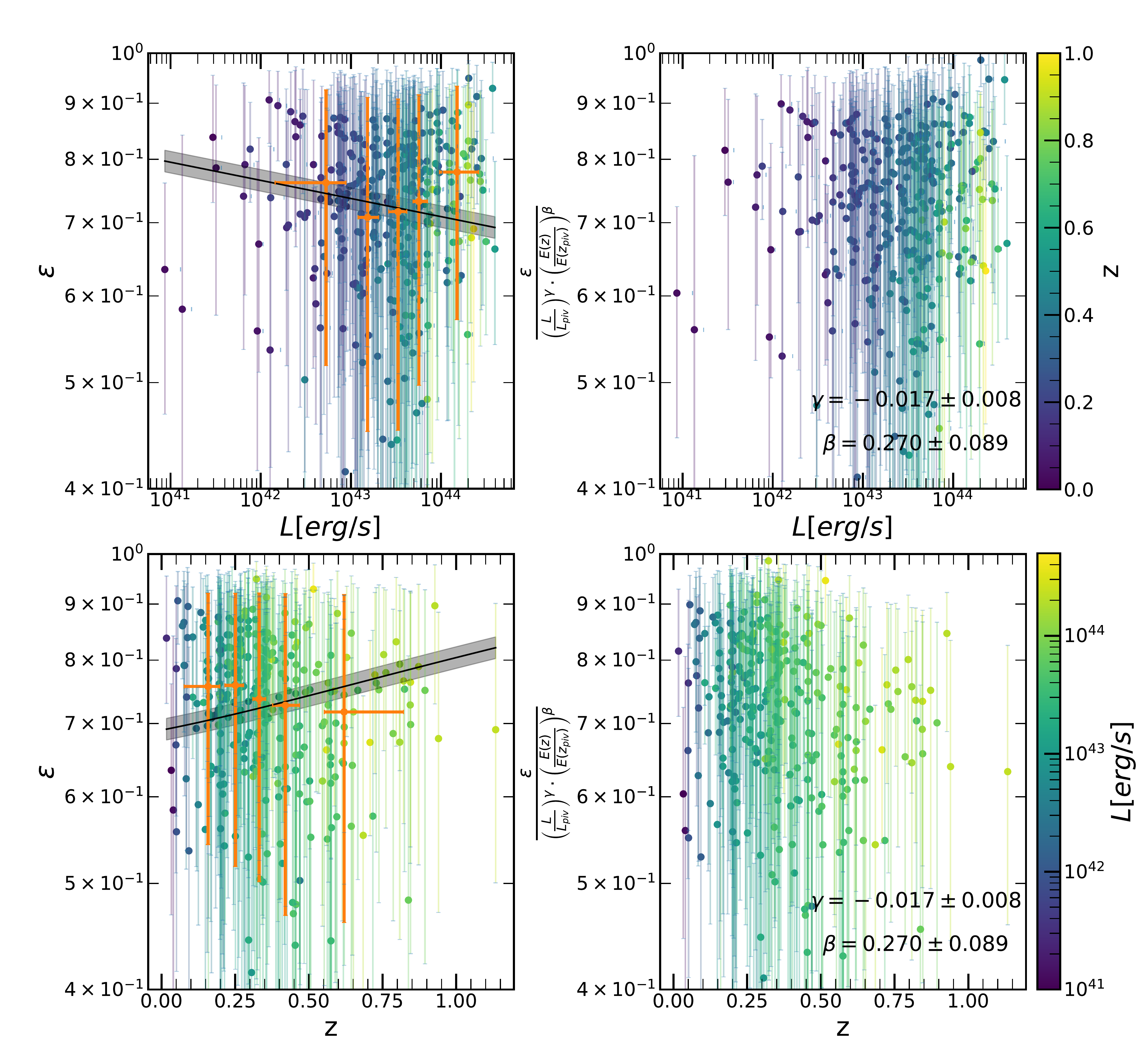}
    \caption{In all four main panels we show in the top left sub-panel the distribution of morphological parameters versus luminosity, color coded with redshift. The black line shows the best fitting line (see Sect.~\ref{sec:Lzevo}); in the bottom left sub-panel we show the distribution of morphological parameters versus redshift, color coded with luminosity, and again the black line shows the best fitting line; shadow grey area represent the scatter around the best fit line.
    The red data point shows (in the cases where a direct comparison can be made) the result value obtained by \cite{lovisari17}, which can be directly compared with the blue data point, that shows our results restricting luminosity and redshift range to the same range as in \cite{lovisari17}.
    Orange data point represent the average morphological parameter values obtained in luminosity or redshift quintiles. 
    We show parameters corrected by redshift and luminosity evolution against luminosity (in the top right sub-panel) and redshift (in the bottom right sub-panel).}
    \label{fig:LZ_evo}
\end{figure*}
\begin{figure*}
    \centering
    \ContinuedFloat    
    \includegraphics[width=0.5\textwidth]{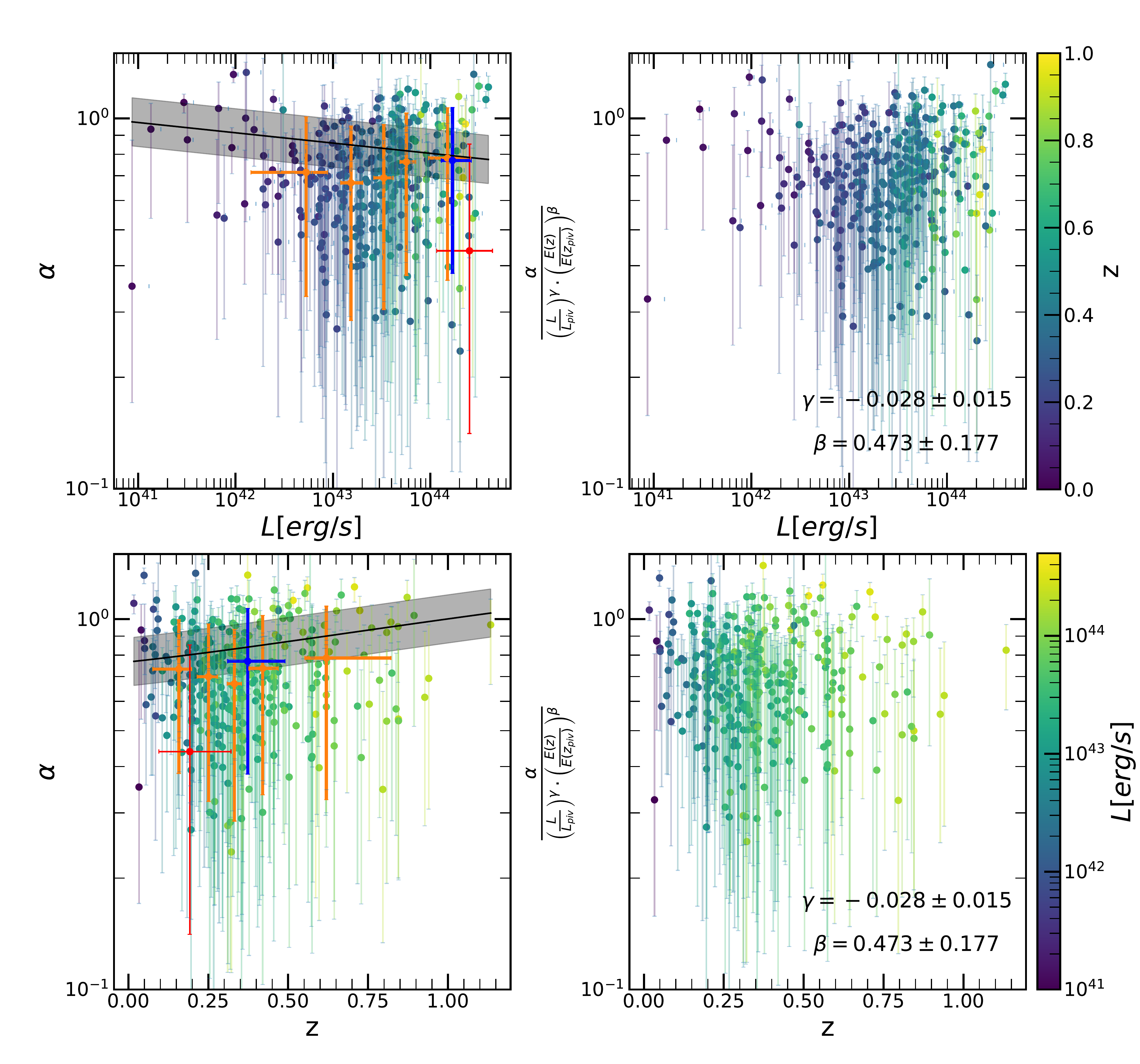}~
    \includegraphics[width=0.5\textwidth]{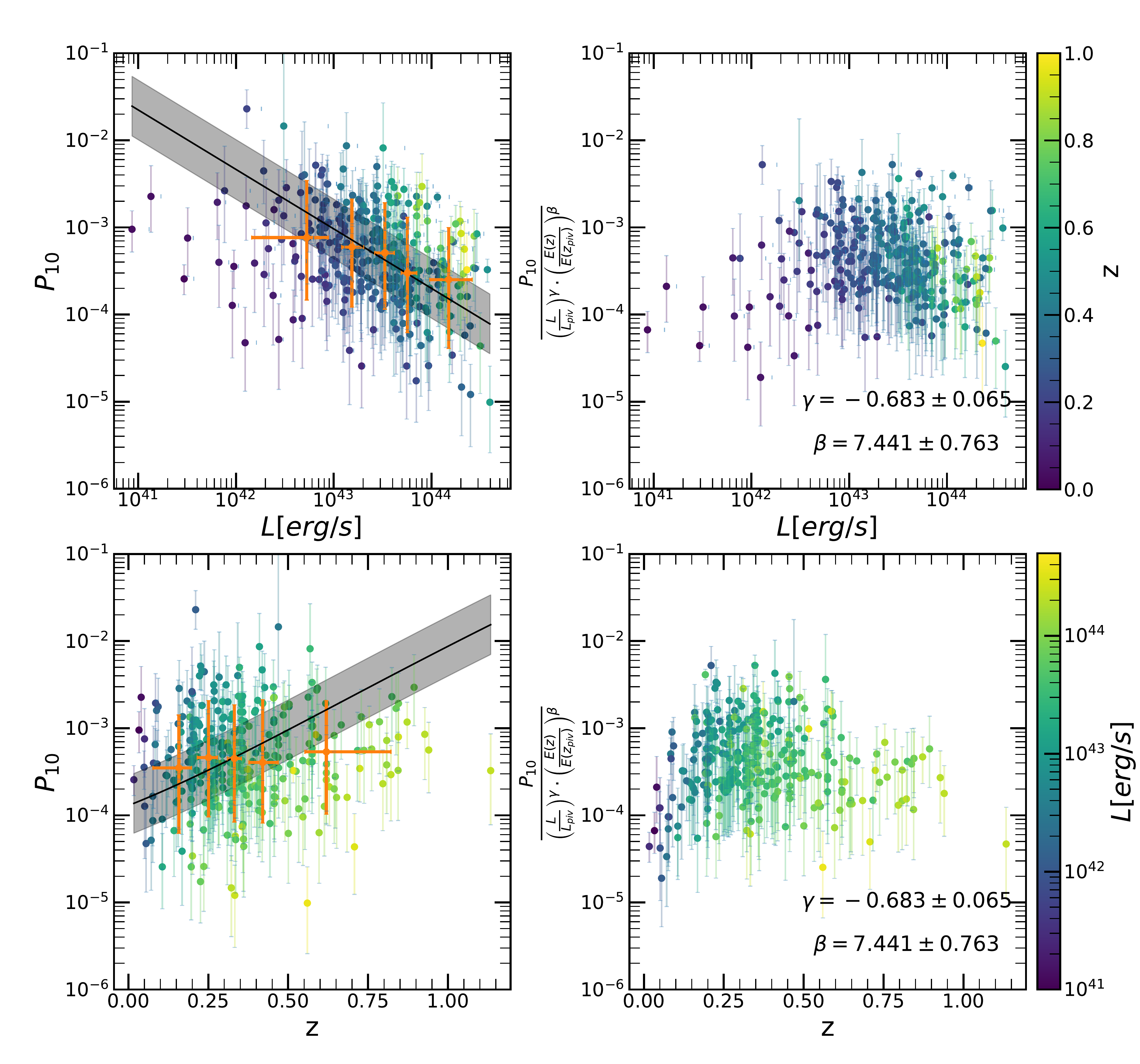}
    
    \includegraphics[width=0.5\textwidth]{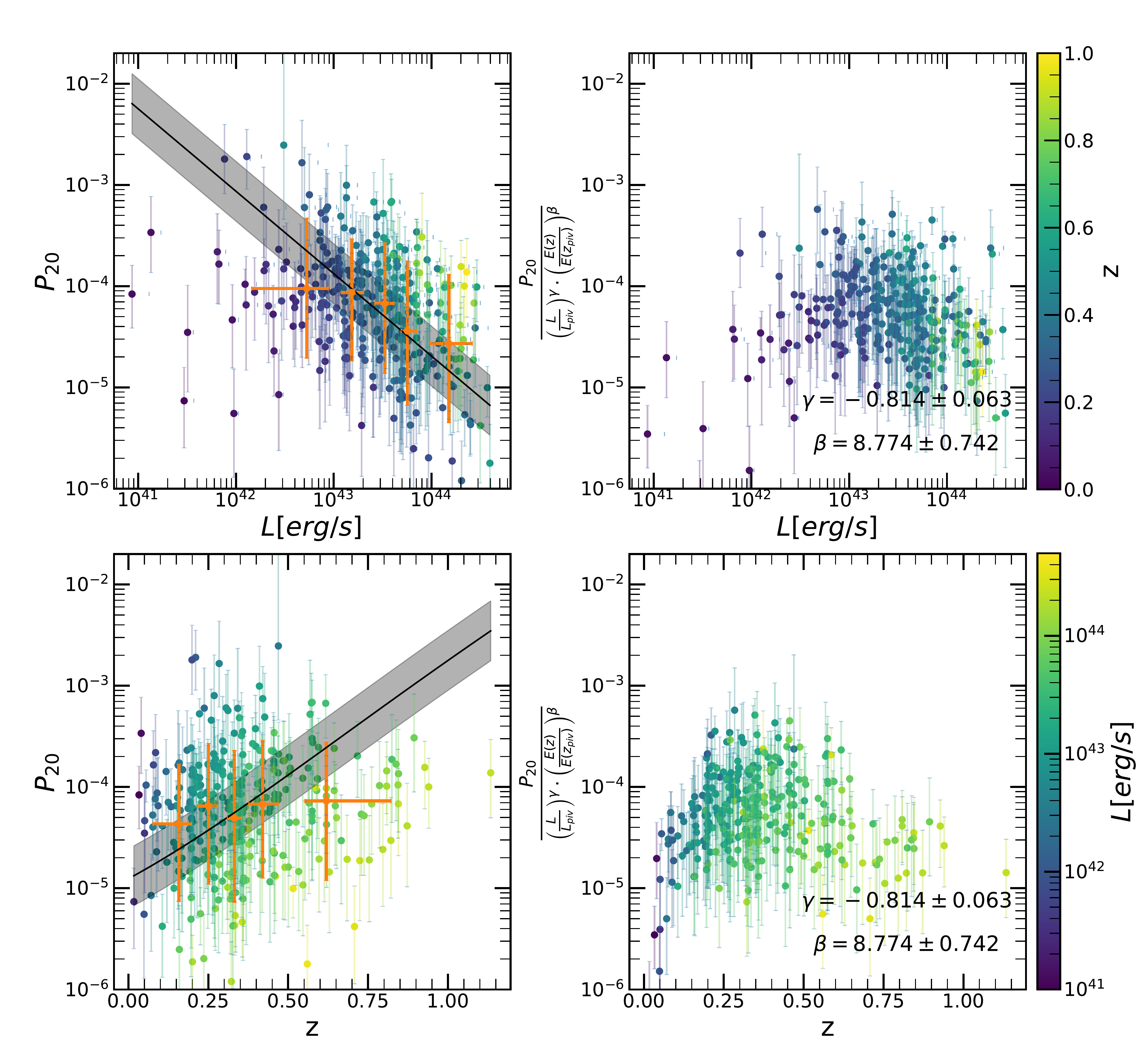}~
    \includegraphics[width=0.5\textwidth]{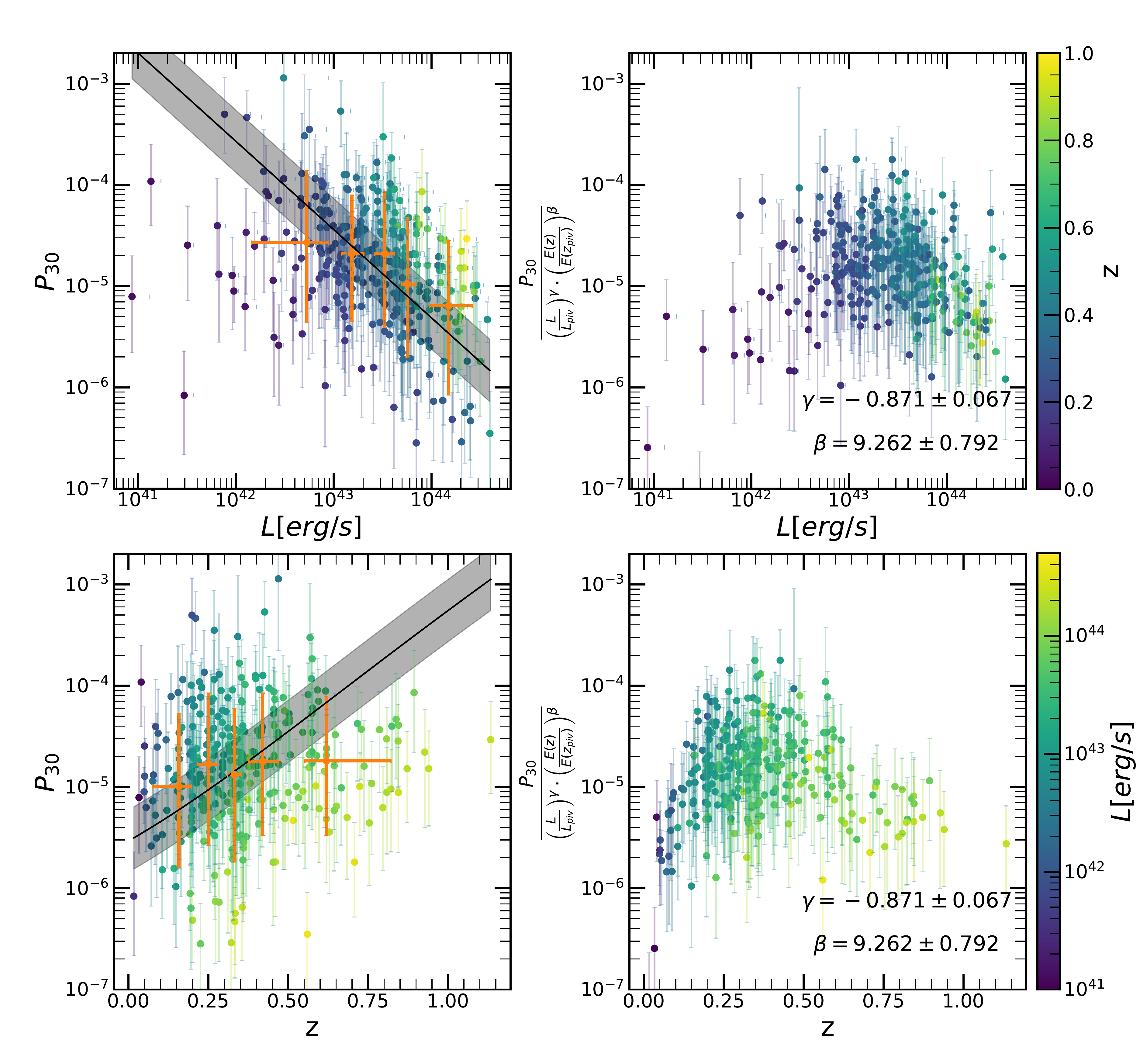}

    \caption{ continued. }
\end{figure*}		
\begin{figure*}
    \centering
    \ContinuedFloat
    \includegraphics[width=0.5\textwidth]{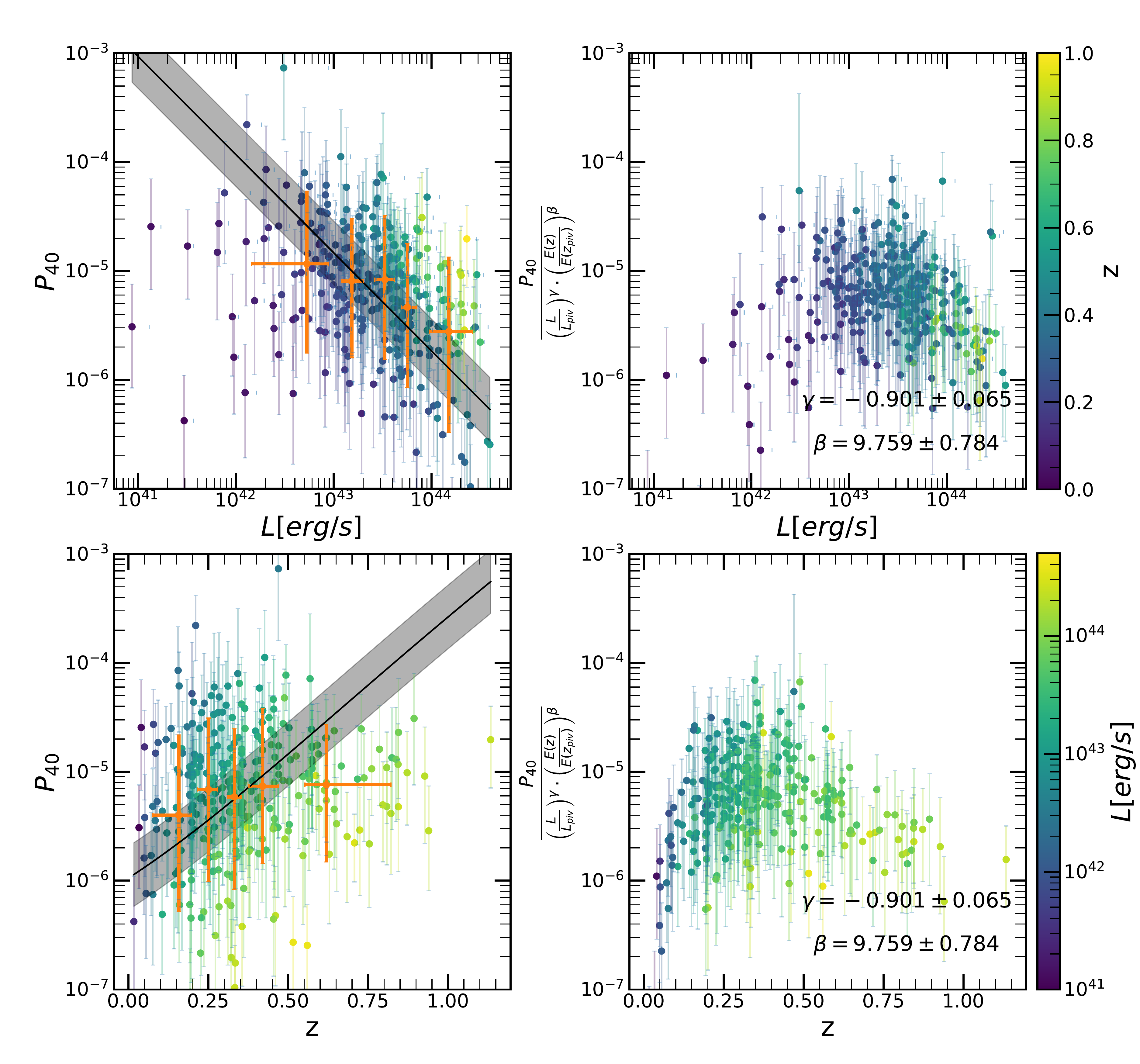}~
    \includegraphics[width=0.5\textwidth]{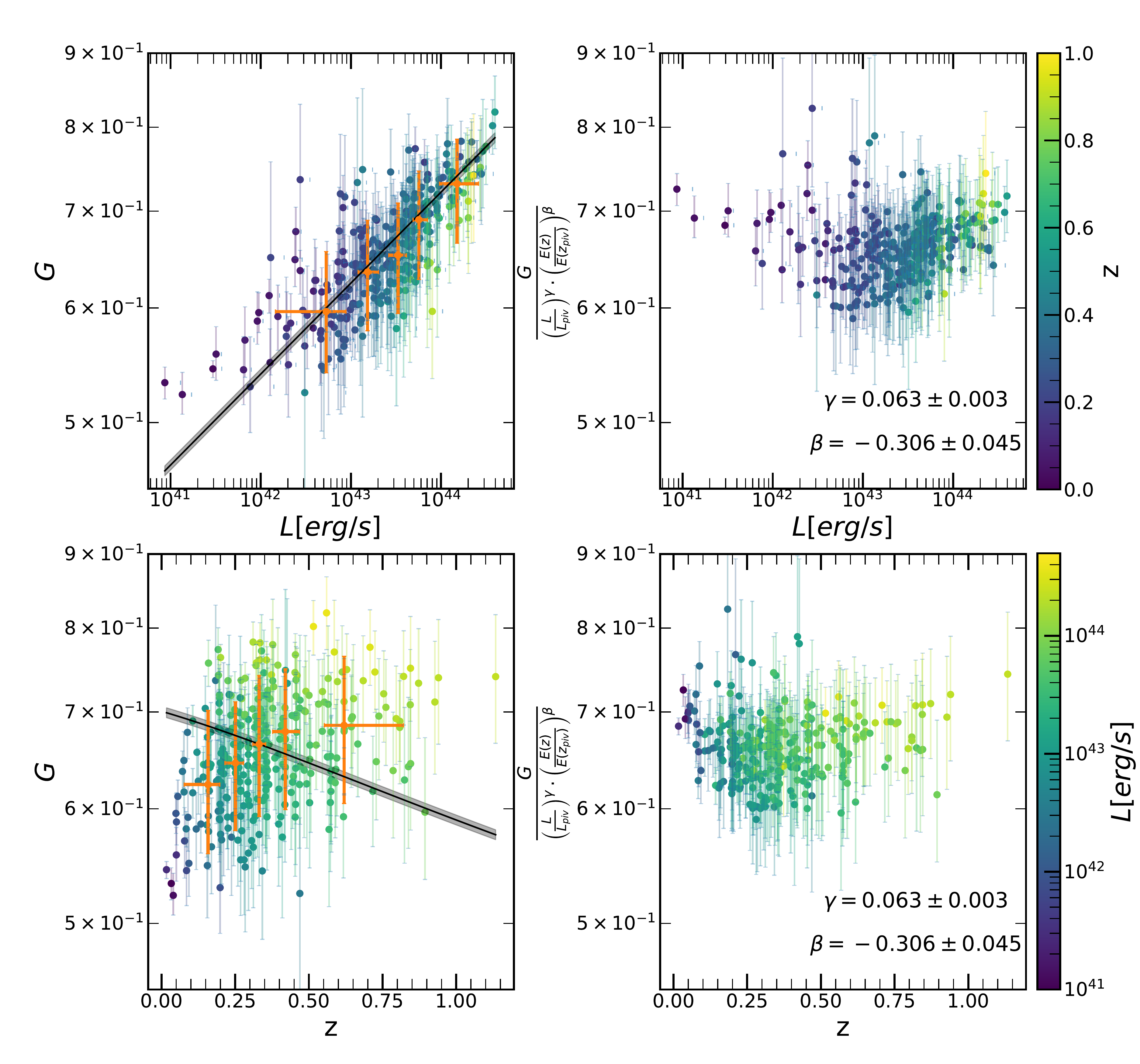}
    
    \includegraphics[width=0.5\textwidth]{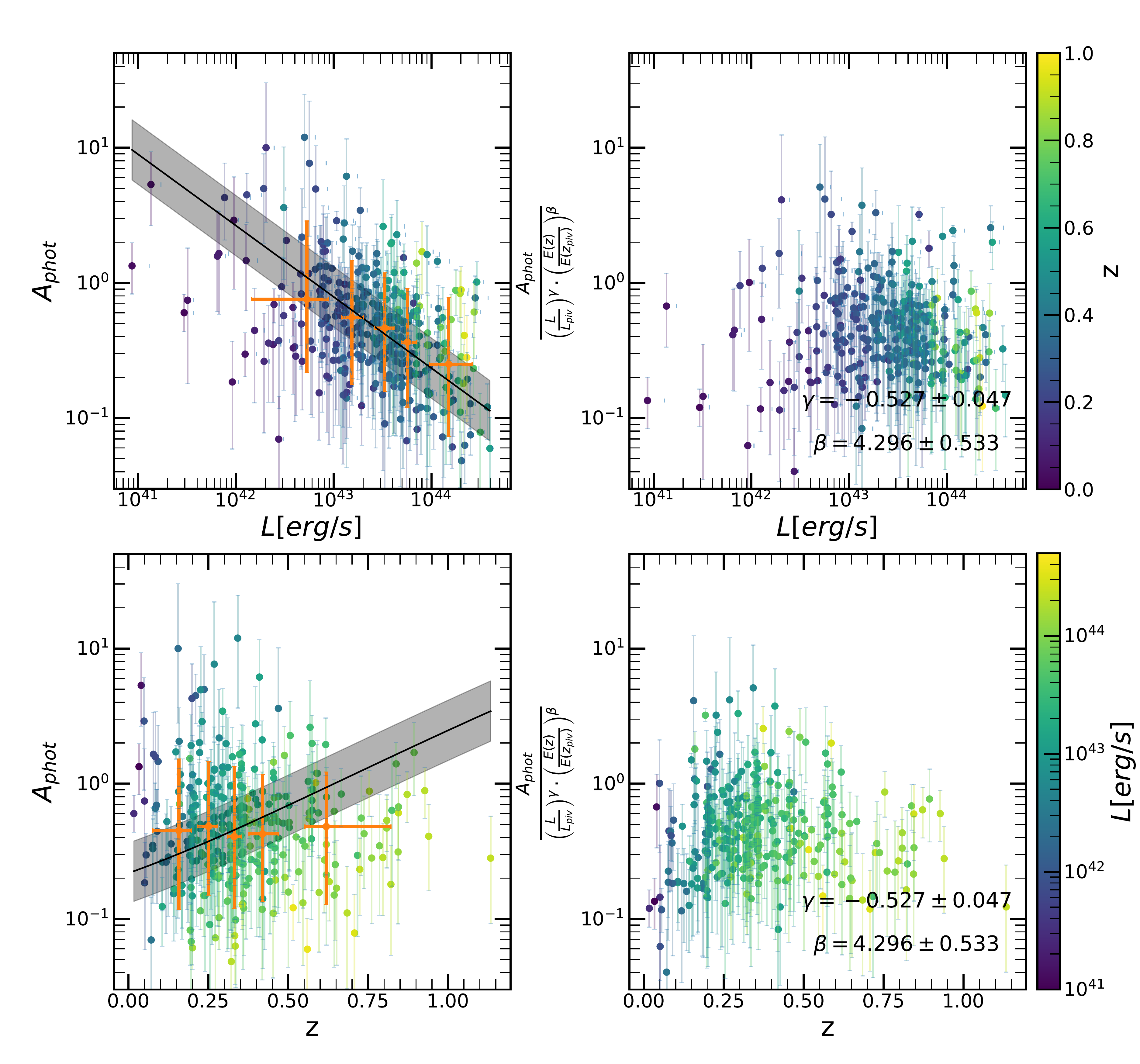}
    \caption{ continued. }
\end{figure*}

\subsection{Constraining the Morphological Parameters and Their Evolution with Redshift and Luminosity}
\label{sec:Lzevo}

eROSITA's great survey capabilities allow to reach down to luminosities as faint as $10^{41}$ erg/s (corresponding approximately to a group-scale mass of $10^{13} M_{\odot}$), and allows to study high redshift clusters up to $z$ = 1.2. For the first time, we are able to constrain both luminosity and redshift evolution of the morphological parameters simultaneously. To understand the underlying evolution in both redshift and luminosity simultaneously in morphological parameters, 
we model a power-law relation between each parameter $\mathcal{M}$, luminosity $L_{500}$, and redshift $z$, taking into account both the selection function (probability of detecting a cluster with given luminosity and redshift) and the luminosity function (the probability of detecting a cluster with a given luminosity and redshift at a fixed cosmology), thus taking into account the observed distribution in luminosity and redshift space of our clusters. 
We note that the scaling relation used to connect intrinsic morphological parameters $\mathcal{M}$ with luminosity and redshift is written as:
        \begin{equation}
\mathcal{M} = \mathcal{M}_0 \cdot \left( \frac{L_{500}}{L_{\rm piv}} \right)^\gamma \left( \frac{E(z)}{E(z_{\rm piv})} \right)^\beta ,
\label{eq:scaling_relation}
\end{equation}
where $L_{\rm piv}=2.6 \times 10^{43}$ erg/s and $z_{\rm piv}=0.35$ are the median luminosity and redshift respectively, of the eFEDS sample. $\gamma$ represents luminosity dependent slope, and $\beta$ represents the slope of the redshift evolution. The likelihood derivation is provided in detail in Appendix~\ref{app:like}. The final likelihood for each cluster is written in Eq.~\eqref{eq:final_like_i}.  The best-fit values of all morphological parameters are given in Table~\ref{tab:alpha_beta_evo}. We show the morphological parameters as a function of luminosity and redshift, while the best-fit models are shown in gray shaded regions in Fig.~\ref{fig:LZ_evo}. We highlight a few important results obtained from our joint modeling: 

\noindent {\bf Central density ($n_{0}$):} We find that the evolution of the central density with luminosity is slightly significant at a 4.2$\sigma$ level with a best-fit value of $0.163_{-0.038}^{+0.039}$. Overall, the central density evolves with redshift in agreement with the self-similar evolution, with the best-fit value of $\beta$ being $1.61_{-0.47}^{+0.49}$. The expected value of the slope is 2 in the self-similar model. On the other hand the significance of our measured evolution is only 3.4$\sigma$, therefore our result is marginally consistent with no redshift evolution. 

The observed increase in central density values with luminosity can be easily explained: first we notice that gas density does not scale with mass according to the self-similar model, however literature results \citep{pratt+09, Lovisari+15} show that gas mass fraction increases with cluster mass, and the only way to obtain a higher gas mass fraction at a fixed scaled radius is by increasing the gas density profile, and therefore increasing the central density as well. 
On the other hand, \cite{mcdonald17} find lack of evolution in the distribution of the central density combining the following three cluster samples: 49 low-$z$ X-ray selected clusters \citep{vikhlinin+09}, 90 SPT selected clusters from $z = 0.2$ to $z = 1.2$ \citep{mcdonald+13}, and the 8 most massive clusters in the SPT cluster survey \citep{bleem15} above $z = 1.2$. They interpret this as lack of evolution in the core properties of clusters in the last $\sim$10 Gyr. A similar result was later reported in \cite{Ghirardini+2020}, when comparing the core properties of the same 8 most massive clusters in SPT \citep{bleem15} above $z = 1.2$, and a \emph{Planck} selected sample of 12 low redshift ($z < 0.1$) clusters \citep[X-COP,][]{xcop}. On the contrary, \cite{Sanders+18} analyzed a sample of 83 SPT selected clusters from $z = 0.2$ to $z = 1.2$ and found no significance difference with respect to the self-similar evolution model. One important remark is that \cite{mcdonald17} selected clusters in order to satisfy evolutionary requirements, meaning that clusters at low redshift where chosen in a mass range expected from the evolutionary scenario \citep{Fakhouri10}, and therefore they might not be representative of the cluster population, but of how single clusters are expected to evolve.

\noindent {\bf Concentration Parameter ($c_{\rm SB}$):} Interestingly, concentration  computed in terms of $R_{500}$, $c_{\rm SB, \ R_{500}}$ as defined in \citet{maughan+12}, is consistent with no evolution with both luminosity, with a slope of $-0.031_{-0.035}^{+0.035}$), and redshift, with a slope of $-0.358_{-0.403}^{+0.392}$. However, when we use the concentration defined with physical distances instead, $c_{\rm SB, \ 40-400 \rm kpc}$ as defined in \citet{santos+08}, we measure a significant luminosity dependence, ($-0.231_{-0.038}^{+0.038}$), and no redshift evolution, ($0.607_{-0.449}^{+0.496}$). This might be a selection effect, where at low luminosity we tend to detect mostly cool-core clusters that present a peaked surface brightness.
However we argue that, since the same trend is not observed for the concentration computed using apertures with respect to $R_{500}$, $c_{\rm SB, \ R_{500}}$, this is not the case, and this trend is caused by the relative ratio of 400~kpc with $R_{500}$. In fact at low luminosity $R_{500}$ becomes smaller than 400~kpc, and this means that a larger fraction of cluster flux will be naturally concentrated within a fixed physical region, and therefore concentration computed at a fixed physical radius will increase significantly.

\noindent {\bf Ellipticity ($\epsilon$) and Cuspiness ($\alpha$):} Ellipticity does not show any significant dependence on luminosity (the slope is $-0.017_{-0.008}^{+0.008}$) and redshift with a slope of $0.270_{-0.096}^{+0.081}$. The change of the ellipticity in the sample with luminosity and redshift is mild (~$2\sigma$). The cuspiness of the ICM changes with luminosity with a slope of $-0.028_{-0.015}^{+0.015}$ and evolves with redshift with a slope of $0.473_{-0.181}^{+0.172}$. Similarly, the dependence of cuspiness on luminosity and redshift is moderate, $<2\sigma$ and $2.8\sigma$, respectively.

\noindent {\bf Power Ratios ($P_{10}, P_{20}, P_{30}, P_{40}$):} All the power ratios indicate significant luminosity and redshift dependence, therefore indicating that these parameters are difficult to interpret without a clear understanding on their evolution. We find that more luminous clusters have smaller power ratio, therefore clusters that look more relaxed, and that clusters at larger redshift  have larger power ratios, therefore less relaxed-looking clusters.

\noindent {\bf Gini coefficient ($G$):} The Gini coefficient shows clear luminosity dependence with a significance of $20\sigma$, where more luminous clusters have larger Gini values. Interestingly, we observe an anti-correlation of $G$ with redshift at a $5\sigma$ confidence level, going in the direction of decreasing with increasing redshift, thus indicating possible presence of more disturbed clusters at high redshift. This is particularly interesting since eROSITA's PSF is expected to produce smoother cluster images at higher redshift, since they appear smaller on angular scales, and smoother images should result in larger Gini coefficients. This shows the great potential of eROSITA in understanding the larger fraction of more disturbed clusters at high redshift, as found in \citet{santos+10}.

\noindent{\bf Photon asymmetry ($A_{phot}$):} The photon asymmetry shows a clear dependence with luminosity with a slope of $-0.527_{-0.049}^{+0.045}$, indicating that more luminous clusters tend to be more spherically symmetric, in other words, larger gravitational potential wells generate rounder objects. Furthermore it shows a significant evolution with redshift with a slope of $4.296_{-0.541}^{+0.525}$. In particular, the plot shows how our analysis is able to disentangle an apparent redshift independence. Evolution in redshift and luminosity conspire to produce an overall unevolving distribution, which starts to become clear when the color coding is added to the figure.

\cite{nurgaliev+17} compared a sample of 36 ROSAT selected clusters \citep[400d,][]{Burenin+07} in redshift range 0.35 $< z <$ 0.9 with a sample of 90 clusters selected from the \emph{SPT} survey in redshift range 0.25 $< z <$ 1.2. Measuring photon asymmetry, they find no statistical difference between the morphological properties of their X-ray and SZ selected cluster samples. Furthermore, they find absence of mass or redshift evolution in photon asymmetry. They also studied a sample of 85 simulated clusters, applying X-ray and SZ selection, and since the measured X-ray morphology is indistinguishable, they conclude that X-ray and SZ surveys are probing the same cluster population. They interpret this lack of evolution in the morphological properties as a lack of direct correlation between dynamical state of the clusters and these properties. Their results are justified by theoretical studies that find that substructures statistics can vary significantly on very short time scales during cluster mergers. The authors claim that this absence of difference between the X-ray and SZ cluster samples indicates that high resolution (1~arcmin) SZ surveys are not biased toward selecting preferentially merging clusters, while other SZ instruments with lower resolution might still have a bias since it is more likely that multiple clusters along the line of sight contribute to the integrated signal.

In summary, it is difficult to properly quantify the effect of PSF on the measured evolution with redshift for some parameters (Gini coefficient, photon asymmetry, and power ratios). The reduction in angular size of clusters at high redshift should imply that cluster images become smoother at higher redshift, and smoother images will result in larger Gini coefficients and smaller photon asymmetry values. The fact that we observe the opposite trend implies that the underlying evolution of these parameters should be even stronger than the measured values. The same logic then can be applied to the power ratios as their evolution should, also in that case, be stronger than the measured ones. The only parameter that would indicate the opposite trend, with more cool cores at high redshift, is the central density. However in this case the expected evolution is not very significant, and at 3$\sigma$ marginally consistent with no evolution.

\subsection{Redshift and Luminosity Independent Morphological Parameters}

The fitting procedure we have applied on our data to recover redshift evolution and luminosity dependence of our clusters, see Sect.~\ref{sec:Lzevo}, can be also exploited to define new morphological parameters which are constructed from the original distributions after the luminosity and redshift dependence are factored out:
\begin{equation}
\mathcal{M}_{new} = \mathcal{M} \cdot \left( \frac{L}{L_{ \rm piv}} \right)^{-\gamma} \left( \frac{E(z)}{E(z_{\rm piv})} \right)^{-\beta} ,
\label{eq:newM}
\end{equation}
using $\gamma$ and $\beta$ from the corresponding row of Table~\ref{tab:alpha_beta_evo}.
By construction, these newly defined parameters lack any redshift evolution or luminosity dependence, and therefore can be used to properly and correctly investigate the presence or absence of bimodality in our parameter distribution. In Fig.~\ref{fig:morph_corr} we show the changes in the distribution of the corrected morphological parameters once this correction is applied. Some correlations become tighter, while others become more loose.
In particular the correlation between core properties, central density, both concentrations, and cuspiness, become tighter, indicating that all these four parameters measure just the cluster core properties, and therefore the connection among them is almost one to one. On the other hand, correlation between parameters sensitive to large scale fluctuations, power ratios, Gini coefficient, and photon asymmetry in particular, become much less significant, indicating that these parameters are measuring complementary properties of the large scale emission of clusters.

\subsection{The New Relaxation Score}

In the literature, it is common to use morphological parameters to characterize the physical state of the ICM, determining which clusters are relaxed and which one are disturbed. In \cite{lovisari17}, the authors use a morphological parameter, which was previously introduced in \cite{rasia+13b}. This parameter combines the concentration and centroid shift values and then is used to distinguish between the relaxed and disturbed systems. The authors measure the deviation from the mean value of concentration and centroid shift, adding these together but changing the sign of the centroid shift deviation to take into account the anti-correlation between these two parameters. However, one important issue is that this parameter does not take into account the strength of the anti-correlation and, since the correlation between these parameters is very strong, this parameter is likely affected by double counting issues.

\begin{figure*}
\includegraphics[width=\textwidth]{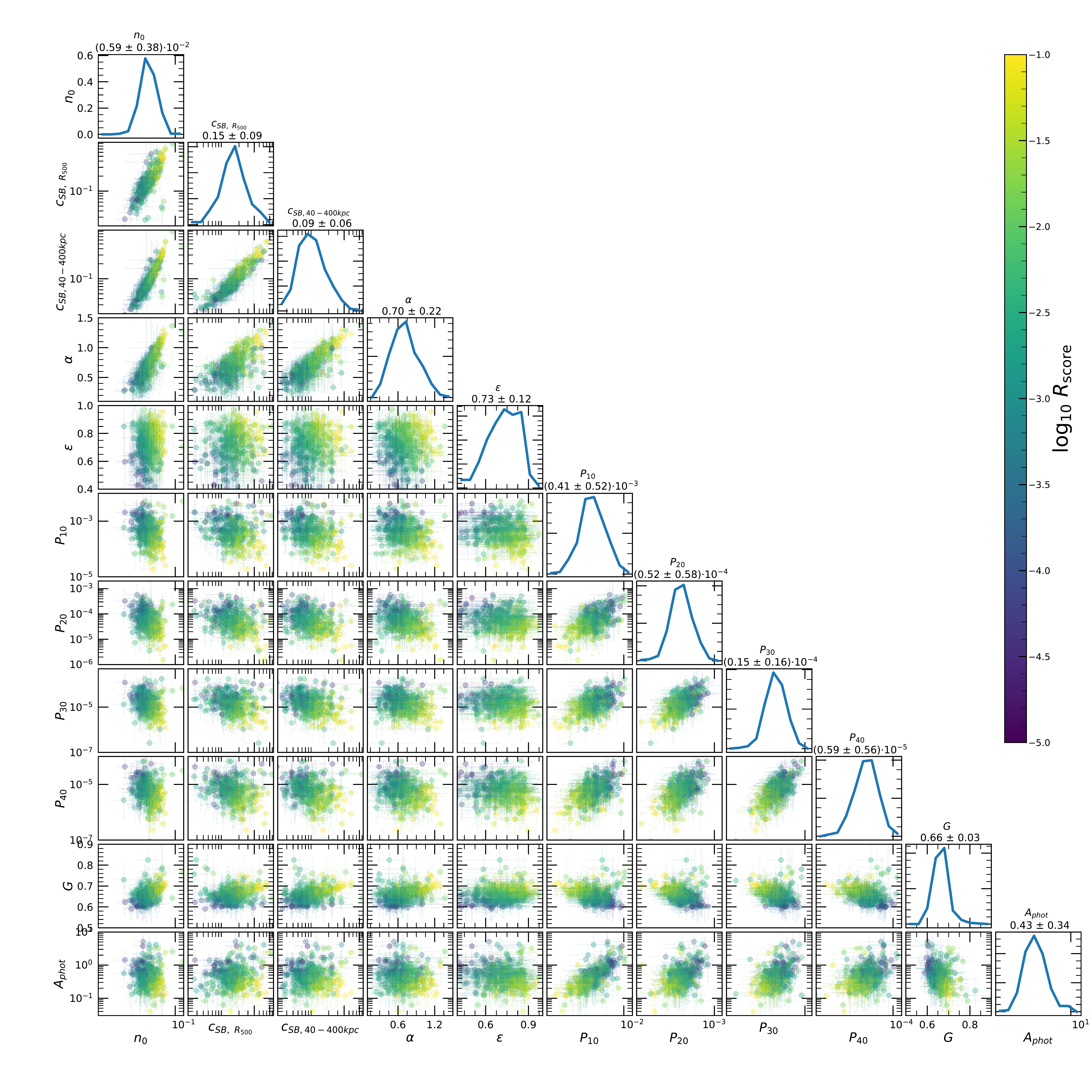}
\caption{Morphological parameters distribution in eFEDS clusters after the redshift and luminosity correction being applied. We color code the points according to their relaxation score, such that the reader can evaluate how the new parameter performs in identifying relaxed clusters.}
\label{fig:morph_corr}
\end{figure*}

Here in this work, we introduce a new parameter, the relaxation score, or $R_{\rm score}$, that contains the information stored in all the morphological parameters by taking into account both the sign (positive or negative) of the correlations of the morphological parameters with respect to concentration as well as the strength of these correlations. The definition of $R_{\rm score}$ is:

\begin{equation}
 R_{\rm score} = \int_{-\infty}^{\mathcal{M}_1}\dots\int_{-\infty}^{\mathcal{M}_n} \mathcal{MN} (\mathbf{\mu}, \mathbf{\Sigma}) d\mathcal{M}_1 \dots d\mathcal{M}_n ,
\label{eq:ccscore}
\end{equation}
where $\mathcal{MN} (\mathbf{\mu}, \mathbf{\Sigma})$ is the multivariate normal PDF in $n$-dimensional space of our parameter distribution, where $\mathbf{\mu}$ is an array containing the mean values for all the morphological parameters $\mathcal{M}_1 \dots \mathcal{M}_n$, and $\mathbf{\Sigma}$ is the covariance matrix computed from our data. In other words, we compute the cumulative distribution function (CDF) in $n$-dimensional space. For the quantities that anti-correlate with concentration, such as power ratios and photon asymmetry, we consider their reciprocal in the calculation of this quantity, Eq.~\eqref{eq:ccscore}, in order to build this parameters from quantities that correlate, not from a mix of correlating and anti-correlating quantities. As noted before, to avoid being biased towards redshift and luminosity dependence in our parameters, we consider the corrected morphological parameters as introduced in Eq.~\eqref{eq:newM} both for computation of means and covariance matrix, and for computation of the $R_{\rm score}$ itself. By construction, the $R_{\rm score}$ should be higher for objects with large concentration, central density, ellipticity, cuspiness, or Gini coefficient, and lower for objects with high photon asymmetry, or high power ratios. This is reflected in Fig.~\ref{fig:morph_corr}.

\begin{figure}
    \centering
    \includegraphics[width=0.5\textwidth]{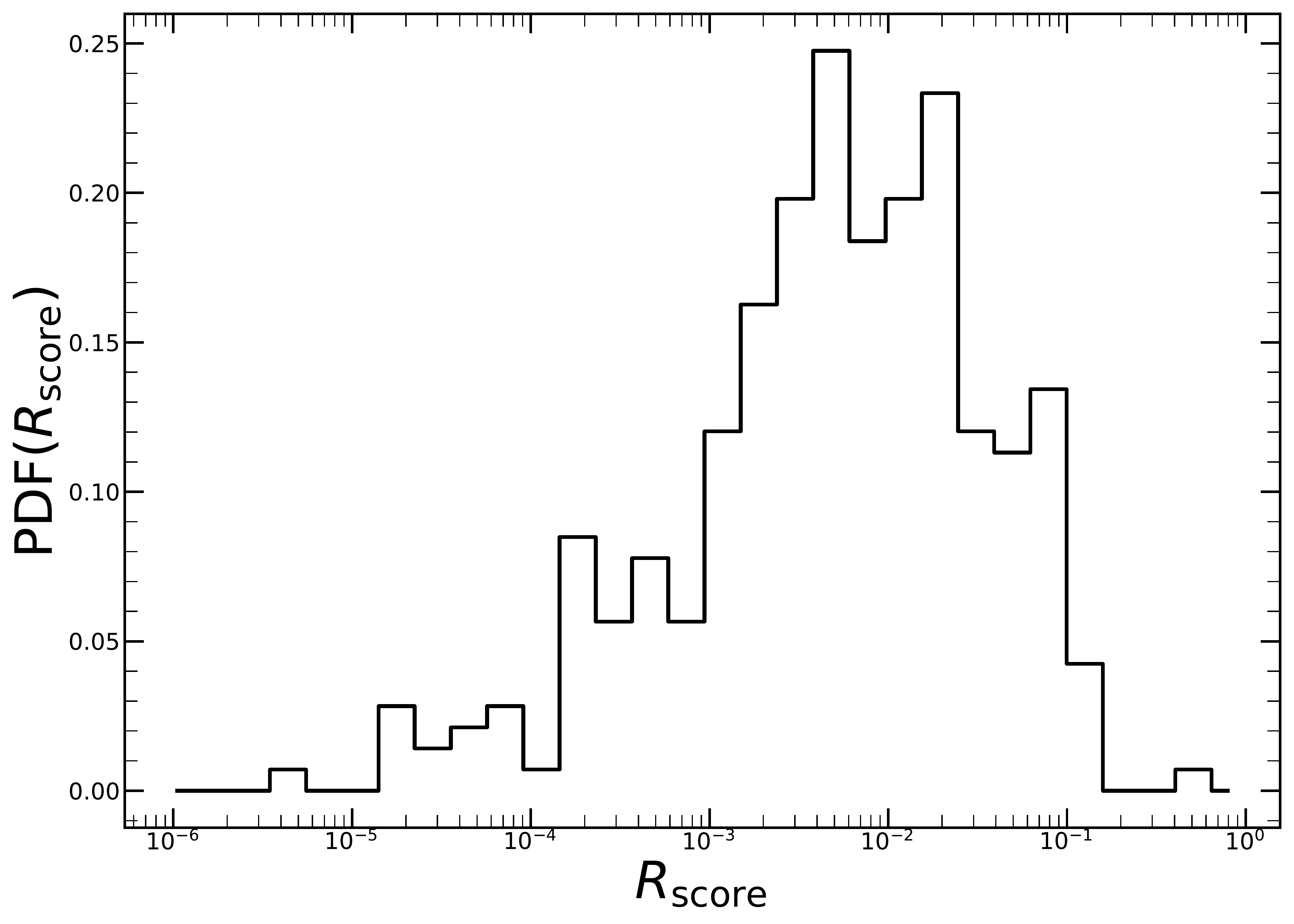}
    \caption{Distribution of the observed $R_{\rm score}$.}
    \label{fig:ccscore}
\end{figure}

\begin{figure}
    \centering
    \includegraphics[width=0.5\textwidth]{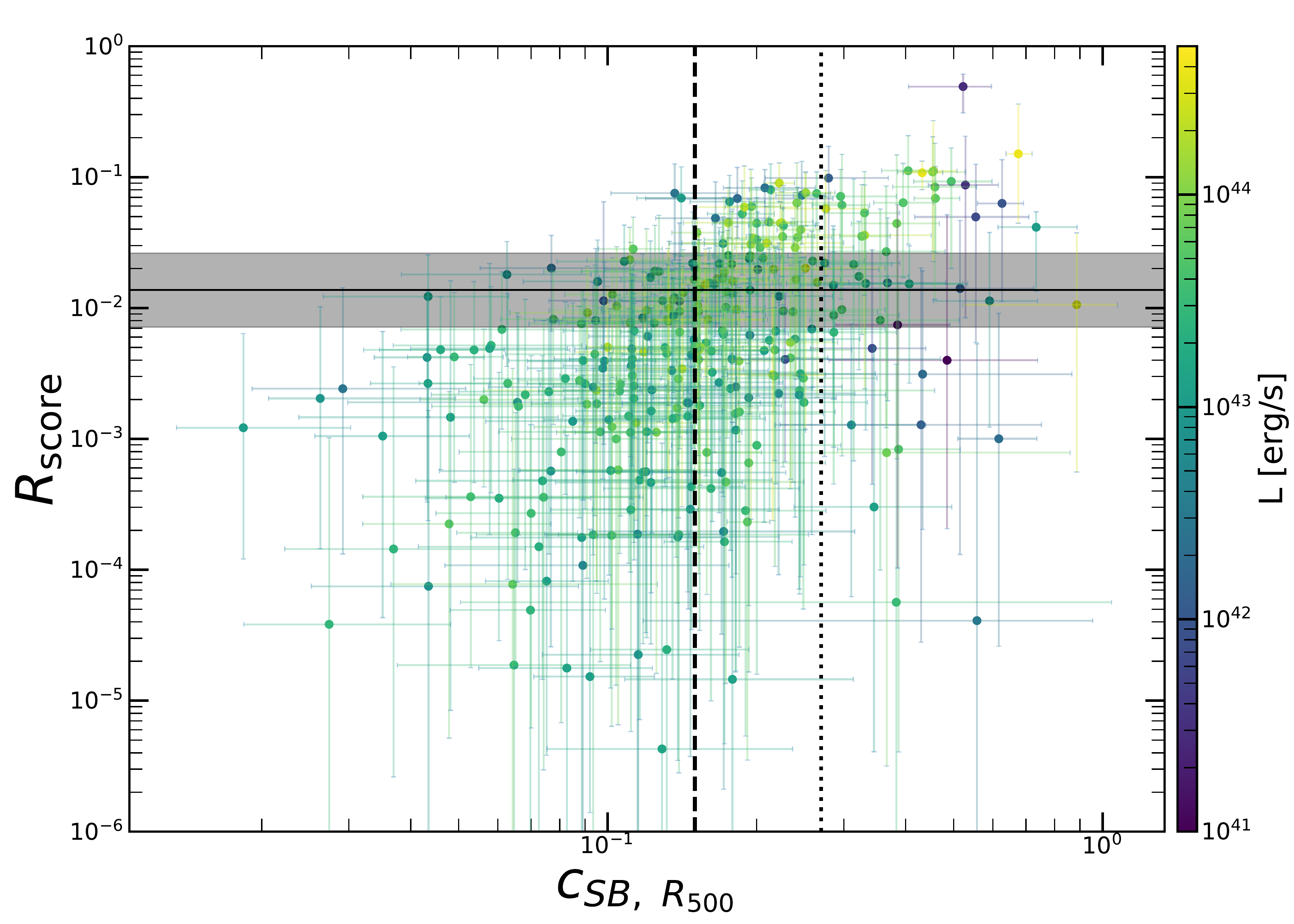}
    \caption{Distribution of the observed $R_{\rm score}$ versus concentration. Horizontal line represents the best fitting value for the log-normal distribution best fit. Dashed and dotted vertical lines indicate the thresholds to distinguish relaxed clusters from disturbed clusters suggested by \cite{lovisari17}, cool-core right of the dotted lines, and-non cool-core left of the dashed line.}
    \label{fig:ccscore_vs_conc}
\end{figure}

\begin{figure}
    \centering
    \includegraphics[width=0.5\textwidth]{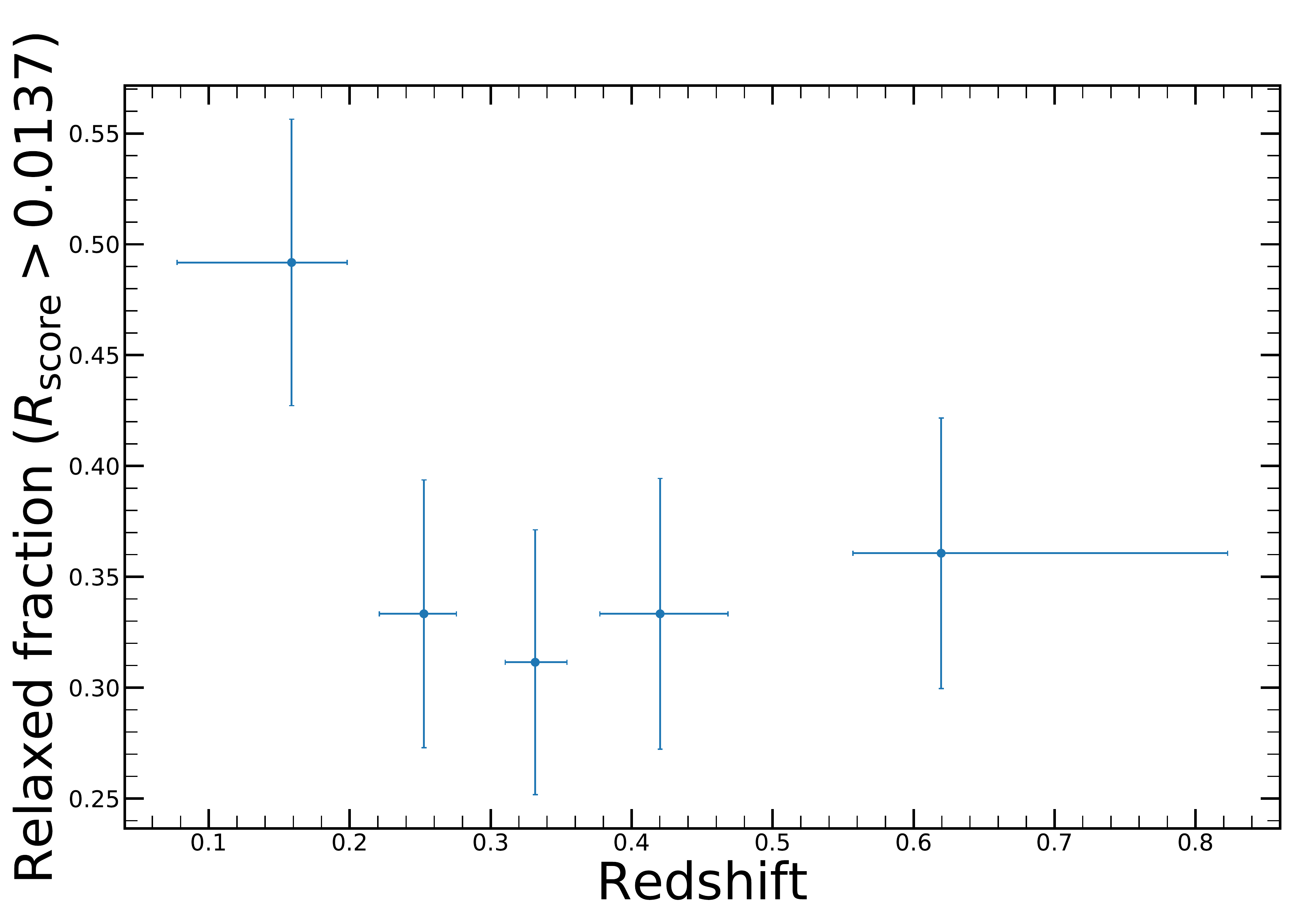}
    \caption{Evolution of the fraction of cool-core clusters as function of redshift. In each bin consists of 65 clusters.}
    \label{fig:ccscore_evo}
\end{figure}

We show the distribution of this newly introduced parameter in Fig.~\ref{fig:ccscore}. At a first glance, it looks unimodal, however we refer to the next section for the appropriate investigation on the best fitting distribution for this new parameter. It is worth to point out that this parameter has been constructed with redshift and luminosity corrected morphological parameters, therefore, by construction, it does not depend on redshift or luminosity. 

Fig.~\ref{fig:ccscore_vs_conc} shows the correlation between the concentration, computed using the \citep{maughan+12}'s definition with $R_{500}$, and the relaxation score. We observe a clear correlation between the two parameters, where relaxation score increases with concentration.
The correlation becomes insignificant above $c_{\rm SB,R_{500}}=0.27$, the threshold adopted by \cite{lovisari17} to identify relaxed clusters.
We interpret this as an indication that concentration is a good indicator to probe the very central state of clusters, but it does not correlate well with the relaxation state. In fact, a relaxed cluster will generally have a cool core in its center, while the opposite may not be true: a merger along the line of sight \citep[e.g.,][]{Dupke+2007}, or a merger in its initial stage, might have a centrally peaked density profile. The mergers whose axis is along the line-of-sight might still not be visible from the cluster center prospective, while it is visible on a larger scale.
Therefore we claim that clusters with large concentration can still be disturbed clusters, as supported by recent theoretical papers \citep[e.g., ][]{Rasia+15,biffi16}.
On the other hand, by construction, the relaxation score is able to capture the ICM dynamics from small to large scales, and therefore can be used to identify the cases where a cool core is still existing in an disturbed cluster.
Furthermore, in Fig.~\ref{fig:ccscore_vs_conc} we observe that the thresholds for disturbed clusters adopted by \cite{lovisari17}, are in agreement with the median of the log-normal best fit shown with the horizontal black line, meaning that almost all the clusters to the left of the dashed line are also below the horizontal line. Therefore we make use of this line to establish a threshold in terms of relaxation score: we define clusters to be relaxed if $R_{\rm score} > 0.0137$, while they are defined to be disturbed if the opposite happens. 

This definition allows us to compute the fraction of relaxed clusters as a function of redshift, by dividing the cluster sample in 5 redshift bins, and in each bin counting clusters based on their relaxation score. We have to add a further complication to this analysis: the relaxation score distribution for most clusters is wide and quite skewed, and therefore computing the fraction using just the median values will bias our results. Therefore we have to count the number of clusters that satisfy our threshold by first randomizing the relaxation score values by using the MCMC chain obtained when estimating it.
In other words we perform 10000 bootstraps iteration of the relaxation score for each cluster, each time calculating the fraction of clusters that are relaxed.

The obtained evolution of the fraction of objects that we classify as relaxed is shown in Fig.~\ref{fig:ccscore_evo}. We observe a fraction of relaxed clusters that is about 50\% in the lowest redshift bin, and at the higher redshift bins it becomes about 30--35\%, consistent with cool-core fraction estimated in previous works \citep[e.g.,][]{rossetti17,Sanders+18}. 
We therefore find a hint (slightly more than 1$\sigma$) of possible evolution with redshift of the fraction of relaxed clusters. However we might argue that the difference in the first bin can also be not a consequence of redshift evolution in the fraction of relaxed clusters, but may be caused by the incompleteness due to our source finding and confirmation technique in the detection of very extended, low-surface-brightness galaxy groups or clusters at low redshift, similarly as it was shown in a focused reanalysis of ROSAT observations \citep{Xu+18}.

\begin{table}[]
\centering
\resizebox{0.5\textwidth}{!}{  
\begin{tabular}{c c c c c c c}
Parameter & $\Delta B_{N}$ & $\Delta B_{2N}$ & $\Delta B_{SN}$ & $\Delta B_{LN}$ & $\Delta B_{2LN}$ & $\Delta B_{SLN}$ \\
\hline
\hline
$n_0$ & 38.44 &  6.37 &  7.82 & 3.13 &  4.23 &  0.00 \\
$c_{\rm SB, \ R_{500}}$ & 40.83 &  2.90 &  7.68 & 0.00 &  1.35 &  0.40 \\
$c_{\rm SB, \ 40-400 \rm kpc}$ & 55.12 &  7.56 &  15.37 & 4.61 &  3.10 &  0.00 \\
$\alpha$ & 1.83 &  3.99 &  0.84 & 0.00 &  3.09 &  0.80 \\
$\epsilon$ & 0.96 &  4.05 &  0.02 & 1.12 &  4.44 &  0.00 \\
P$_{10}$ & 51.24 &  8.64 &  14.39 & 0.00 &  3.02 &  1.32 \\
P$_{20}$ & 12.48 &  3.19 &  0.00 & 3.28 &  0.90 &  2.91 \\
P$_{30}$ & 3.36 &  6.83 &  0.00 & 8.99 &  4.36 &  4.08 \\
P$_{40}$ & 1.12 &  4.46 &  0.00 & 11.57 &  2.18 &  4.79 \\
G & 1.04 & 6.02 & 0.00 & 1.00 & 5.12 & 0.68 \\
A$_{phot}$ & 48.85 &  3.03 &  15.47 & 3.05 &  1.10 &  0.00 \\
\hline
$R_{\rm score}$ & 25.13 & 7.41 & 6.34 & 2.23 & 4.14 & 0.00 \\

\hline
\hline
\end{tabular}
}
\caption{Bayes factor with respect to the best fit model for single normal distribution (N) fit, double normal distribution (2N), skewed normal distribution (SN), log-normal distribution (LN), double log-normal distribution (2LN), and skewed log-normal distribution (SLN). 
A zero indicates that it is the best fitting model, a value larger than 5 indicates a strongly disfavoured model.
}
\label{tab:gaussian_fit}
\end{table}

\subsubsection{Investigating the bimodality}

We investigate whether in our eFEDS cluster subsample we observe an indication for bimodality in the morphological parameter distribution, and for the relaxation score as well. Our strategy consists in fitting a normal, a log-normal, a double normal, a double log-normal, a skewed normal, and a skewed log-normal distribution to our data. For each distribution we estimate the bayesian evidence, a quantity that is widely used in bayesian statistic  when doing model comparison, thus it is a very good indicator in our specific case. In particular, we compute the Bayes factor, obtained by the difference between the bayesian evidences of the compared models. This is a quantity that directly relates to the relative odds of the two models compared, by telling which of the two models is more likely, and how much more probable it is. We make use of the Jeffreys scale \citep{jef61} to estimate the model preference and its relative odds. In particular, a model is strongly preferred if the Bayes factor is at least 5; we say it is weakly preferred if the Bayes factor is more than 2.5 but less than 5. These differences in Bayesian evidences correspond approximately to 3$\sigma$ and 2$\sigma$ significance level.
We remind the reader that, since the bayesian evidence, being defined as the integral of the likelihood over the prior, takes into account by construction the size of the parameter space, the Bayes factor penalizes by definition a model with more parameters or with larger parameter space. 

However we cannot apply this fitting to the observed morphological parameters, since they are affected by redshift and luminosity dependence, as characterized in Sect.~\ref{sec:Lzevo}, and therefore we have to first correct these parameters by this evolution, and then characterize which is their best fitting parent distribution.
Additionally, we note that the low photon statistics in our data naturally cause large uncertainties in the measured morphological parameters, therefore we need to consider these uncertainties when fitting the data distribution, building a likelihood as the following:
\begin{equation}
P(\hat{\mathcal{M}} | \theta, \mathcal{D}) = \int P(\hat{\mathcal{M}} | \mathcal{M}) \cdot \mathcal{D}(\mathcal{M} | \theta) d\mathcal{M} ,
\end{equation}
where $P(\hat{\mathcal{M}} | \mathcal{M})$ takes into account uncertainties in the morphological parameter estimation, and $\mathcal{D}$ represents the distribution fitted to the data (normal, log-normal, double normal, double log-normal, skewed normal, or skewed log-normal) that depends on the parameters $\theta$.
As stated above we are not really interested in the values of the fitted model parameters, but in the value of the bayesian evidences $E$:
\begin{equation}
E =  P(\hat{\mathcal{M}} | \mathcal{D}) = \int P(\hat{\mathcal{M}} | \theta, \mathcal{D}) P(\theta | \mathcal{D}) d\theta ,
\end{equation}
which is easy to derive the model posterior, since it is linked  to the previous equation by the Bayes theorem:
\begin{equation}
P(\mathcal{D} | \hat{\mathcal{M}}) = \frac{P(\hat{\mathcal{M}} | \mathcal{D}) P(\mathcal{D})} { P(\hat{\mathcal{M}})} .
\end{equation}

The fit of the model to the data, and the evidence estimation is performed using the Bayesian nested sampling algorithm MultiNest \citep{multinest}, and we compare the values of the Bayes factor for the six fitted distributions against the best fitting distribution in Table~\ref{tab:gaussian_fit}. Overall, a skewed distribution is always preferred by our data, except for concentration, cuspiness, and $P_{10}$, which favor a log-normal distribution.

We point out that a bimodal distribution, either double normal or double log-normal, is never the preferred model to our data. On the other hand, a double log-normal has always Bayes factor difference with respect to the best fitting model smaller than 5, hence we cannot exclude the possibility that the observed preference for a single peaked distribution is due to random fluctuation.
Interestingly, we point out that for concentration we find that the best fitting distribution is log-normal, as already found in a previous study using a \emph{Planck} selected sample \citep{rossetti17}. This analysis has been performed on the $R_{\rm score}$ as well, and the best fitting distribution is the skewed log-normal distribution. 

It is worth to point out that none of the morphological parameters analyzed in this work prefers significantly a skewed distribution (we mean that Bayes factor of the best non-skewed distribution versus best skewed distribution is always smaller than 5). As an exercise, we remove the fit performed using a skewed distribution from our analysis, and compare single versus double peaked distributions in order to understand whether there would be a clear preference toward a double peaked distribution. We find that to be the case only for concentration when computed between 40 and 400~kpc, and for photon asymmetry, but we point out that even for these two cases, the significance of this preference is so small that it is just barely worth mentioning. We can conclude that in our eFEDS cluster subsample there is no indication for the presence of bimodality in the distribution of our morphological parameters, and even for the $R_{\rm score}$, the parameter that should have separated best the two cluster populations, the best fitting distribution is either skewed log-normal or log-normal (depending on if we allow for a skewed distribution or not).

In the last years there have been several studies that indicate that the observed cluster population can be split into two categories, cool-core and non-cool-core clusters  \citep[e.g.,][]{Sanderson+09,cavagnolo+09,hudson+10,liu2020}. 
These studies find that some morphological indicators have a characteristic bimodal distribution, and therefore they conclude that clusters can be easily divided into two distinct populations, cool cores and non cool cores. Typically, clusters, observed by \emph{Chandra} and \xmm, due to the particular trends in cluster research in the last two decades, tend to be extreme clusters and are not representative of the cluster population in the Universe. These clusters have been observed either for constraining AGN feedback, therefore the AGN contribution to ICM is quite strong, or clusters with an major merger with significant non-thermal pressure support. Therefore, it is clear that such selection would bias cluster population toward the atypical clusters and the observed bimodality could easily be explained by the sample selection. For instance, \cite{cavagnolo+09} studied a cluster sample of 239 objects. Selection was based purely on the presence of clusters in the  \emph{Chandra} archive \citep[ACCEPT, see ][]{Donahue+06}, therefore they have a large variety of clusters that spans very wide redshift and mass range. They find that the distribution of the values of the central entropy is bimodal, and they interpret the results by stating that this indicates a clear dichotomy between cool-core and non-cool-core clusters. \cite{Sanderson+09} investigated a sample of 20 clusters at very low redshift ($z<0.1$) selected from HIFLUGCS \citep{Reiprich+02}. They studied the entropy slope distribution at fixed radius, a morphological parameter that is quite useful in understanding cooling in the cores of galaxy clusters. We cannot compute this parameter, since it requires higher data quality than eFEDS to compute the temperature profiles needed for entropy profile estimation. They find significant presence of a bimodal distribution which they attribute to the presence of two distinct cluster populations, cool cores and non cool cores. We argue that this sample was quite limited in the number of cluster involved, and therefore the bimodal distribution is not statistically significant, or that the cluster population studied is not the same as the one studied here, given the large flux used in HIFLUGCS to select the cluster sample.
\cite{hudson+10} followed up \cite{Sanderson+09} by enlarging the cluster sample used for the analysis: they have also selected clusters from  HIFLUGCS cluster sample, however they more than triple the number of studied clusters, reaching 64, and double the redshift range, reaching up to z $<$ 0.2, however only three clusters in the redshift range between $z = 0.1$ and $z = 0.2$ were included in this study. They studied several more morphological indicators, including central density, central entropy, cuspiness, core temperature, finding indication for bimodality in several of these parameters. Similarly to \cite{Sanderson+09}, this sample could be either affected by low numbers statistics, or, being based on clusters with very high fluxes, they are selecting an entire different cluster population.

On the other hand, several other studies \citep[e.g.,][]{pratt+10,santos+10,ghirardini_a+17,Yuan+20}, investigated similar morphological indicators but do not find presence of bimodality, consistent with our results. For instance, \cite{santos+10} studied the distribution of the concentration parameter in three different cluster samples: a low and an intermediate redshift subsamples of the 400 Square Degree ROSAT PSPC survey \citep[][]{Burenin+07} with median redshift of $z = 0.08$ (28 clusters) and $z = 0.59$ (20 clusters), 15 high redshift clusters from both ROSAT Deep Cluster Survey \citep[RDCS,][]{Rosati+98}, and the Wide Angle ROSAT Pointed Survey\citep[WARPS,][]{Jones+98} with median redshift of $z = 0.83$ (15 clusters). 
Their cluster sample was one of the first that was able to study cluster morphological parameters from low to high redshifts. In their sample they did not find any indication of bimodal distribution of the concentration parameter, thus refuting claims of a cool core / non cool core bimodality. They conclude that cluster population cannot be easily divided in two populations because transition between cool cores and non cool cores happens smoothly, since the observed distribution is unimodal. Nevertheless, they find hints of evolution of the concentration parameter, which indicates, similarly to our results, that at high redshift there are fewer cool cores than at low redshift. \cite{pratt+10} studied the entropy in the REXCESS cluster sample \citep{Bohringer+07}, a local luminosity limited cluster sample comprised of 33 clusters at $z < 0.2$ drawn from the REFLEX cluster sample \citep{Bohringer+01}. They observe two peaks in the distribution of the central entropy value and slope, however even tough they estimate that a single Gaussian is the worst description of their data, they cannot distinguish between a skewed normal distribution and a bimodal Gaussian distribution. \cite{Yuan+20} analyzed the entire \emph{Chandra} archive, which contains 964 clusters up to $ z = 1.5$. They studied the distribution of concentration and P30, finding absence of bimodality in the distribution of these morphological indicators. \cite{ghirardini_a+17} investigated the population of clusters observed by \emph{Chandra} at $z > 0.4$, selecting only those clusters with at least 20 ks observing time; they find no indication for preference for bimodal distribution in central entropy distribution.

Recently, there have been great efforts in characterizing the role of the selection effects in determining the distribution of the objects using morphological parameters as it has become increasingly important to disentangle the selection bias from the measurements. In particular, it was proposed that X-ray cluster samples selecting clusters according to their flux were affected by cool-core bias \citep{Eckert+11}. 
\cite{santos+17} compared the fraction of cool-core clusters, estimated using threshold in concentration, cuspiness, and central density, in \emph{Planck} ESZ cluster sample \citep{PESZ} of 164 clusters, with a flux limited X-ray selected sample \citep{Voevodkin+04} which is a subsample of the HIFLUGCS \citep{Reiprich+02} cluster sample. They find that X-ray selected clusters contain a significantly larger fraction of cool-core clusters compared to the sample of SZ selected clusters. \cite{rossetti17} studied the evolution and mass dependence of concentration parameter in the \emph{Planck} cosmology sample \citep[][]{PSZ1}, containing 189 clusters, and compared their results with same properties measured in ME-MACS X-ray selected sample \citep{ME-MACS}, containing 103 clusters. They investigated the cool core fraction by dividing their sample in two different redshift bins, and in two mass bins. They find a 1.5$\sigma$ indication that low mass systems have smaller cool core fraction, and they find no evolution with redshift.
Furthermore, they investigate presence of bimodal distribution in the concentration parameter. Interestingly, they find that \emph{Planck} selected cluster do not show any indication of a bimodal distribution, and on the contrary the best fitting distribution is a log-normal. However the X-ray selected ME-MACS cluster sample is best fitted by a bimodal distribution. They, assuming that \emph{Planck} sample is a representative sample of cluster population, simulate detection of a mock ME-MACS-like sample, and observe an appearance of a double peaked distribution when applying ME-MACS selection criteria.

It is quite challenging to compare our results with most of the past studies mentioned above, as the eFEDS cluster sample we use in this work is different in luminosity, mass, and redshift, with respect to these works. 
It is possible to extrapolate our best fitting results, however this would ignore possible real difference in the underlying cluster population between low and high luminosity clusters. The only study which provides such an opportunity is \citet{lovisari17}. The authors studied a sample of 150 clusters selected from the \emph{Planck} early cluster catalog \citep{PESZ}. These clusters extend in redshift up to $z = 0.55$, and have median luminosity of $2.5_{-1.4}^{+1.8} \times 10^{44}$  erg/s\footnotemark[1]. In comparison, the eFEDS sample involving 325 clusters extends out to redshifts of $z=1.2$ with a median luminosity of $2.6_{-1.8}^{+4.9} \times 10^{43}$ erg/s\footnotemark[1].\footnotetext[1]{The numerical values after the median value are not its errors but indicate difference with the 16$^{\rm th}$ and 84$^{\rm th}$ percentiles in the distribution.} It is clear that our redshift range is twice as wide and our median luminosity is factor 10 lower than in L17.
They investigated the thresholds that yield the most complete and the most pure sample of relaxed and disturbed clusters. However, in their work, they use the visual inspection method with few astronomer to classify relaxed and non relaxed clusters. This method obviously could be biased and is very difficult to reproduce or apply in our sample, where photon statistic is low, and the number of cluster is higher. They report that it is generally better to use more than one single parameter in categorizing clusters as relaxed and disturbed. In particular, they mention that the use of a new parameter obtained as a combination of concentration and centroid shift \citep{rasia+13b} is very powerful in discriminating cool cores and non cool cores as we suggest in this work. They find that central density and Gini coefficient distribution are statistically different when splitting their sample in two according to the luminosity such that more luminous objects tend to be more concentrated, to have larger Gini coefficient, and to have larger gas central densities, similarly to our results.
In Fig.~\ref{fig:LZ_evo} we add the median value and the scatter observed in L17 for central density, concentration (the one defined with respect to $R_{500}$), and cuspiness, since these are the parameters for which a direct comparison can be made.
We also calculate the median morphological parameters for the eFEDS clusters that fall within the L17 luminosity and redshift range. We notice that for concentration and central density we have a very good agreement in both the median value and the scatter of the parameters, indicating that eROSITA and \emph{Planck} selected clusters have similar morphological properties. For the cuspiness instead we generally measure density profiles that are quite steeper on average. However, we remind that cuspiness is a morphological parameter that has the very large and skewed  uncertainty in our work, therefore it is not a very significant difference.

\section{Conclusions}
\label{sec:conclusion}

In this paper we characterize the distribution of the morphological parameters in a sample of galaxy clusters and groups detected in the  eROSITA Final Equatorial-Depth Survey (eFEDS). Thanks to the great survey science capabilities of eROSITA, we are able to map these properties in a parameter space region previously unexplored, from low mass and low redshift groups, to high redshift clusters. 

The eFEDS surveys includes a total of 542 candidate clusters and groups of galaxies, down to the flux limit of $\sim10^{-14}~{\rm erg}~{\rm s}^{-1}~{\rm cm}^{-2}$ in the soft band (0.5--2~keV) within 1\arcmin (see Liu A. et al., submitted). We apply a selection cut on the extent and detection likelihoods of 12 to reduce the total contamination in the sample. This selection reduces the contamination to 14\% leaving a total 325 clusters in the final sample. This sample covers a luminosity range from $9 \times 10^{40}$ erg/s to $4 \times 10^{44}$ erg/s and a redshift range from 0.017 out to 1.2. Our conclusions obtained from this analysis are summarized below.

In this work, covering a large luminosity and redshifts range, our sample allows to constrain the evolution of morphological parameter with both redshift and luminosity. The evolution in the central density is slightly significant both for luminosity and redshift. Overall, the central density evolves with redshift in agreement with the self-similar model and is consistent with the previous studies. Similarly, most morphological indicators, in particular Gini coefficient, power ratios, and photon asymmetry, show significant evolution with redshift and luminosity. 

The evolution constrained in these parameters are then modelled to convert the evolution-independent morphological parameters into a new morphological parameter, the relaxation score.
Using this new parameter, we look for indication of bimodal distribution in our parameters, however, we find that skewed distribution or single peaked normal or log-normal distribution provide a better representation of our observations. By defining a threshold in relaxation score we are able to determine whether a cluster is relaxed or not, and we found that the fraction of relaxed objects does not evolve with redshift.
Based on the correlation between concentration and relaxation score, we define a threshold to distinguish relaxed clusters from disturbed clusters by converting concentration threshold used in \cite{lovisari17} into a threshold in relaxation score. Using  this threshold on a redshift independent parameter we find that at low redshift the fraction of relaxed clusters are about 50\%. This fraction decreases to  30--35\% at higher redshifts at $z>$ 0.2. However, we argue that the high fraction of relaxed clusters at the lowest redshift bin could be explained by the incompleteness in the sample. Due to our detection strategy, we are more sensitive to detect relaxed groups and clusters with beta model profiles. The surface brightness distributions of disturbed clusters tend to be different than beta model profiles, flatter and more extended therefore the detection algorithm might miss these nearby objects. Another challenge is to confirm the redshifts of these groups and cluster using optical surveys. Disturbed clusters and groups at low redshifts ($z<$ 0.2) are likely to be more extended compared to relaxed clusters and groups. As they appear to be so extended on the sky that red-sequence based algorithms struggle to select the correct optical center and the correct members galaxies. Therefore, the optical incompleteness in the low-redshift regime might affect the disturbed clusters more than relaxed clusters.
However in our eFEDS cluster sample we do not select our clusters upon any optical properties, therefore this deficit of disturbed clusters is not caused by optical cleaning of the sample.

This work bridges the gap between characterization of cluster samples detected by the previous flux limited X-ray surveys, while setting the stage for the upcoming eROSITA All-Sky survey. Cluster surveys performed with ROSAT \citep{Voges+99}, REXCESS \citep{Bohringer+07}, HIGFLUGS \citep{Reiprich+02}, and ME-MACS \citep{ME-MACS}, have successfully covered clusters with high-luminosities $>5\times 10^{44}$ erg/s out to redshifts of 0.5. Ground based SZ selected cluster surveys with SPT \citep{Bleem+20} and ACT \citep{Hilton+21}, on the other hand, being sensitive to the most massive and high luminosity clusters, extended our knowledge of underlying cluster populations out to high redshifts of 1.7 \citep{nurgaliev+17}. For instance, The highest luminosity we reach in the eFEDS sample is about $4 \times 10^{44}$ erg/s. The eRASS1 will explore the cluster samples with luminosities down to $\sim 10^{40}$ erg/s and redshifts out to 1.5, probing a completely new cluster population with eROSITA complementary to previous flux limited X-ray surveys and mass limited SZ surveys.

\begin{acknowledgement}
This work is based on data from eROSITA, the soft X-ray instrument aboard SRG, a joint Russian-German science mission supported by the Russian Space Agency (Roskosmos), in the interests of the Russian Academy of Sciences represented by its Space Research Institute (IKI), and the Deutsches Zentrum f{\"{u}}r Luft und Raumfahrt (DLR). The SRG spacecraft was built by Lavochkin Association (NPOL) and its subcontractors, and is operated by NPOL with support from the Max Planck Institute for Extraterrestrial Physics (MPE).

The development and construction of the eROSITA X-ray instrument was led by MPE, with contributions from the Dr. Karl Remeis Observatory Bamberg \& ECAP (FAU Erlangen-Nuernberg), the University of Hamburg Observatory, the Leibniz Institute for Astrophysics Potsdam (AIP), and the Institute for Astronomy and Astrophysics of the University of T{\"{u}}bingen, with the support of DLR and the Max Planck Society. The Argelander Institute for Astronomy of the University of Bonn and the Ludwig Maximilians Universit{\"{a}}t Munich also participated in the science preparation for eROSITA.

The eROSITA data shown here were processed using the eSASS software system developed by the German eROSITA consortium.

E.B. acknowledges financial support from the European Research Council (ERC) Consolidator Grant under the European Union’s Horizon 2020 research and innovation programme (grant agreement No 101002585).
DNH acknowledges support from the ERC through the grant ERC-Stg DRANOEL n. 714245.
This work was supported in part by the Fund for the Promotion of Joint International Research, JSPS KAKENHI Grant Number 16KK0101. 

\end{acknowledgement}

\bibliographystyle{aa} 
\bibliography{references} 

\begin{appendix}
\onecolumn
\section{Likelihood derivation}
\label{app:like}

The likelihood is the probability distribution of the data given parameters in the model we use when seen as a function of the parameters.

\begin{equation}
\mathcal{L} (\theta | {\rm data}) = P({\rm data} | \theta) = P(\hat{L},\hat{\mathcal{M}}, \hat{z}, \hat{t}_{\rm exp} , \mathcal{M}, L, z, t_{\rm exp}, {\rm det = 1} | \theta, C)
\end{equation}
where with $\hat{x}$ we indicate the measured quantities, and with $x$ the intrinsic quantities. Here $x$ can be the morphological parameter of interest $\mathcal{M}$, luminosity $L$, exposure time $t_{\rm exp}$, or redshift $z$. With ${\rm det = 1}$ we indicate that the detection of the cluster has happened. $\theta$ indicates scaling relation parameters: the normalization $\mu$, luminosity slope $\gamma$, redshift dependence $\beta$, and intrinsic scatter $\sigma$. We further point out that our model depends on cosmology $C$, however a cosmology fit is not one of the goals of this paper, hence the cosmology is assumed to be fixed to a concordance cosmology with $H_0 = 70 $ km/s, $\Omega_{\rm m}$ = 0.3, $\Omega_\Lambda$ = 0.7.

Furthermore we assume that exposure time $t_{\rm exp}$ and redshift $z$ are perfect measurements, meaning that they do not have any uncertainty associated to them (in principle they have extremely small uncertainties, but including them in our fit will only slow down the computation without any significant improvement in the quality of the fit). This is mathematically expressed as 
\begin{align}
&P(\hat{z} | z) = \delta (z-\hat{z}) \nonumber \\
&P(\hat{t}_{\rm exp} | t_{\rm exp}) = \delta (t_{\rm exp}-\hat{t}_{\rm exp})
\end{align}
where with $\delta_x(y)$ we indicate Dirac's delta function. This means that we can marginalize over $t_{\rm exp}$ and $z$, thus our likelihood becomes the marginal likelihood as follows:
\begin{align}
\mathcal{L} (\theta | {\rm data}) = P(\hat{L},\hat{\mathcal{M}}, \hat{z}, \hat{t}_{\rm exp} , \mathcal{M}, L, {\rm det = 1} | \theta, C) &= \int P(\hat{L},\hat{\mathcal{M}}, \hat{z}, \hat{t}_{\rm exp} , \mathcal{M}, L, {\rm det = 1} | z, t_{\rm exp}, \theta, C) P(z | C) P(t_{\rm exp}) dz \ dt_{\rm exp} \nonumber \\  &=  \int P(\hat{L},\hat{\mathcal{M}}, \mathcal{M}, L, {\rm det = 1} | z, t_{\rm exp}, \theta, C) P(z | C) P(t_{\rm exp}) P(\hat{z} | z) P(\hat{t}_{\rm exp} | t_{\rm exp}) dz \ dt_{\rm exp} \nonumber \\&= \int P(\hat{L},\hat{\mathcal{M}}, \mathcal{M}, L, {\rm det = 1} | z, t_{\rm exp}, \theta, C) P(z | C) P(t_{\rm exp}) \delta (z-\hat{z}) \delta (t_{\rm exp}-\hat{t}_{\rm exp}) dz \ dt_{\rm exp} \nonumber \\&=  P(\hat{L},\hat{\mathcal{M}}, \mathcal{M}, L, {\rm det = 1} | \hat{z}, \hat{t}_{\rm exp}, \theta, C) \cdot P(\hat{z} | C) \cdot  P(\hat{t}_{\rm exp}).
\end{align}
\noindent Then in the following we break down the likelihood in several pieces (selection function, errors on data, scaling relation, and cosmology dependence) by exploiting even more the Bayes theorem $P(A,B) = P(A|B) P (B) = P(B|A) P(A)$

\begin{equation}
\mathcal{L} (\theta | {\rm data}) = P(\hat{L},\hat{\mathcal{M}}, \hat{z}, \hat{t}_{\rm exp}, \mathcal{M}, L, {\rm det = 1} | \theta, C) = P({\rm det = 1} | \hat{L},\hat{\mathcal{M}}, \hat{z}, \hat{t}_{\rm exp}, \mathcal{M}, L, \theta, C) \cdot P(\hat{L},\hat{\mathcal{M}}, \mathcal{M}, L | \hat{z}, \theta, C) \cdot P(\hat{z} | C) \cdot  P(\hat{t}_{\rm exp})
\end{equation}
where we have noticed that theselection function is the only term that depends on exposure time, and the selection function was isolated from this equation. The selection function is written as
\begin{equation}
f_{sel} = P({\rm det = 1} | \hat{L},\hat{\mathcal{M}}, \hat{z}, \hat{t}_{\rm exp}, \mathcal{M}, L, \theta, C) = P({\rm det = 1} | L, \hat{z}, \hat{t}_{\rm exp}, C) \ ,    
\end{equation}
In this equation we have assumed that the selection function is independent of the measured quantity $\hat{L}$ but depends only on the corresponding intrinsic quantity $L$, and that morphological parameters do not enter directly into the selection function. 

\noindent Next we split errors on the data from the intrinsic quantities:
\begin{equation}
\mathcal{L} (\theta | {\rm data})  = f_{sel} \cdot P(\hat{L},\hat{\mathcal{M}}, \mathcal{M}, L | \hat{z}, \theta, C) P(\hat{z} | C) P(\hat{t}_{\rm exp}) = f_{sel} \cdot P(\hat{L},\hat{\mathcal{M}} | \mathcal{M}, L, \hat{z}, \theta, C) \cdot P(\mathcal{M}, L| \hat{z}, \theta, C) \cdot P(\hat{z} | C) \cdot P(\hat{t}_{\rm exp})
\end{equation}
thus defining the modeling of of our data as:
\begin{equation}
f_{err} = P(\hat{L},\hat{\mathcal{M}} | \mathcal{M}, L, \hat{z}, \theta, C) = P(\hat{L},\hat{\mathcal{M}} | \mathcal{M}, L) \ ,
\end{equation}
where in the last equation we assume independence of the errors in our data from redshift, scaling relation parameters, and cosmology.

\noindent Then we isolate the dependent quantity $\mathcal{M}$ from the independent quantity in the scaling relation:
\begin{equation}
\mathcal{L} (\theta | {\rm data}) = f_{sel} \cdot f_{err} \cdot P(\mathcal{M}, L | \hat{z}, \theta, C) = f_{sel} \cdot f_{err} \cdot P(\mathcal{M} | L, \hat{z}, \theta, C) \cdot P(L | \hat{z}, C) \cdot P(\hat{z} | C) \cdot P(\hat{t}_{\rm exp}) 
\end{equation}

\noindent thus defining 
\begin{equation}
f_{sr} = P(\mathcal{M} | L, \hat{z}, \theta, C) = P(\mathcal{M} | L, \hat{z}, \theta) = \mathcal{LN} \left( \mu \cdot (L/L_{piv})^\gamma \cdot (\hat{z}/\hat{z}_{piv})^\beta, \sigma \right) \ ,
\end{equation}
where we have assumed that intrinsic morphological parameters are cosmology independent, and we have explicitly written the scaling relation in terms of the parameters $\theta$.

\noindent Finally we can write:
\begin{equation}
\mathcal{L} (\theta | {\rm data}) = f_{sel} \cdot f_{err} \cdot f_{sr} \cdot P(L | \hat{z}, C) \cdot P(\hat{z} | C) \cdot P(\hat{t}_{\rm exp}) \ ,
\end{equation}
where we can assume a scaling relation between mass and luminosity, which allows us to compute $P(L | \hat{z}, C)$ directly from the cluster mass function, since:
\begin{equation}
f_{c} = P(L | \hat{z}, C) = P(M | \hat{z}, C) \frac{dM}{dL} \ ,
\end{equation}

\noindent Therefore in the end we can summarize our likelihood as 
\begin{equation}
\mathcal{L} (\theta | {\rm data}) = P(\hat{L},\hat{\mathcal{M}}, \hat{z}, \hat{t}_{\rm exp}, \mathcal{M}, L, {\rm det = 1} | \theta, C) = f_{sel} \cdot f_{err} \cdot f_{sr} \cdot f_{c} \cdot P(\hat{z} | C) \cdot P(\hat{t}_{\rm exp}) \ .
\end{equation}
We finally point out that the $P(\hat{t}_{\rm exp})$ and $P(\hat{z} | C)$ terms are only a function of the data, and since we are not fitting cosmology, this is just a multiplicative factor in our likelihood, which therefore can be removed, since the fit is insensitive to such multiplicative factors independent of model parameters.

\noindent As a last step, we can marginalize over the intrinsic properties $L$ and $\mathcal{M}$ since we are not interested in estimating them, therefore 
\begin{equation}
\mathcal{L} (\theta | {\rm data})  \propto P(\hat{L},\hat{\mathcal{M}}, {\rm det = 1} | \hat{z}, \hat{t}_{\rm exp} \theta, C) =  \int_\mathcal{M}  \int_L 
P(L | \hat{z}, C) \cdot
P(\mathcal{M} | L,\hat{z},\theta) \cdot
P(\hat{L},\hat{\mathcal{M}} | L, \mathcal{M}) \cdot 
P({\rm det = 1} | L, \hat{z}, \hat{t}_{\rm exp}) d\mathcal{M} dL \ .
\label{eq:final_like_i}
\end{equation}
We point out that this is an improper likelihood, since it is not normalized, therefore we have to normalize it by calculating the denominator $\mathcal{D}$ by integrating the previous equation over $\hat{L}$ and $\hat{\mathcal{M}}$. However we immediately notice that such an integral applies only to the $f_{err}$ term, which is built to be normalized:

\begin{equation}
\int P(\hat{L},\hat{\mathcal{M}} | L, \mathcal{M}) d\hat{L} d\hat{\mathcal{M}} = 1 \ .
\end{equation}

\noindent This simplification of the likelihood allows us to isolate in the denominator the integral over $\mathcal{M}$:

\begin{align}
 \mathcal{D} = \int P(\hat{L},\hat{\mathcal{M}}, {\rm det = 1} | \hat{z}, \hat{t}_{\rm exp} \theta, C)  d\hat{L} d\hat{\mathcal{M}} &=    \int_L 
P(L | \hat{z}, C) \cdot
P({\rm det = 1} | L, \hat{z}, \hat{t}_{\rm exp})\cdot \left( \int_\mathcal{M}
P(\mathcal{M} | L,\hat{z},\theta) d\mathcal{M} \right) dL \nonumber \\
&= \int_L 
P(L | \hat{z}, C) \cdot
P({\rm det = 1} | L, \hat{z}, \hat{t}_{\rm exp}) dL \ .   
\end{align}

\noindent Therefore our denominator does not depend on the scaling relation parameters $\theta$. This implies that we don't need to compute this because it would just be a multiplicative constant in the likelihood, to which our fit is insensitive.

\noindent To conclude, in the previous calculation we have omitted the fact that this is the likelihood for a single cluster with observed quantities $\hat{L},\hat{\mathcal{M}}, \hat{z} = \hat{L}_{i}, \hat{Q}_{i}, \hat{z}_{i}$, where $i$ is a running index that goes through all detected clusters.
Therefore the total likelihood is:
\begin{equation}
\mathcal{L} = \prod_{i=1}^{N_{\rm clusters}}\mathcal{L}_i ( \theta | \hat{L}_{i}, \hat{Q}_{i}, \hat{z}_{i}) \ .
\label{eq:final_like}
\end{equation}

\end{appendix}
\end{document}